\documentclass[review]{elsarticle}

\usepackage{graphicx}
\usepackage[T1]{fontenc}
\usepackage{dsfont}
\usepackage{mathrsfs}
\usepackage{slashed}
\usepackage{amsmath}
\usepackage{amsfonts}
\usepackage{amssymb}
\usepackage{amsbsy}
\usepackage{booktabs}
\usepackage[table, xcdraw]{xcolor}
\usepackage{xspace}
\usepackage[colorlinks,citecolor=blue,linktoc=all,linkcolor=cyan]{hyperref}
\usepackage[capitalise]{cleveref}
\usepackage{threeparttable}
\usepackage{multirow}
\usepackage{enumerate}
\usepackage{textgreek}
\usepackage{threeparttable}
\usepackage{braket}
\usepackage[inline,shortlabels]{enumitem}
\usepackage{floatrow}
\usepackage{subfig}
\usepackage{caption}
\biboptions{sort&compress}
\usepackage{indentfirst}
\usepackage{rotating}

\usepackage{titlesec}
\usepackage{sectsty}
\titleformat{\section}{\normalfont\Large\bfseries}{\thesection}{1em}{}
\titleformat{\subsection}{\normalfont\large\bfseries}{\thesubsection}{1em}{}
\titleformat{\subsubsection}{\normalfont\normalsize\bfseries}{\thesubsubsection}{1em}{}

\numberwithin{equation}{section}
\numberwithin{table}{section}
\numberwithin{figure}{section}

\begin{document}

\begin{frontmatter}
    \title{Muon beams towards muonium physics: progress and prospects}

    \author[mymainaddress,mysecondaryaddress]{Siyuan Chen\fnref{fn1}}
    \author[mymainaddress,mysecondaryaddress]{Mingchen Sun\fnref{fn1}}
    \fntext[fn1]{These authors contributed equally to this work.}
    \author[mymainaddress,mysecondaryaddress]{Jian Tang\corref{mycorrespondingauthor}}
    \cortext[mycorrespondingauthor]{Corresponding author}
    \ead{tangjian5@mail.sysu.edu.cn}
    \address[mymainaddress]{School of Physics, Sun Yat-sen University, Guangzhou 510275, China}
    \address[mysecondaryaddress]{Platform for Muon Science and Technology, Sun Yat-sen University, Guangzhou 510275, China}

    \begin{abstract}
        Advances in accelerator technology have led to significant improvements in the quality of muon beams over the past decades. Investigations of the muon and muonium enable precise measurements of fundamental constants, as well as searches for new physics beyond the Standard Model. Furthermore, by utilizing muon beams with high intensity and polarization, studies of the dynamics of the muon and muonium at the atomic level can offer valuable insights into materials science. This review presents recent progress and prospects at the frontiers of muon beams and high-precision muonium physics. It also provides an overview of novel methods and detection techniques to achieve high sensitivities in different areas, including particle physics, nuclear physics, materials science and beyond.
    \end{abstract}


\end{frontmatter}

\newpage

\thispagestyle{empty}
\tableofcontents
\newpage

\section{Introduction}

Muon was first observed as an unexpected component of cosmic rays by P. Kunze in 1937~\cite{Kunze_1933}, and was later identified by C.~D.~Anderson and S.~H.~Neddermeyer in 1936~\cite{Anderson:1936zz,Neddermeyer:1937md}. Since then, research on muon physics has remained active ever since. It is a fermion with spin 1/2 and a unitary negative electric charge. In the Standard Model, the muon (symbol $\mu$) is the second-generation charged lepton, along with the electron ($e$, first generation) and the tau ($\tau$, third generation). With a mass of 105.6583755(23)~MeV$/c^2$~\cite{Tiesinga:2021myr}, the muon is about 207 times larger than the electron. Due to the reduced bremsstrahlung emission compared to electrons, it can pass through the materials more easily.
Moreover, it is an unstable particle with a lifetime of about 2.1969811(22) $\mu$s~\cite{MuLan:2012sih}, shorter only than that of the neutron ($\sim 878.4(5)$~s)~\cite{ParticleDataGroup:2024cfk}.
The relative long lifetime enables the exploration of muon decay processes with high-precision detectors.
These properties make the muon an excellent probe for frontier studies in both fundamental physics and applications. The muon predominantly decays through the weak interaction into an electron and two neutrinos, namely Michel decay:
\begin{equation}
    \begin{aligned}
        \mu^+\to e^+ + \nu_e + \overline{\nu_\mu}~, \\
        \mu^-\to e^- + \overline{\nu_e} + \nu_\mu~,
    \end{aligned}
\end{equation}
where the energy distribution of decay positrons follows the Michel spectrum with a maximum energy of 52.8 MeV.
Due to the V--A structure of the weak interaction, which leads to maximal parity violation~\cite{Lee:1956qn,Wu:1957my}, the angular distribution of decay positrons is asymmetrical and concentrated along the muon's spin direction. In 1957, R.~L.~Garwin \textit{et al.} observed parity violation in muon decay and suggested that the resulting asymmetry could serve as a powerful tool to probe magnetic fields in nuclei, atoms, and interatomic regions~\cite{Garwin:1957hc}. This work effectively predicted the wide applications of the muon spin rotation/relaxation/resonance ($\mu$SR) technique nowadays~\cite{amato2024introduction,Nuccio2014,Hillier2022}.
In addition to the standard Michel decay, the higher-order quantum electrodynamics (QED) corrections---the radiative decay $\mu^+\to e^+\nu_e\overline{\nu_\mu}\gamma$ and the internal conversion decay $\mu^+\to e^+\nu_e\overline{\nu_\mu} e^+ e^-$---have also been precisely measured~\cite{MEG:2016leq,SINDRUM:1985vbg}.
Interestingly, some neutrinoless decay or conversion modes of muon, such as $\mu^+\to e\gamma$~\cite{MEGII:2025gzr}, $\mu^+\to e^+e^-e^+$~\cite{Mu3e:2020gyw}, $\mu^+\to e^+\gamma\gamma$~\cite{Grosnick:1986pr}, $\mu^- N\to e^- N$~\cite{Mu2e:2014fns,COMET:2018auw}, and $\mu^+e^-\to\mu^-e^+$~\cite{Bai:2024skk}, which are also known as the charged lepton flavor violation processes, are predicted by various new physics models~\cite{Calibbi:2017uvl}.

There are two approaches by which one can obtain muons---the cosmic-ray showers and the accelerators. The dominant muon-producing channel is the two-body decay of a pion:
\begin{equation}
    \begin{aligned}
        \pi^+\to\mu^+ +\nu_{\mu}~, \\
        \pi^-\to\mu^- +\overline{\nu_{\mu}}~.
    \end{aligned}
\end{equation}
In the decay of a pion at rest, the spin of $\mu^\pm$ is anti-parallel to its momentum direction. This is a direct consequence of angular momentum conservation and the V--A structure of the weak interaction, which produces only left-handed neutrinos. This intrinsic property enables the production of muons with nearly 100\% polarization under appropriate conditions.
Because of the absorption and decay in the atmosphere, the cosmic-ray muon have a low flux ($\sim 1~\text{cm}^{-2}\cdot\text{min}^{-1}$), and a broad energy spectrum at sea level. Their polarization is also substantially reduced during propagation through the atmosphere~\cite{sun2025cosmicraymuonpolarization}. With the development of accelerator technology, accelerators first reached the energy threshold for pion production in 1948~\cite{Gardner:1948pkt}. To meet the rapidly increasing demand from the muon science community, muon beamlines based on high-intensity accelerators have been developed over the past decades. Taking advantage of the trend toward more focused energy and beam spot distribution, higher fluxes and polarization, research involving muon beams has been further expanded to areas beyond particle physics~\cite{Crivelli:2018vfe,Mu2e:2014fns,Abe:2019thb,Mu3e:2020gyw,Accettura:2023ked,Bai:2024skk,MuSEUM:2025cmo,MEGII:2025gzr,Muong-2:2025xyk,Adelmann:2025nev}, including applications in materials science~\cite{ito2019investigation, nuccio2014muon, mcclelland2020muon, blundell2025muon}, tomography~\cite{Nagamine:1995np,Nagamine2005MuonRadiographyBlastFurnace,MORRIS_2008,yu2024proposed, gamage2022portable, paccagnella2025exploring,Xu:2024btf,Yu:2025ufj,Ning:2025hai}, and others~\cite{Gorringe:2015cma,An:2025lws}. Evidently, the development of muon beams has served as an essential gateway to muon science. For a comprehensive review of muon physics, the authoritative three-volume series edited by V. W. Hughes and C. S. Wu can be consulted~\cite{HughesWu1975MuonPhysicsV1,HughesWu1975MuonPhysicsV2,HughesWu1975MuonPhysicsV3}.

\begin{figure}[t!]
    \centering
    \includegraphics[width=0.8\textwidth]{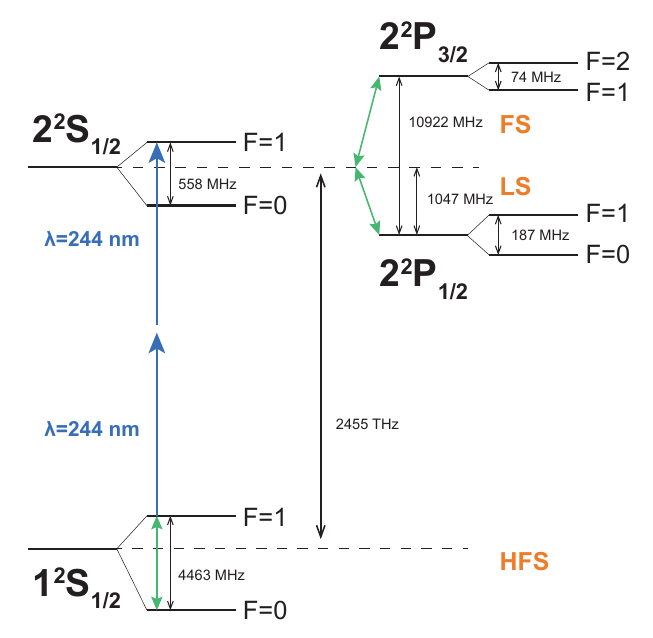}
    \caption{Schematic of the muonium energy levels (not to scale), illustrating the hyperfine structure (HFS), fine structure (FS), Lamb shift (LS) of $n=1, n=2$ states and related transition frequencies.}
    \label{fig:energy_level}
\end{figure}

Muonium ($\text{M}\equiv\mu^+e^-$) is a unique system in muon physics~\cite{hughes2012muon}. As a coulombic bound state consisting of two point-like leptons---a positive muon ($\mu^+$) and an electron ($e^-$)---it can be regarded as a purely leptonic analogue of hydrogen.
In 1960, V.~W.~Hughes \textit{et al.} first observed the formation of muonium in pure argon gas via its characteristic Larmor precession frequency~\cite{Hughes:1960zz}.
To date, a broad platform for muonium physics has been established.
Compared to positronium ($e^+e^-$), muonium offers advantages of a higher mass and a longer natural lifetime.
As a purely electromagnetic two-body system free from any hadronic components, it avoids the complexities associated with hadronic and nuclear structures present in ordinary atoms.
A complete QED calculation of its energy levels is allowed, as shown in \cref{fig:energy_level}.
Transitions between certain quantum states can be induced experimentally using microwave or laser spectroscopy.
Therefore, it is an ideal probe in cutting-edge research areas, including atomic spectroscopy, precision tests of QED and searches for physics beyond the Standard Model (see also comprehensive reviews of K.~P.~Jungmann~\cite{Jungmann:2004sa,Jungmann:2016gak} and T.~P.~Gorringe~\cite{Gorringe:2015cma}).
Beyond fundamental physics, muonium finds diverse applications.
In condensed matter physics, muonium is more likely to be formed in semiconductors or insulators, enabling $\mu$SR investigation of weak magnetic fields and hydrogen impurity in such materials~\cite{PhysRevB.92.081202}. In physical chemistry, muonium is crucial for characterizing free radicals by $\mu$SR~\cite{McKenzie2013}.

The continuous improvement of muon beamline technology has advanced not only fundamental research on muons and muonic atoms but also the development of muon-based experimental techniques.
Among these, $\mu$SR and muon-induced X-ray emission (MIXE) are notable for exploiting the high polarization and intensity of modern muon beams.
Originating from early studies of weak-interaction symmetry and muonic atoms in the 1950s--1970s~\cite{Garwin:1957hc, Reidy:1975xr}, these methods have since evolved into versatile probes of condensed matter and materials.
Recent advances in beam intensity, detector resolution, and electronics have transformed $\mu$SR into a sensitive tool for exploring magnetic ordering~\cite{ ruderman1954indirect, kasuya1956theory, yosida1957magnetic, ruegg1981muon, le1993searching, blundell1995mu+, andreica2001magnetic, aeppli1988magnetic}, superconductivity~\cite{bauer2014absence, bauer2010unconventional,sonier2000musr, maisuradze2009comparison,sanna2004nanoscopic, luetkens2009electronic}, spin dynamics~\cite{frandsen2016volume, amato2004weak, bourdarot2005hidden, aoki2003time, spehling2012magnetic, hartmann1991asymmetric}, and hydrogen behavior~\cite{cox2003shallow,cox2009muonium,patterson1988muonium}, while MIXE has become increasingly effective for non-destructive elemental and structural analysis.
Compared with conventional techniques such as Nuclear magnetic resonance spectroscopy (NMR), Electron Paramagnetic Resonance (EPR), X-Ray fluorescence (XRF), and Energy Dispersive Spectroscopy (EDS), $\mu$SR and MIXE provide superior sensitivity to weak magnetic fields, dynamic processes, and elemental composition, extending experimental access to regimes previously inaccessible.

This article provides a brief review of the current status of experimental studies on muonium and its detection. It begins by introducing the global muon beam facilities, including their key parameters and scientific applications, followed by a summary of the next-generation muon sources and a discussion of future prospects of muon beam techniques. In the context of muonium physics, major experimental topics include searches for new physics via muonium-to-antimuonium conversion,high-precision measurements of the muonium Lamb shift, the 1S-2S transition, and the hyperfine structure, as well as investigations of antimatter gravity. For each area, recent advances in experimental techniques and measurement precision are discussed, and future proposals are also reviewed.
Finally, technologies for various muon- or muonium-based applications (such as $\mu$SR and MIXE techniques) are summarized, along with an outline of their basic principles and recent progress.

\section{Status of muon beams}

\subsection{Overview}
\begin{figure}[t!]
    \centering
    \includegraphics[width=0.7\textwidth]{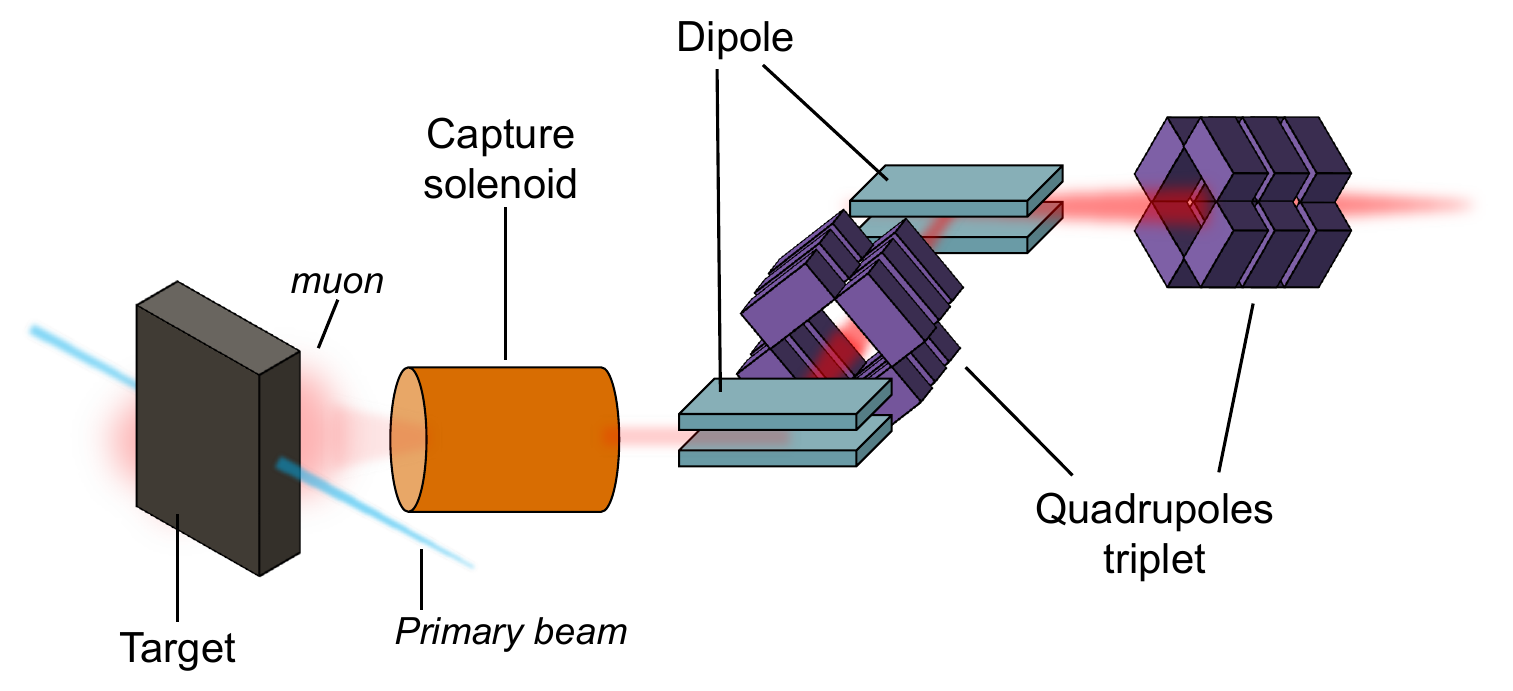}
    \caption{Schematic of the main components in a muon beamline. The muons produced from the target are first captured and focused by the capture solenoid. The dipole magnet bends and selects muons with specific momenta. The quadrupole triplets are used to further focus the beam and transport muons to the experimental stations.}
    \label{fig:muon_beamline}
\end{figure}

In most of the current and planned accelerator muon sources, muons are typically produced via pion decays, where pions are generated by the interaction of high-energy (over hundreds of MeV) protons with a target~\cite{nagamine2003introductory}. Considering a $\pi^+$ decaying at the surface of the target, the escaping $\mu^+$ is called a surface muon, which is monoenergetic ($\sim$29.8~MeV/$c$) and almost 100\% polarized. By further moderating surface muons using a solid noble gas mixture~\cite{Prokscha:2008zz} or laser ionization~\cite{Nagatomo:2014bja}, slow muons with energies of a few keV can be produced. Another scenario is that a pion escapes the target and decays in flight. The produced muon is called a decay muon or cloud muon, which has a higher energy (over 4.1 MeV) than the surface muon. The typical setup of a muon beamline can be described by~\cref{fig:muon_beamline}. After a high-energy, high-intensity primary beam hits the production target, muons are produced with a large emittance. Therefore, it is essential to design an appropriate capture solenoid to achieve the required muon intensity. The captured muons are then transported to the experimental stations by dipole magnets for the bending purpose and quadrupole-magnet triplets for transverse focusing. According to the time structure of the primary beam, muon beams can be classified as continuous-wave (CW) or pulsed-wave (PW).

\begin{figure}[t!]
    \centering
    \includegraphics[width=0.5\textwidth]{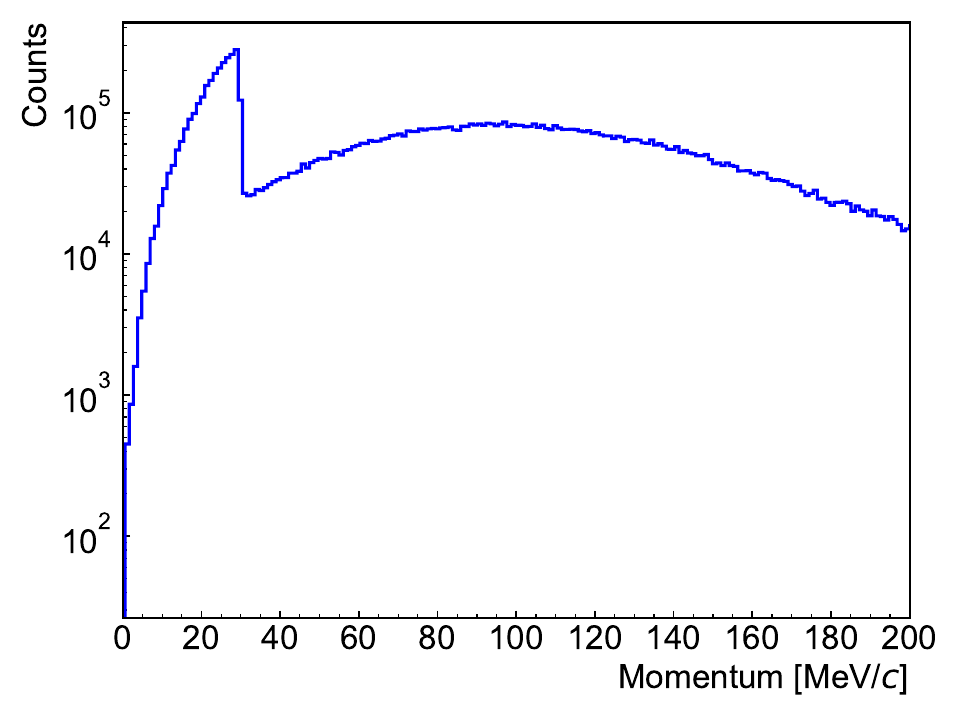}
    \caption{Simulated momentum spectrum of muons detected by a virtual detector near the production target. The peak at about 30 MeV is from the surface $\mu^+$s.}
    \label{fig:muon_spectra}
\end{figure}

In the field of muon physics, surface muon beams have found the widest range of applications. The invention of the surface muon beam dates back to the 1970s. At that time, T.~Bowen \textit{et al.} stood out to search for the muonium-to-antimuonium conversion process suggested by G.~Feinberg and S.~Weinberg~\cite{Feinberg:1961zz,Feinberg:1961zza}. To form muonium, muons have to be stopped in the production target. However, only decay muons with large momenta were available in the original beamline design. T.~Bowen proposed that muons produced from pions decaying at rest in a thin target are monoenergetic and almost 100\% polarized, carrying low momenta as shown in~\cref{fig:muon_spectra}. Eventually, they succeeded in constructing the first surface muon beamline at Lawrence Berkeley National Laboratory (LBNL)~\cite{Pifer:1976ia,Bowen:1985qf}.

\begin{figure}[t!]
    \centering
    \includegraphics[width=\textwidth]{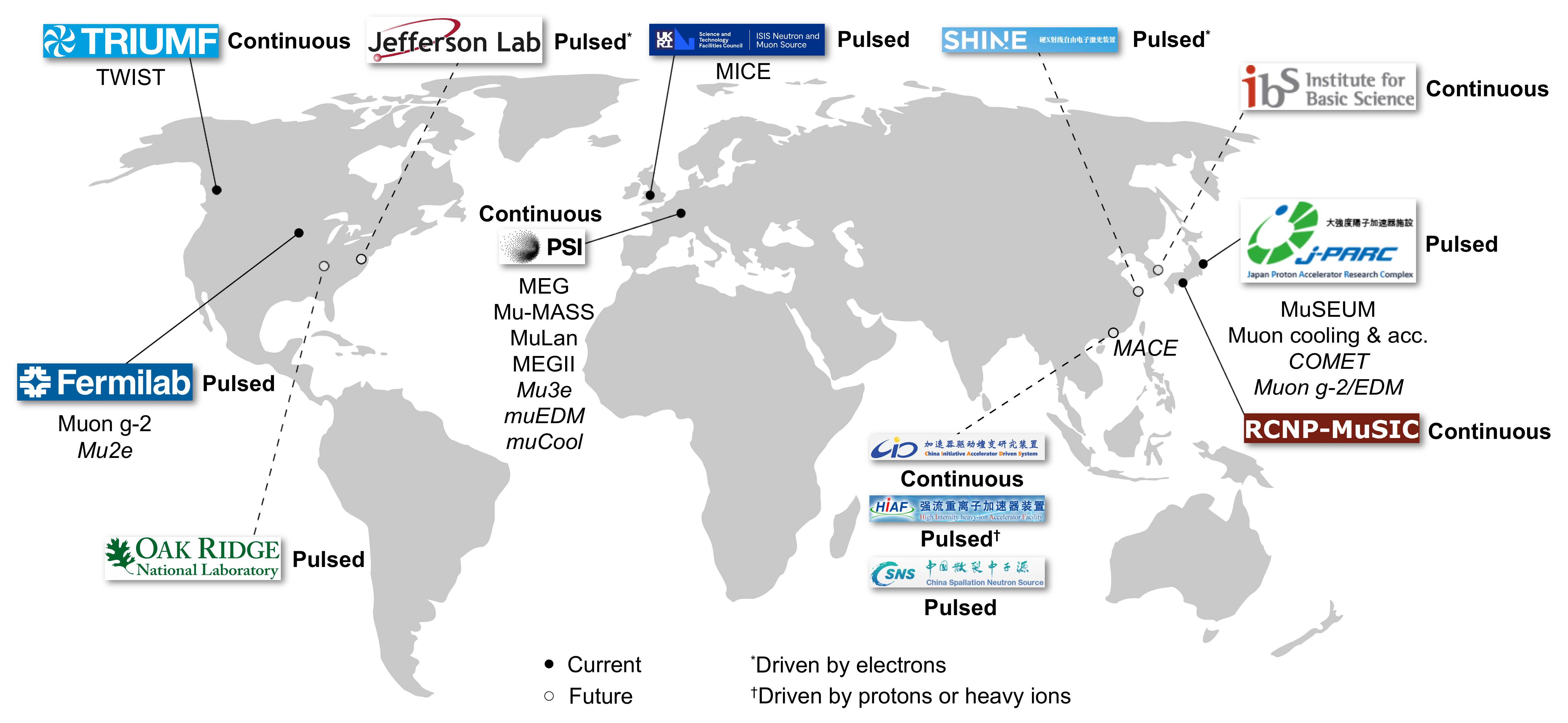}
    \caption{Map of current or future muon facilities around the world, which shows their rough locations, with the operating modes highlighted in \textbf{bold}. Solid circles with solid lines stand for existing facilities, while open circles with dashed lines indicate facilities under construction or in the planning stage. Representative experiments conducted at these muon sources are labeled in small text, with next-generation experiments in \textit{italics}.}
    \label{fig:facilities}
\end{figure}

Accelerator facilities capable of delivering high-intensity protons or other particles are essential for the construction of muon beamlines. The major facilities currently in operation worldwide include Paul Scherrer Institute (PSI) in Switzerland, J-PARC (Japan Proton Accelerator Research Complex) in Japan, Research Center for Nuclear Physics (RCNP) in Japan, Rutherford Appleton Laboratory (RAL) in the UK, TRIUMF in Canada, and Fermi National Accelerator Laboratory (FNAL or Fermilab) in the USA, all of which are supported by advanced accelerator complexes. After years of progress in muon science, efforts are now underway to develop next-generation muon beam facilities in this century. Ongoing construction projects for new accelerator complexes include the China Spallation Neutron Source (CSNS), the China initiative Accelerator Driven System (CiADS), High Intensity heavy-ion Accelerator Facility (HIAF) in China, Rare isotope Accelerator complex for ON-line experiment (RAON) in Korea, Spallation Neutron Source (SNS) in the USA, and future muon collider concept at European Spallation Source (ESS) in Sweden~\cite{Abele:2022iml,Alekou:2022emd}. All of these facilities have proposed plans to construct dedicated muon sources, advancing the frontiers of muon science. Geographical locations and general features of these muon facilities are plotted on a world map in~\cref{fig:facilities}. This section provides a brief review of the key parameters and scientific cases of some major muon sources, spanning developments from the recent past to the foreseeable future.

\subsection{S$\mu$S at PSI}\label{sec:smus}
\begin{figure}[t!]
    \centering
    \includegraphics[width=\textwidth]{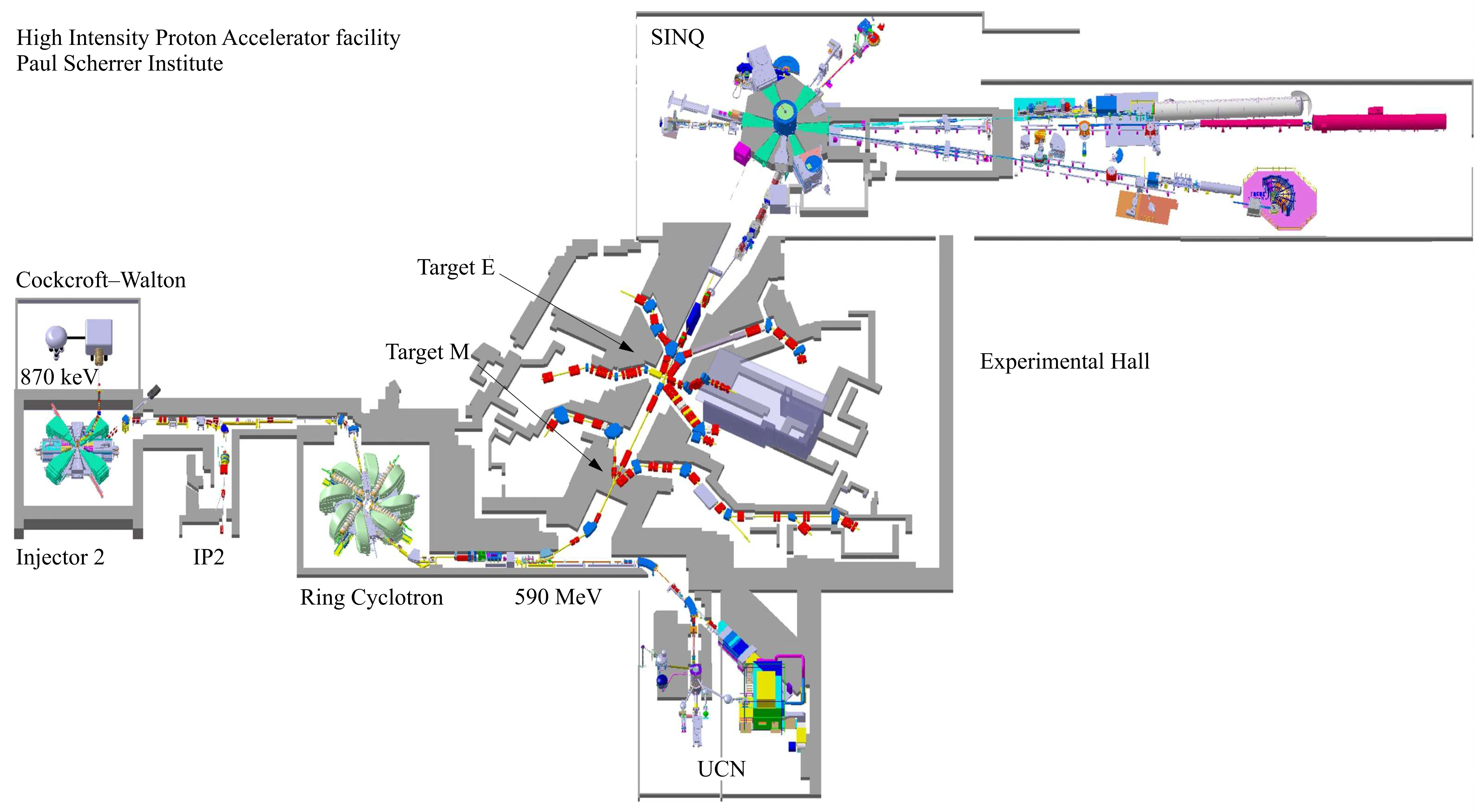}
    \caption{Layout of the High Intensity Proton Accelerator at PSI (reproduced from Ref.~\cite{Grillenberger:2021kyv}).}
    \label{fig:hipa}
\end{figure}

The Swiss Muon Source (S$\mu$S) is based on the PSI High Intensity Proton Accelerator (HIPA), which operates at a beam current of up to 2.4 mA with a power of up to 1.4~MW~\cite{Grillenberger:2021kyv}. The protons injected into the Ring cyclotron are accelerated by 50.6~MHz cavities to 590~MeV and finally sent to the meson production target stations, Target M and Target E, as plotted in \cref{fig:hipa}~\cite{Kiselev:2021pwl}. Both targets are made of polycrystalline graphite and rotate at 1 turn per second for heat dissipation. Target M measures 320 mm in diameter and 5.2 mm thick, while Target E is 450~mm in diameter and 40~mm or 60 mm thick. These target stations provide pions or muons for seven secondary beamlines, including $\pi$M1, $\pi$M3, $\pi$E1, $\pi$E3, $\pi$E5, $\mu$E1, and $\mu$E4. The detailed performances of these beamlines are listed in \cref{tab:smus}.

\begin{table}[!h]
    \caption{Performance of S$\mu$S beamlines~\cite{Grillenberger:2021kyv}.}
    \centering
    \label{tab:smus}
    \begin{tabular}{lccccccc}
        \hline
                                                 & $\pi$M1       & $\pi$M3       & $\pi$E1       & $\pi$E3       & $\pi$E5       & $\mu$E1       & $\mu$E4       \\ \hline
        Target                                   & M             & M             & E             & E             & E             & E             & E             \\
        Particle                                 & $e/\pi/\mu/p$ & $\mu$         & $\pi/\mu/p$   & $\mu$         & $\pi/\mu$     & $\mu$         & $\mu$         \\
        Momentum (MeV/$c$)                       & 10--450       & 10--40        & 10--450       & 10--40        & 10--120       & 60--120       & 10--40        \\
        Max Rate ($\text{s}^{-1}\text{mA}^{-1}$) & $2\times10^8$ & $3\times10^6$ & $1\times10^9$ & $3\times10^7$ & $5\times10^8$ & $6\times10^7$ & $4\times10^8$ \\ \hline
    \end{tabular}%
\end{table}

A total of five beamlines and six instruments are allocated for $\mu$SR studies, providing a wide range of experimental conditions including muon energy (0.5~keV--60 MeV), temperature (0.01--1200 K), magnetic field ($\leq9.5$ T), and pressure ($\leq2.5$~GPa). Furthermore, a low-energy muon beam (LEM) can be generated at the $\mu$E4 beamline by moderating the surface muon beam down to keV energies~\cite{Prokscha:2008zz,Harshman:1987zz}. This enables depth-selective ($ < $ 200 nm) $\mu$SR investigations of material surfaces and interfaces. Studies on muonium spectroscopic measurements are also ongoing at the LEM facility~\cite{Crivelli:2018vfe}. The $\pi$E5 beamline is notable for being the most intense CW surface muon beam in the world, making it popular for particle physics research. Representative experiments include SINDRUM II~\cite{Ahmad:1988ur,SINDRUMII:1993gxf,SINDRUMII:1996fti,SINDRUMII:1998mwd,SINDRUMII:2006dvw}, MACS~\cite{Willmann:1998gd,Willmann:2021boq}, MEG~\cite{MEG:2009vff,MEG:2011naj,MEG:2013oxv,MEG:2016leq}, MEG II~\cite{MEGII:2018kmf,MEGII:2023ltw,MEGII:2025gzr}, and Mu3e~\cite{Mu3e:2020gyw}, which have established, or are about to establish, increasingly stringent limits on cLFV processes. To improve the phase space quality of muon beams, the muCool project~\cite{Antognini:2021fae} is being carried out at the $\pi$E5 beamline, aiming to serve next-generation muon experiments and $\mu$SR instruments~\cite{Accettura:2023ked}.

\begin{figure}[t!]
    \centering
    \includegraphics[width=0.9\textwidth]{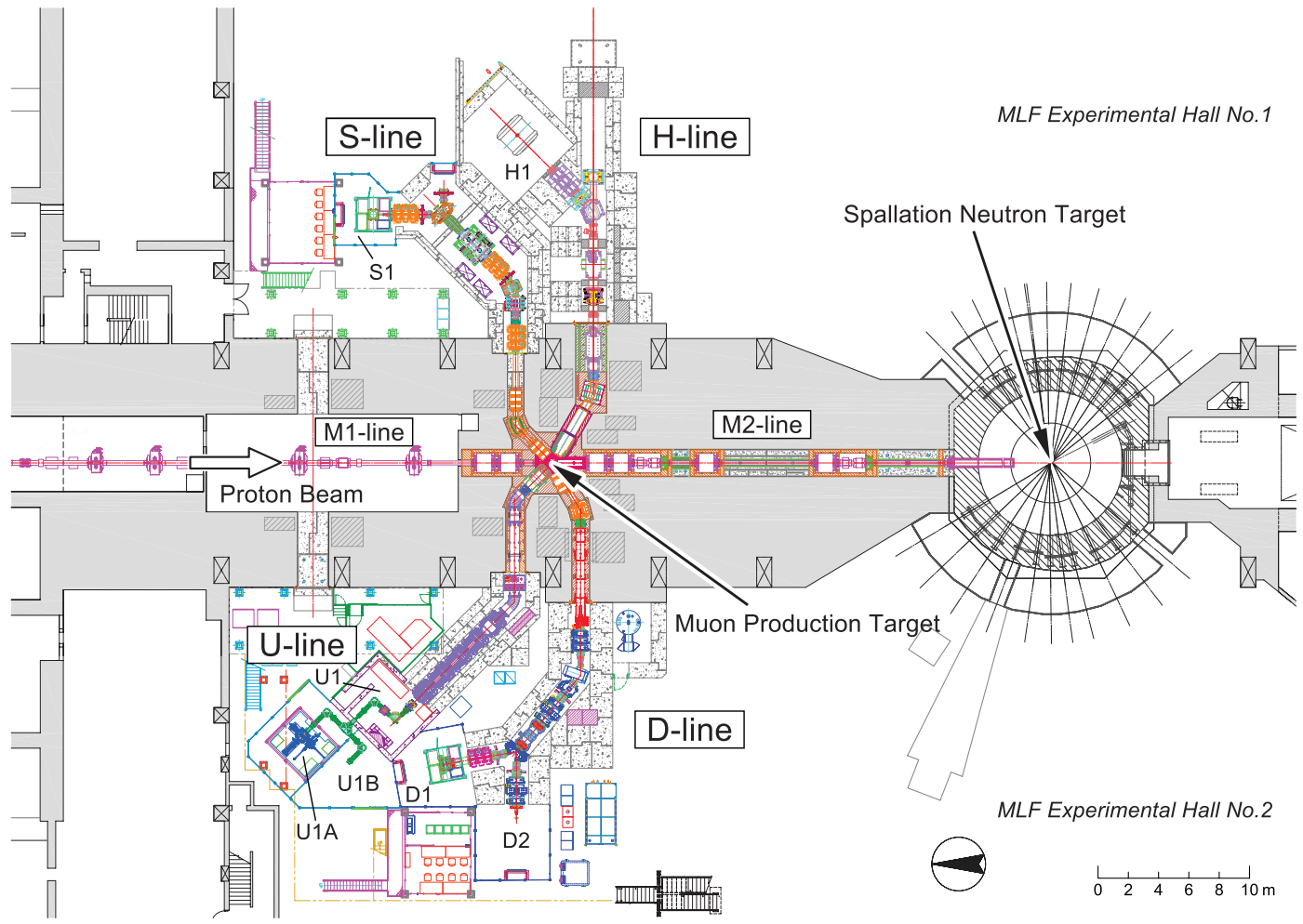}
    \caption{Layout of the MUon Science Establishment at J-PARC (reproduced from Ref.~\cite{Kawamura:2018apy}).}
    \label{fig:muse}
\end{figure}

\subsection{MUSE at J-PARC}
The J-PARC consists of a linear accelerator, a rapid-cycling synchrotron (RCS), and a main ring (MR)~\cite{Nagamiya:2012tma}. The RCS finally provides protons with an energy of 3 GeV at a 25 Hz repetition rate for the Materials and Life Science Experimental Facility (MLF), where the MUon Science Establishment (MUSE) is located. MUSE utilizes a rotating graphite target similar to the S$\mu$S's to produce muons, and contains four muon beamlines, including D-line, U-line, S-line, and H-line~\cite{Shimomura:2024puh}, as plotted in \cref{fig:muse}.

The D-line delivers both positive/negative decay muons and surface muons to D1 and D2 areas. The high-energy (3 GeV) proton beam produces a high yield of negative muons. Recent applications include non-destructive elemental analysis, precision spectroscopy of muonic atoms, and related areas. Slow muons with energies ranging from 50 eV to 30~keV can be produced at the U-line through laser ionization of muonium, which is formed by stopping surface muons in materials~\cite{Kanda:2023gqp}. This opens up the opportunity for conducting research in areas similar to those pursued at the LEM facility. At the S-line, the PW surface muon beam with a flux of $4.6\times10^5~\mu^+/\text{s}$ is used for $\mu$SR studies. It is worth mentioning that the first RF acceleration of muons has been achieved at this beamline~\cite{Aritome:2024rlu}. In the other area of S-line, the measurement of muonium 1S-2S transition using laser spectroscopy will also be carried out~\cite{Zhang:2021cba}. The H-line offers higher beam transport efficiency through the use of solenoids, making it suitable for the future cLFV search experiment---Direct Emission of Electrons from Muon to Electron conversions (DeeMe)~\cite{Natori:2014yba}, Muonium Spectroscopy Experiment Using Microwave (MuSEUM)~\cite{MuSEUM:2025cmo}, muon cooling and acceleration~\cite{Kamioka:2023xob}, muon $g-2$ measurement~\cite{Abe:2019thb}, and the world's first muon accelerator~\cite{Aritome:2024rlu}.

\begin{figure}[t!]
    \centering
    \includegraphics[width=0.8\textwidth]{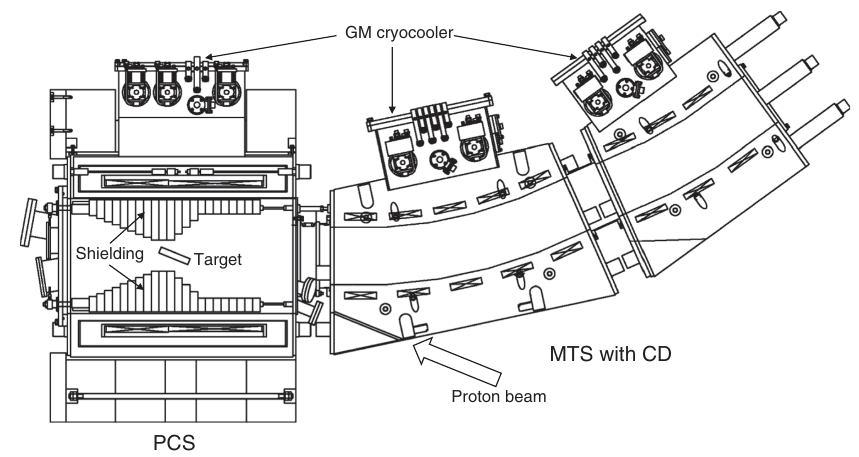}
    \caption{Layout of the MUon Science Innovative Channel at RCNP (reproduced from Ref.~\cite{Cook:2016sfz}). Secondary particles are collected by a pion capture solenoid (PCS). The muons are then transported through two sets of muon transport solenoids (MTS), with each MTS coil integrating a correction dipole (CD). All superconducting coils along the beamline are cooled using Gifford-McMahon (GM) cryocoolers.}
    \label{fig:music}
\end{figure}

\subsection{MuSIC at RCNP}\label{sec:music}
The MUon Science Innovative Channel (MuSIC) facility was constructed at the RCNP, the University of Osaka, to demonstrate the superconducting solenoid magnets capture scheme~\cite{Hino:2014bpx,Cook:2016sfz}, as shown in~\cref{fig:music}.
At MuSIC, a 400~W proton cyclotron accelerator accelerates the protons to 400 MeV, which are then delivered to a graphite cylindrical target with a length of 20 cm and a radius of 2 cm. The target is located at the center of the capture solenoid (PCS).

Pions and muons emitting backward are captured using a 3.5 T magnetic field. The captured muons are subsequently focused by a graded magnetic field and injected into a curved muon transport solenoid (MTS), which selects muons in the 20--60~MeV$/c$ momentum range. Measurements of muon lifetime and muonic X-rays indicate an available muon yield of approximately $4\times10^8~\mu^+/\text{s}$ with a 400 W proton beam. This high intensity mainly results from the excellent collection efficiency of the PCS. However, the spin polarization rate of muons could only reach 60\% due to the large acceptance of the solenoid~\cite{Tomono:2018lar}. The capture scheme of MuSIC is also applicable to future high-intensity muon facilities, including $\mu-e$ conversion experiment, muon collider, and neutrino factories.

\begin{figure}[t!]
    \centering
    \includegraphics[width=0.9\textwidth]{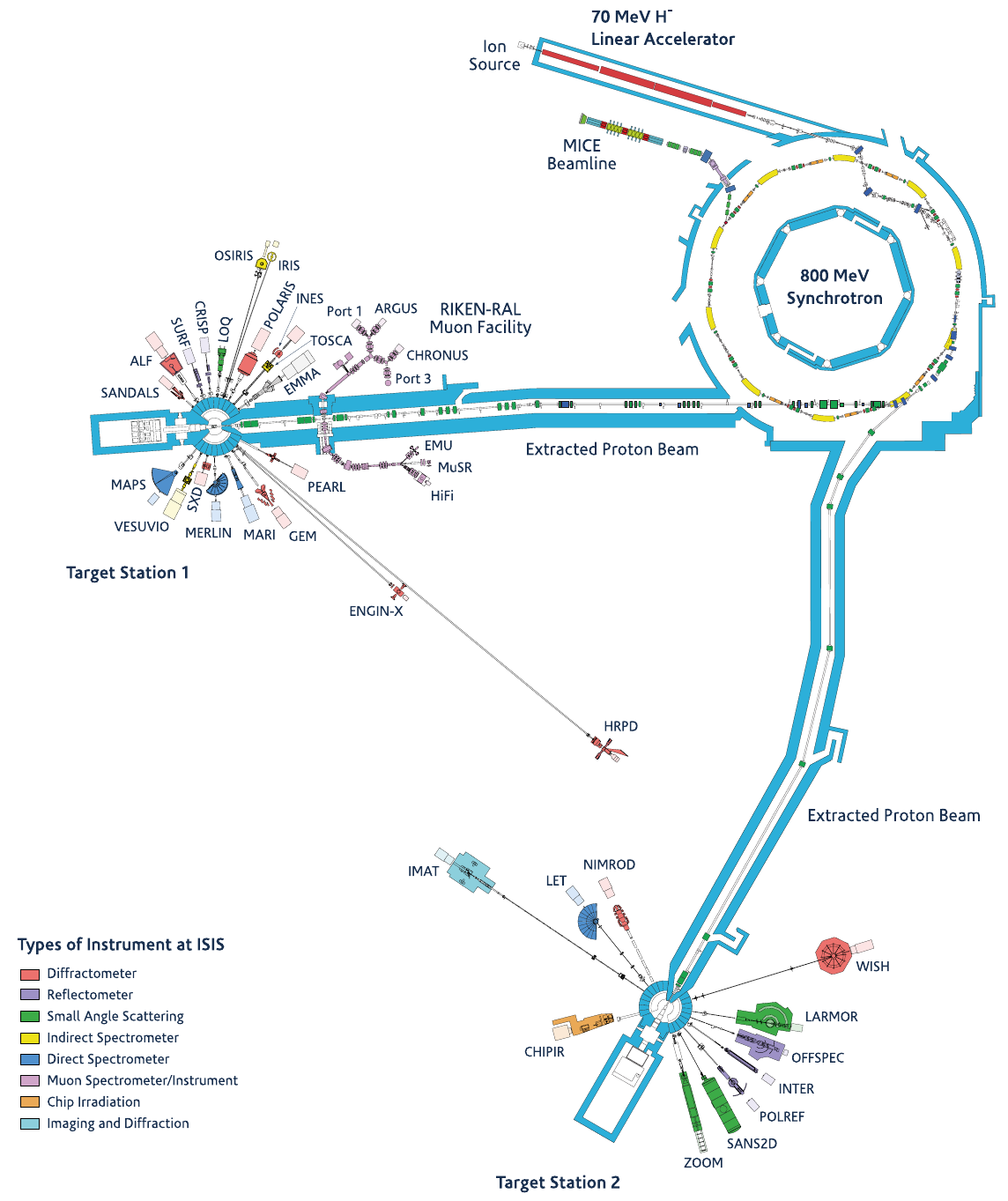}
    \caption{Layout of the ISIS Pulsed Neutron and Muon Source at RAL (reproduced from Ref.~\cite{Thomason:2019fwe}).}
    \label{fig:isis}
\end{figure}

\subsection{ISIS at RAL}
The ISIS Pulsed Neutron and Muon Source, located at the RAL, has operated muon facilities for over 30 years~\cite{Thomason:2019fwe}. The proton accelerator consists of a Radio Frequency Quadrupole (RFQ), a linear accelerator, and a synchrotron, delivering 800 MeV protons to the muon target station at 50 Hz. The target is made of graphite and has a thickness of 10~mm along the beam direction. The muon target station provides 28 MeV/$c$ surface muons to the European Commission facility, as well as decay muons (positive or negative) with momenta ranging from 17 to 120 MeV/$c$ to the RIKEN-RAL facility~\cite{Matsuzaki:2001kj,Hillier:2019szm}. There is also a dedicated port for slow muon research at the RIKEN-RAL facility, but the intensity is rather low~\cite{Matsuda:2001vhp}. The layout of the instruments at ISIS Pulsed Neutron and Muon Source is shown in \cref{fig:isis}.

The EC muon facility is equipped with three $\mu$SR instruments, EMU ($5\times10^5~\mu^+/\text{s}$, 5000~G, 50~mK--1500 K), HiFi ($5\times10^5~\mu^+/\text{s}$, 5 T, 30 mK--1500 K), and MuSR ($1\times10^6~\mu^+/\text{s}$, 3000 G, 40 mK--1000 K). These instruments offer a variety of experimental conditions for general-purpose $\mu$SR studies. The RIKEN-RAL facility includes four ports used for different scenarios, including muon catalyzed fusion, proton radius experiment, elemental analysis, and $\mu$SR. Notably, another muon production target is installed in the synchrotron, with a dedicated beamline directing muons to the Muon Ionization Cooling Experiment (MICE) hall for the demonstration of ionization cooling~\cite{MICE:2019jkl}.

\begin{figure}[t!]
    \centering
    \includegraphics[width=0.9\textwidth]{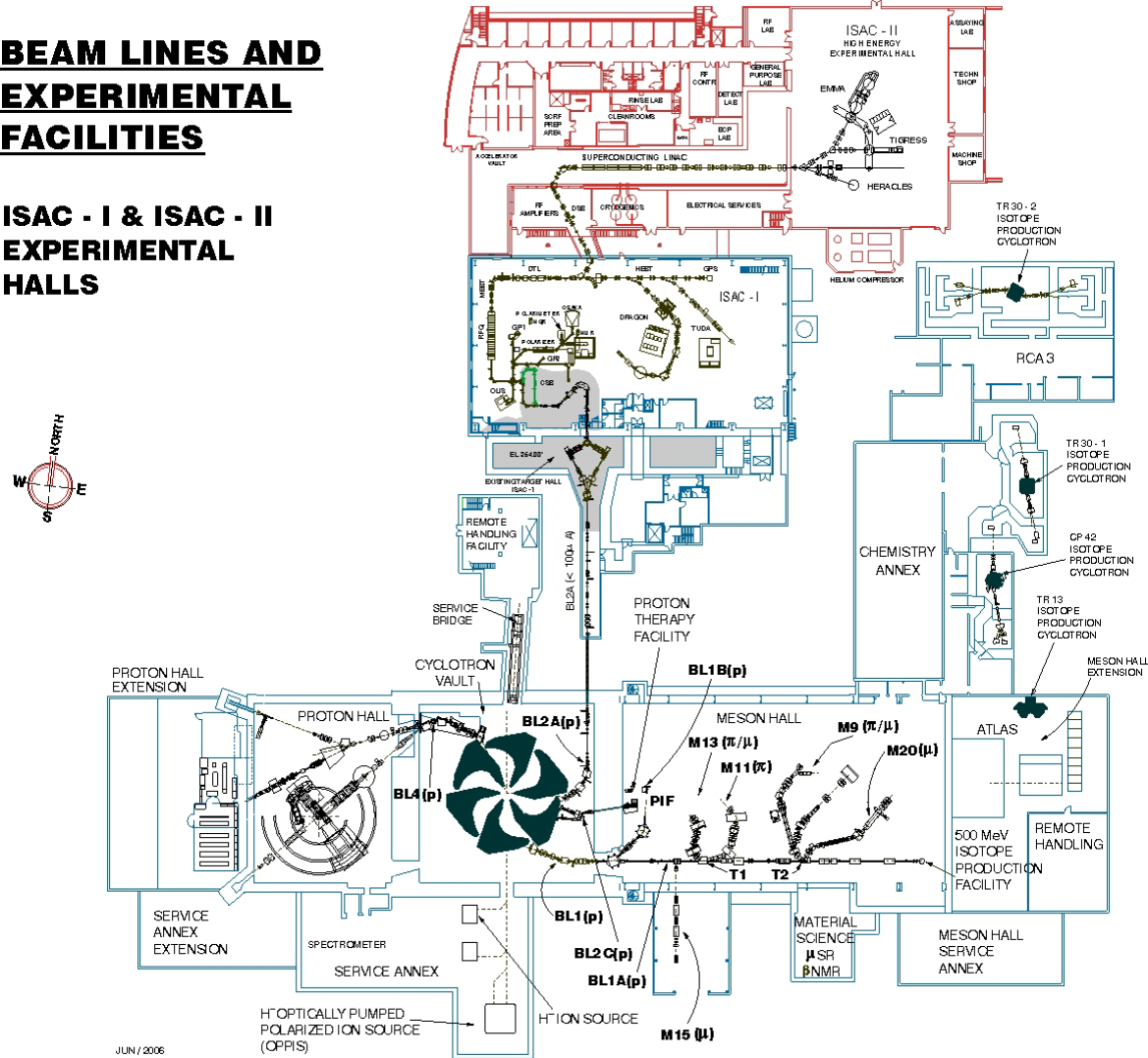}
    \caption{Layout of beamlines and instruments at TRIUMF (reproduced from Ref.~\cite{Marchetto:2011kh}).}
    \label{fig:cmms}
\end{figure}

\subsection{CMMS at TRIUMF}
The first extraction of the beam from the TRIUMF cyclotron took place in 1974. It delivers 520 MeV protons to the muon production targets T1 and T2 with a beam power of 75~kW~\cite{beveridge1986muon,987839} to the Centre for Molecular and Materials Science (CMMS). Detailed beamline layout can be seen in \cref{fig:cmms}. The M15 beamline provides surface muons from the 1AT1 target, typically made of 1 cm graphite or beryllium. A permanent quadrupole doublet located close to the target enables efficient muon collection, but limits the beam momentum range to approximately 19--40 MeV/$c$. M20 extracts surface muons from a 10~cm thick beryllium block target called 1AT2. The M9 beamline branches from the 1AT2 target into two legs. M9A, equipped with an ultra-fast electrostatic, is being rebuilt as a surface muon beamline dedicated to $\mu$SR experiments. M9H will set up a persistent superconducting solenoid where pions decay in flight to produce high-energy $\mu^+$ and $\mu^-$ beams. It is optimized for transverse-field $\mu$SR measurements across a wide energy range. It will also support high-pressure liquid or gas sample environments under extreme conditions ($\leq$1000~K and $\leq$ 0.6~GPa).

\begin{figure}[t!]
    \centering
    \includegraphics[width=0.9\textwidth]{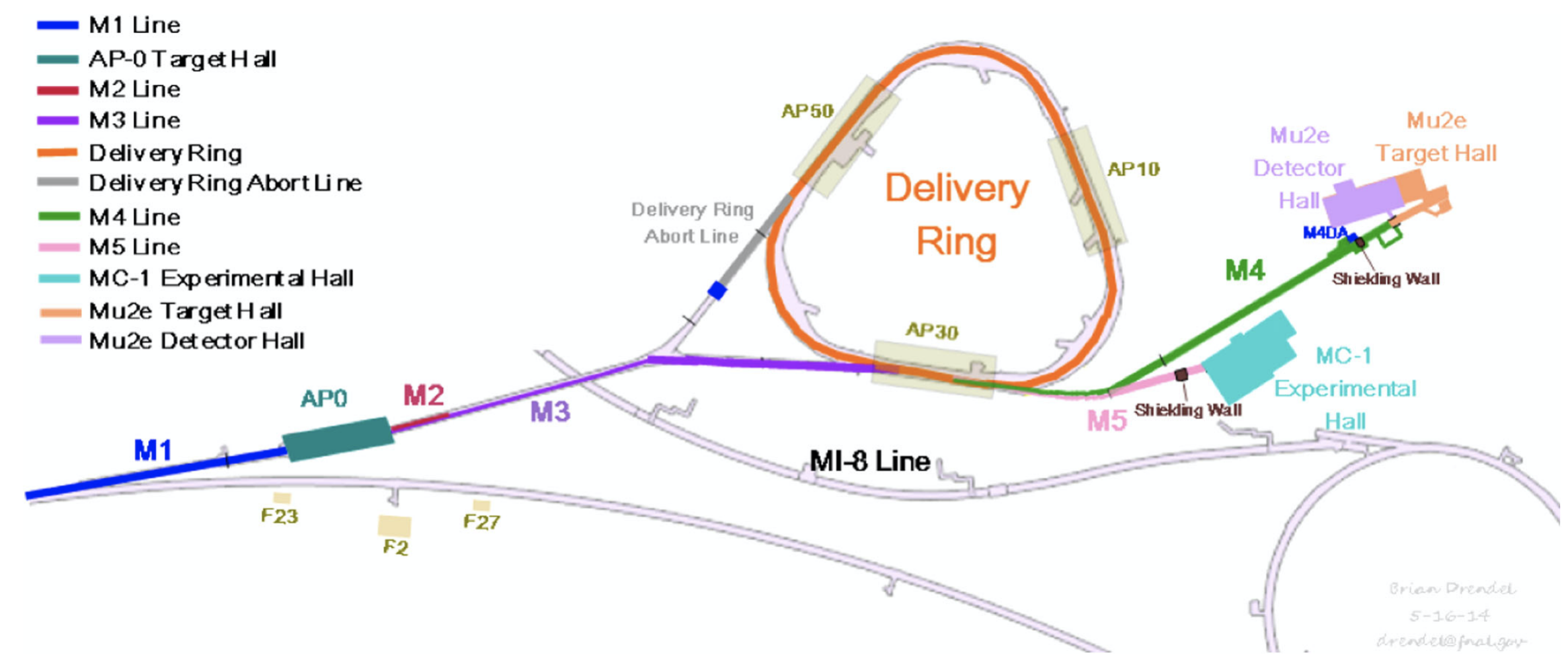}
    \caption{Layout of the Muon Campus at Fermilab (reproduced from Ref.~\cite{Aiba:2021bxe}).}
    \label{fig:campus}
\end{figure}

\subsection{Muon Campus at FNAL}
The proton accelerator system at the Muon Campus (as shown in \cref{fig:campus}) delivers an 8.89 GeV/$c$ proton beam supplied by the Recycler ring to the AP0 target station initially~\cite{Ganguly:2022ufq}. The Muon Campus hosts both the Muon $g-2$~\cite{Muong-2:2025xyk} and Mu2e~\cite{Mu2e:2014fns} experiments, aiming at the discovery of BSM physics. For the Muon $g-2$ experiment, the 3.1 GeV muon beam is injected into the Delivery Ring (DR). After 4 turns in the DR, protons are kicked out to the beam dump, and the remaining muon beam is extracted to the Muon $g-2$ storage ring. For the Mu2e experiment, the proton beam passes the AP0 target and enters the DR directly. The DR will resonantly extract the proton beam to the Mu2e target, producing a PW muon beam. Additionally, several potential experiments at FNAL, including fixed-target muon experiments and muon EDM studies, are also proposed, considering the Proton Improvement Plan-II (PIP-II)~\cite{Stanek:2023bae}.

\subsection{Next-generation muon facilities}
\subsubsection{HIMB at PSI}

\begin{figure}[t!]
    \centering
    \includegraphics[width=0.9\textwidth]{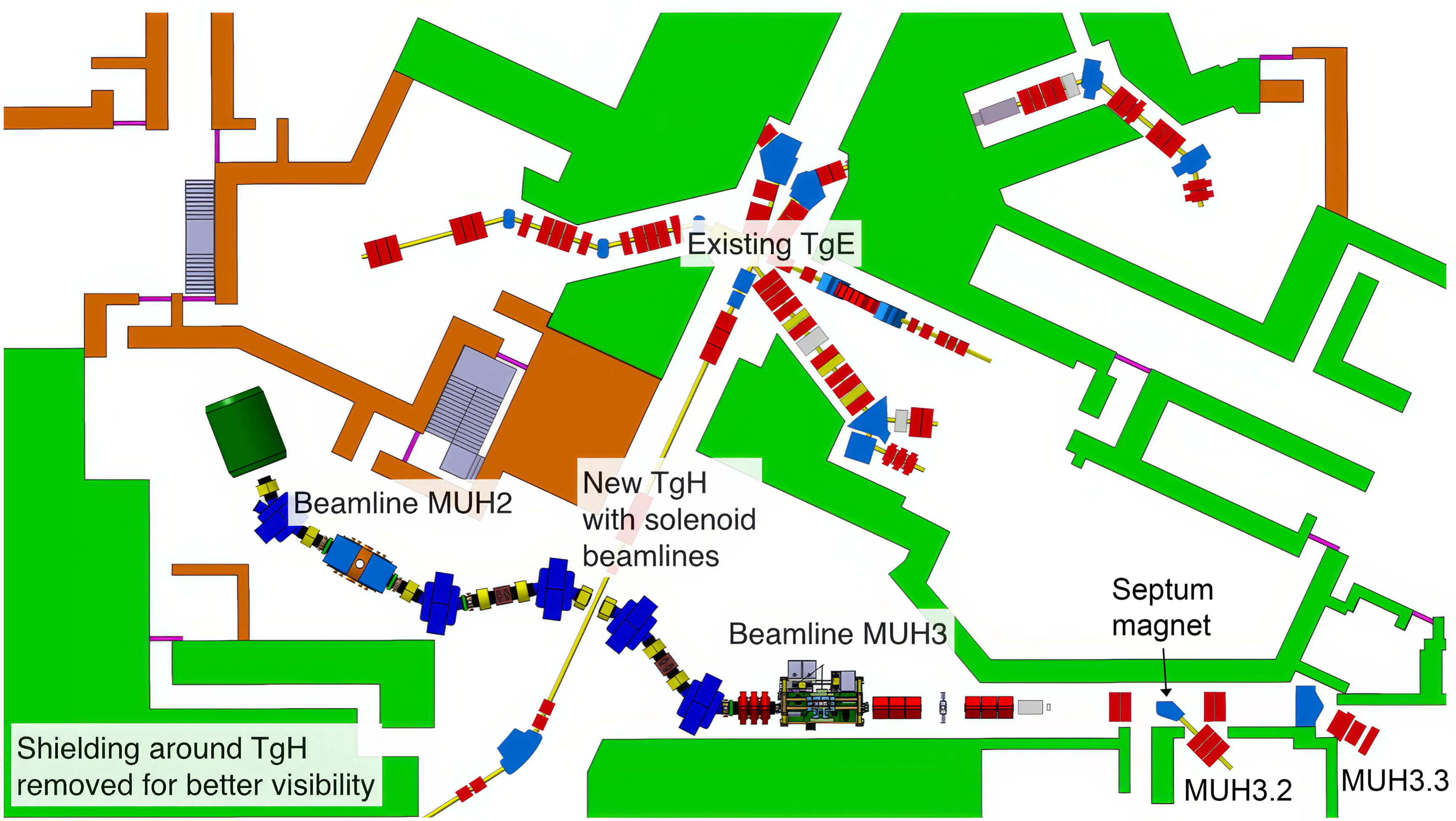}
    \caption{Layout of upgrade plans of the HIMB at PSI (reproduced from Ref.~\cite{Aiba:2021bxe}).}
    \label{fig:himb}
\end{figure}

The High-Intensity Muon Beams (HIMB) will be constructed with the Isotope and Muon Production using Advanced Cyclotron and Target (IMPACT) project at PSI~\cite{impact2022}. Based on the original proton beam, the Target M station, including its beamlines $\pi$M1 and $\pi$M3 mentioned in \cref{sec:smus}, will be replaced by a new Target H station with MUH2 and MUH3 beamlines as shown in \cref{fig:himb}. The goal of HIMB is to deliver the most intense CW surface muon beam at a flux of $\mathcal{O}(10^{10})~\mu^+$/s for both particle physics and materials science experiments~\cite{Aiba:2021bxe}.
This rate is expected to be achieved using a thicker target along with capture solenoids positioned near the target. The thickness of the new muon production target is optimized to be 20~mm, corresponding to effectively 100 mm along the beam direction. The capture solenoids are similar to the ones of the $\mu$E4 beamline at PSI, and are expected to be resistant to radiation damage.

\subsubsection{MELODY at CSNS}
The CSNS accelerator provides 1.6 GeV protons at a repetition rate of 25 Hz. Taking advantage of the CSNS-II upgrade plan, which increases the beam power from 100~kW to 500 kW, the construction of the Muon station for sciEnce, technoLOgy and inDustry (MELODY) has been proposed, with operation expected to begin by 2029~\cite{Hong:2024gnp,Bao:2023nup}.
The muon target station consists of two beamlines for surface muons and decay muons. Muons will be produced at the rate of $10^5$--$10^7~\mu^+$/s using a copper target with dimensions of $24\times24\times1.24~\mathrm{cm}^3$ to improve heat dissipation performance~\cite{Qiang:2024knc}. However, the designed repetition frequency of 1 Hz is relatively low, and is expected to be increased to 5 Hz in the future upgrade plan~\cite{Chen:2023uvp}.

\subsubsection{MuST at CiADS}
CiADS, currently under construction, presents an opportunity for a high-intensity muon source in China~\cite{Cai:2023caf}. By 2027, the CiADS accelerator is expected to deliver a 600 MeV proton beam with 300 kW beam power and reach 3 MW by 2030.
The construction of the Muon Science and Technology application platform at CiADS (MuST) will be carried out in two phases. Phase I will consist of a target station (TSa), which will receive a 300~kW proton beam, and two muon beamlines (MuB-R1 and MuB-L1) arranged on either side. Upon completion of Phase I, the terminal muon intensity is expected to exceed $\mathcal{O}(10^{8})~\mu^+$/s, reaching a world-leading level~\cite{Cai:2023caf}. After 2029, Phase II is planned to add a second target station (TSb) and two additional muon beamlines (MuB-R2 and MuB-L2). The beam power will be upgraded to 3 MW, with the terminal muon intensity expected to exceed $\mathcal{O}(10^{9})~\mu^+$/s~\cite{Cai:2023caf}.

\subsubsection{HIAF}

HIAF can provide high-intensity protons or ions to extract muons~\cite{Xu:2025spd}. Current feasibility studies have proposed a 9.3~GeV proton beam operating at 3 Hz at HIAF. The High Energy Fragment Separator (HFRS), a 192 m beamline, is capable of producing and transporting a PW muon beam at a momentum up to 7.5 GeV/$c$. Simulations show that muon fluxes can reach $\mathcal{O}(10^{6})$/s for both positive and negative muons; however, the corresponding muon purities are approximately 2\% and 20\%, respectively~\cite{Xu:2025spd}. By optimizing the magnetic rigidity values of the HFRS sections, a 100\% pure muon beam with a momentum of 1.5~GeV/$c$ is expected to be achieved~\cite{Xu:2025spd}.

\subsubsection{HEF at J-PARC}
In addition to MUSE, J-PARC is preparing a dedicated muon beamline for the COMET experiment, which searches for $\mu-e$ conversion~\cite{COMET:2018auw,COMET:2018wbw}. This beamline is located at the Hadron Experimental Facility (HEF), which operates with a 30 GeV proton beam extracted from the Main Ring accelerator at a maximum power of 84~kW~\cite{Takahashi:2025wqw}. The COMET experiment requires low-energy negative muons to be stopped in a target to form muonic atoms. Therefore, an 8 GeV operation mode of the proton beam will be used. The ``bunched slow extraction'' technique is employed to maintain a well-defined time structure, with a pulse width of about 100 ns and a separation greater than 1.17 $\mu$s. For muon production and transport, COMET introduces a superconducting solenoid system demonstrated at the MuSIC facility (described in \cref{sec:music}) to obtain an intense low-momentum negative muon beam. Muons with momenta of about 40~MeV/$c$ will be delivered to the detector region and stopped at the center of the aluminum target. The experiment will be carried out in two stages, Phase-I and Phase-II, requiring beam powers of 3.2 kW and 56 kW, respectively, and is expected to achieve a maximum muon intensity of up to $10^{11}~\mu^-$/s.

\subsubsection{RAMIS at RAON}

Based on the RAON at the Institute for Rare Isotope Science (IRIS) in Korea, the RAON Muon Irradiation System (RAMIS) muon beamline has been proposed and completed its installation in 2022~\cite{kim2020current,kim2025current}. The beamline is initially designed to operate with 600 MeV protons at a beam power of 100 kW, with plans to upgrade to 400 kW. A rotating IG-430U polycrystalline graphite target with a thickness of 5 mm is used for muon production. According to G4Beamline simulations, a muon rate of $5.2\times10^6~\mu^+$/s can be achieved. The system has already undergone integration test and will be used for future $\mu$SR studies~\cite{jeong2021design}.

\subsubsection{ORNL/SNS}
The SNS at Oak Ridge National Laboratory (ORNL) plans to upgrade its proton accelerator beam power from 1.4~MW to 2.8 MW in the next decade, along with the construction plan of a muon source~\cite{Williams:2022ypo}. For current facilities, the muon target station is typically located upstream of the neutron target. However, less than 5\% of the protons contribute to muon production, while the rest are used for neutron generation. The scattering effect of the muon target also reduces neutron yield by about 20\%. Taking advantage of the laser stripping technology, the proposed muon source separates muon and neutron production. It also helps to generate extremely narrow proton pulses.

A proof-of-principle experiment~\cite{Liu:2020hcu} conducted in 2019 at SNS demonstrated that the laser stripping system can operate stably and continuously with a 30 ns pulse width at a 50~kHz repetition rate. Preliminary studies indicate a stripping efficiency of up to 90\%, corresponding to a muon beam pulse width of 49.7 ns. Simulation shows that adopting a copper target can increase the muon yield to $10^9~\mu^+$/s, while decay muons can be produced ten times more than this rate. A low-energy muon beamline is also proposed. These beamlines will be mainly dedicated to $\mu$SR research.

\subsection{Summary of global muon beams}

\begin{sidewaystable}
    \caption{Reported specifications of global muon beamlines based on proton accelerators.}
    \label{tab:facilities}
    \resizebox{\linewidth}{!}{
        \begin{tabular}{clccccccccccccc}
            \hline
            Mode                & \multicolumn{1}{c}{Facility}    & Name        & \begin{tabular}[c]{@{}c@{}}Proton\\ energy\\ (GeV)\end{tabular} & \begin{tabular}[c]{@{}c@{}}Proton\\ power\\ (kW)\end{tabular} & \multicolumn{3}{c}{\begin{tabular}[c]{@{}c@{}}Rate\\ ($\mu/$s)\end{tabular}} & \multicolumn{3}{c}{\begin{tabular}[c]{@{}c@{}}Momentum\\ (MeV$/c$)\end{tabular}} & \multicolumn{3}{c}{\begin{tabular}[c]{@{}c@{}}Momentum\\ distribution\\ (\%)\end{tabular}} & \begin{tabular}[c]{@{}c@{}}Polarization\\ Rate\\ (Surface)\end{tabular}                                                       \\ \cline{6-14}
                                &                                 &             &                                                                 &                                                               & Surface                                                                      & Decay                                                                            & Slow                                                                                       & Surface                                                                 & Decay & Slow & Surface & Decay & Slow &             \\ \hline
            \multirow{8}{*}{PW} & \multirow{2}{*}{J-PARC/Japan}   & MUSE        & 3/8                                                             & 1000                                                          & $10^7$                                                                       & $10^7$                                                                           & $10^2$                                                                                     & 28                                                                      & 120   & 2.5  & 5       & 3     & --   & --          \\ \cline{3-15}
                                &                                 & HEF         & 8                                                               & 3.2/56                                                        & --                                                                           & $10^{11}$                                                                        & --                                                                                         & --                                                                      & 40    & --   & --      & --    & --   & --          \\ \cline{2-15}
                                & \multirow{2}{*}{ISIS/UK}        & EC          & \multirow{2}{*}{0.8}                                            & \multirow{2}{*}{180}                                          & $10^6$                                                                       & --                                                                               & --                                                                                         & 28                                                                      & --    & --   & 2       & --    & --   & $\sim99\%$  \\ \cline{3-3} \cline{6-15}
                                &                                 & RIKEN-RAL   &                                                                 &                                                               & $10^5$                                                                       & $10^5$                                                                           & $10^{-2}$                                                                                  & 28                                                                      & 120   & 2    & 4       & 4     & --   & $\sim99\%$  \\ \cline{2-15}
                                & FNAL/USA                        & Muon Campus & 8                                                               & 25                                                            & --                                                                           & $10^{10}$                                                                        & --                                                                                         & --                                                                      & 3094  & --   & --      & 10    & --   & $\sim95\%$  \\ \cline{2-15}
                                & CSNS/China                      & MELODY      & 1.6                                                             & 20                                                            & $10^7$                                                                       & $10^6$                                                                           & --                                                                                         & 28                                                                      & 120   & --   & 6       & 8     & --   & $\sim95\%$  \\ \cline{2-15}
                                & HIAF-HIRIBL/China               & --          & 9.3                                                             & --                                                            & --                                                                           & $10^6$                                                                           & --                                                                                         & --                                                                      & 1500  & --   & --      & 3.6   & --   & $\sim85\%$  \\ \cline{2-15}
                                & ORNL-SNS/USA                    & --          & 1.3                                                             & 1400/2800                                                     & $10^9$                                                                       & $10^{10}$                                                                        & $10^5$                                                                                     & --                                                                      & --    & --   & --      & --    & --   & --          \\ \hline
            \multirow{6}{*}{CW} & \multirow{2}{*}{PSI/Switzerland} & S$\mu$S     & \multirow{2}{*}{0.59}                                           & \multirow{2}{*}{1400}                                         & $10^8$                                                                       & $10^7$                                                                           & $10^4$                                                                                     & 28                                                                      & 500   & 2.5  & 8       & 8     & --   & $\sim95\%$  \\ \cline{3-3} \cline{6-15}
                                &                                 & HIMB        &                                                                 &                                                               & $10^{10}$                                                                    & --                                                                               & $10^5$                                                                                     & 28                                                                      & --    & 1.5  & 10      & --    & 7    & --          \\ \cline{2-15}
                                & RNCP/Japan                      & MuSIC       & 0.4                                                             & 0.4                                                           & $10^8$                                                                       & $10^7$                                                                           & --                                                                                         & 28                                                                      & 110   & --   & 10      & 10    & --   & $\sim60\%$  \\ \cline{2-15}
                                & TRIUMF/Canada                   & CMMS        & 0.52                                                            & 75                                                            & $10^6$                                                                       & $10^6$                                                                           & --                                                                                         & 29.4                                                                    & 173   & --   & 7.1     & 6     & --   & $\sim100\%$ \\ \cline{2-15}
                                & CiADS/China                     & MuST        & 0.6                                                             & 300/3000                                                      & $10^{10}$                                                                    & $10^7$                                                                           & --                                                                                         & 28.5                                                                    & 120   & --   & 5       & --    & --   & $\sim96\%$  \\ \cline{2-15}
                                & RAON/Korea                      & RAMIS       & 0.6                                                             & 100/400                                                       & $10^{6}$                                                                     & --                                                                               & --                                                                                         & 28.5                                                                    & --    & --   & 3       & --    & --   & --          \\ \hline
        \end{tabular}
    }
\end{sidewaystable}

Over the past few decades, the development of high-power proton accelerators has provided opportunities for the construction of muon beam facilities. Taking advantage of high flux, high polarization, and controllable beam parameters, accelerator-based muon sources have become a fundamental platform for modern muon science and its applications, producing muons that cover an energy range from keV to GeV. \cref{tab:facilities} lists the global muon facilities that have been constructed, are under construction, or are being planned.

These facilities can be categorized into continuous-wave (CW) and pulsed-wave (PW) muon sources according to their time structures. In particular, the cyclotrons can deliver proton beams with high repetition rates. Due to the relatively long pion lifetime (26 ns), the emitted secondary muons are time-smeared, resulting in a CW beam. For CW beams, muons can be approximately regarded as entering the detector system and decaying individually. Therefore, under a good time resolution and an appropriate beam intensity, the ``pile-up'' effect can be effectively avoided. However, some degree of ``pile-up'' effect will also occur when the muon intensity becomes too large ($>1/\tau_\mu$). In the case of PW beams, muons arrive in bunches, each containing numerous muons, which can significantly enhance the statistics of the experiment. Moreover, the time structure of the beam can provide an external trigger for event selection. However, ``pile-up" effects may occur during the beam-on period.

Surface muons are the most widely used type in both particle physics and materials science. Their main advantages include low energy, high polarization ($>90\%$), and monoenergetic. Slow muons can be produced by further moderating surface muons, and are applied in fields such as muon cooling, muonium physics, and surface physics studies. However, only a few beamlines are currently available for actual experiments. Related works mainly conducted in Switzerland, the UK, and Japan, but further optimization and testing are still needed. On the other hand, decay muons have high energy and stronger transmitting capabilities, making them suitable for muon scattering processes and muon tomography studies.

In the rare process searches and spectroscopic studies, the muon beam intensity is one of the most critical parameters. The reported beam intensities of existing and proposed muon facilities are summarized in \cref{tab:facilities}. In future upgrades and construction plans for muon sources, several approaches are under consideration to enhance muon beam intensity, including increasing the proton beam power on target, optimizing the muon production targets, and improving the design of capture solenoids. The next-generation muon facilities, such as HIMB and MuST, are expected to increase the continuous surface muon intensity to the order of $10^{10}\mu^\pm$/s, which will significantly improve the precision of muon physics experiments.

\section{Future prospects of muon beams}
With the constant development of muon beams and the growing demands of experimental muon physics, future muon beams are expected to move toward more ambitious parameters: higher intensity, extreme high or low momentum region, micro phase space, and beyond~\cite{Nagamine:2003vx,Cywinski:2009zz}. High-energy muon beams are also considered in the neutrino factories for an advanced source of muon neutrinos, providing a powerful tool to the current accelerator-based neutrino experiments such as T2K, NOvA, DUNE, and HyperK~\cite{Bogomilov:2014koa,T2K:2025wet,DUNE:2020jqi,Hyper-KamiokandeProto-:2015xww,An:2025lws}. The beamlines themselves are expected to be more cost-effective and more compact. Meanwhile, the study of advanced target materials could lead to a fundamental enhancement in muon yield and a reduction in positron contamination. For example, Ref.~\cite{Cai:2023caf} reports a liquid lithium target scheme that achieves a substantial increase in the muon rate. For the state-of-the-art muon collider prospect, significant challenges remain in the cooling and acceleration of muons. Here, we provide a brief review and discussion of several novel methods and techniques that are currently gaining attention.

\subsection{Muon production and transport}

Most of the muon source facilities mentioned above utilize protons as their primary driver beam. To date, a variety of alternative drivers have been investigated, including heavy-ion beams, accelerator-based electron beams, and even laser-based schemes. The primary beams with even higher energies are also expected to be employed to serve future neutrino factories and muon colliders.

\subsubsection{Higher energy ion or proton driven production}
The heavy-ion case, represented by HIAF, has been introduced above. It is expected to deliver muon beams with a momentum up to 7.5 GeV$/c$, significantly much higher than those available in common facilities, thereby enabling muon-scattering and tomography studies~\cite{Xu:2025spd}. However, the muon purity decreases substantially at higher momentum in the recent report of the HIAF muon beamline. In addition, this muon source relies on the High energy FRagment Separator (HFRS), which is primarily designed for nuclear physics studies. The performance may therefore be less promising than that of dedicated muon beamlines.

The proposed Neutrinos from Stored Muons (nuSTORM) facility plans to produce and store muon beams in a ring with central momenta between 1--6~GeV$/c$, taking advantage of the 100 GeV proton beam from the CERN Super Proton Synchrotron (SPS)~\cite{Ahdida:2020whw,nuSTORM:2025tph}. The storage ring scheme is also adopted by the Muon Campus at FNAL~\cite{Ganguly:2022ufq}.
The M2 beamline at the CERN SPS delivers high energy muons with momenta up to 280 GeV$/c$~\cite{Bernhard:2019jqz}. This beamline hosts two fixed-target experiments: NA64$\mu$, which searches for the invisible states via muon scattering processes~\cite{Crivelli:2025pjb}, and MUonE, which aims to precisely measure the hadronic component of the muon $g-2$ anomaly~\cite{MUonE:2019qlm}.

The mentioned facilities above can easily produce multi-GeV muons using the powerful heavy-ion or proton accelerators. However, due to the extended lifetime of pions, these facilities require large-scale beamline designs to ensure that pions can finally decay into muons. The contamination of muon beams could also be a possible challenge for both the beamline design and the experimental applications. The application of this type of muon source is therefore limited. Notably, a new approach has emerged in which muons are first ``cooled" to significantly reduce their phase space, known as muon cooling. Then, the muons will be re-accelerated to high energies. This is now believed to be a critical step in the journey to future multi-TeV muon colliders.

\subsubsection{Accelerator-based electron driven production}

In order to reduce the scale and cost of muon sources and to broaden their application scenarios, K.~Nagamine \textit{et al.} proposed a compact muon source design based on a 300 MeV, 10~$\mu$A electron accelerator for $\mu$SR applications~\cite{Nagamine:2009zz}. When high-energy electrons interact with the target material nuclei, muons can be produced through the following processes~\cite{Blomqvist:1976mq,Nagamine:2009zz}:
\begin{itemize}
    \item Photo-production---high-energy electrons emit bremsstrahlung photons, and these photons subsequently interact with nuclei via photonuclear reaction to produce $\pi^{\pm}$s,
          \begin{equation}
              e^-+Z_1\to\gamma+Z_1,\gamma+Z_2\to\pi^\pm+Z_3~.
          \end{equation}
    \item Electro-production---electrons generate pions through inelastic scattering with nuclei via virtual-photon exchange,
          \begin{equation}
              e^-+Z_1\to(e^-)^{\prime}+Z_1+\pi^\pm+Z_2~.
          \end{equation}
    \item Bethe--Heitler pair production---bremsstrahlung photons create a $\mu^+\mu^-$ pair in the Coulomb field of a nucleus,
          \begin{equation}
              \gamma+Z\to\mu^++\mu^-+\mathrm{anything}~.
          \end{equation}
    \item Direct annihilation---high-energy positrons annihilate with electrons in the target to directly produce a $\mu^+\mu^-$ pair,
          \begin{equation}
              e^++e^-\to\mu^++\mu^-~.
          \end{equation}
\end{itemize}

A high-energy electron linac is vital for particle physics, materials science, and biological science, especially as a key component of synchrotron radiation and X-ray free electron laser (XFEL) facilities~\cite{He_2014,Zhao_2017}. In recent years, there have been several feasibility studies for muon beamlines driven by high-energy electron beams through the mentioned processes.
The Shanghai High repetition rate XFEL and Extreme Light facility (SHINE) currently under construction will deliver an 8 GeV electron beam operating at a 1 MHz repetition rate. A high-repetition-rate muon beam based on SHINE has been proposed, which is expected to generate $\sim10^4~\mu^{\pm}$/bunch using a thick tungsten target~\cite{Liu:2025ejy}. Similarly, the Continuous Electron Beam Accelerator Facility (CEBAF) at Jefferson Lab has proposed a baseline design of a beamdump facility, which would enable the construction of a 6 GeV muon beamline by exploiting its electron beam up to 12 GeV~\cite{Achenbach:2025ynn}. The muon rate is expected to reach $\mathcal{O}(10^8)~\mu^{\pm}$/s level.
For the direct annihilation mechanism, the Low EMittance Muon Accelerator (LEMMA) concept was proposed in Ref.~\cite{Boscolo:2020xfv}. In this scheme, 45 GeV positrons are used to reach the muon pair production threshold, and a thin liquid lithium target is employed. By simulation study, an emittance of $\mathcal{O}(10^{-6})$~m~rad and a rate of $\mathcal{O}(10^9)~\mu^{\pm}$/bunch could be obtained.

In summary, the muon production driven by an accelerator-based electron beam can not only help reduce the overall size and cost of the facility but also suppress radiation contamination, including neutrons. However, such sources inherently suffer from a lower muon rate compared with proton-driven facilities due to the much smaller production cross sections. At present, no dedicated beamline of this type exists worldwide, and substantial development is still required.

\subsubsection{Laser-based electron driven production}

All of the accelerator facilities discussed above are based on conventional radio-frequency (RF) acceleration technology. The accelerating gradient is limited to $\mathcal{O}(100)$ MeV/m, which results in large scales and high costs. With the rapid development of laser technology in recent decades, ultra-short and ultra-intense laser systems have emerged. Moreover, there have been laser-driven radiation sources, including X-rays, neutrons, electrons, and protons. Among these advances, the laser wakefield acceleration (LWFA) technique has attracted significant attention~\cite{Tajima_2020}. LWFA can achieve accelerating gradients of $\mathcal{O}(100)$ GeV/m, which is three orders of magnitude higher than those of RF accelerators. It greatly enhances the capability for particle acceleration.

\begin{figure}[t!]
    \centering
    \includegraphics[width=\textwidth]{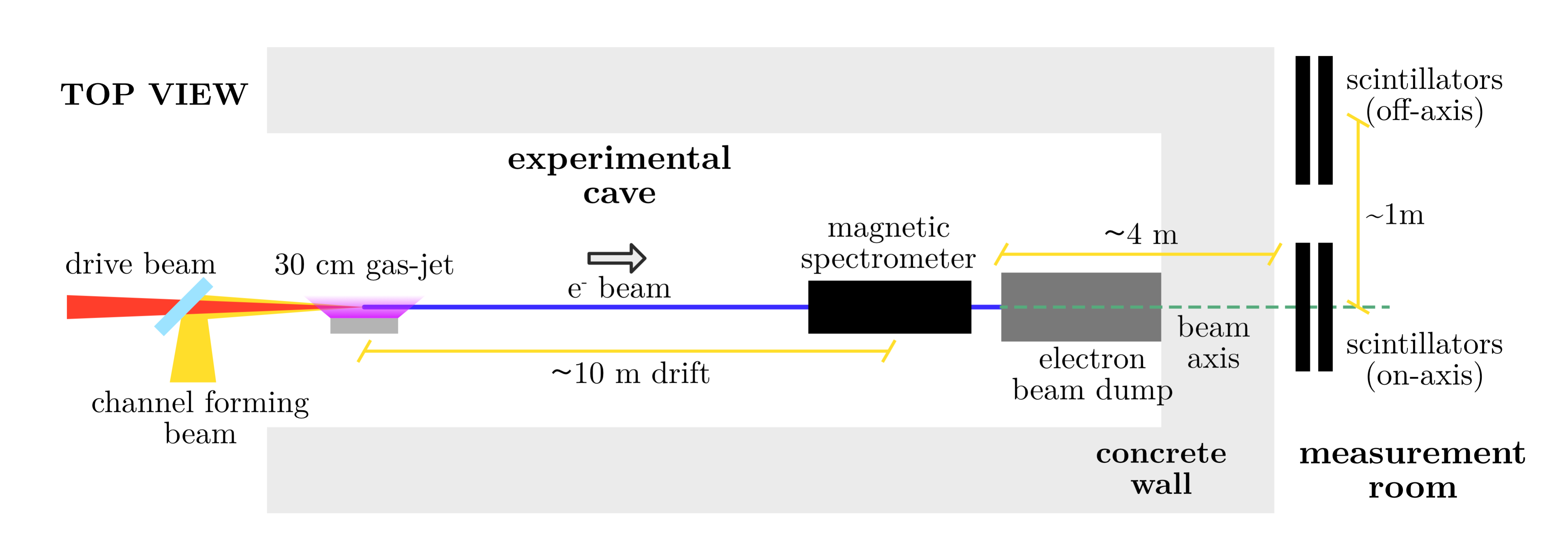}
    \caption{Schematic diagram of the experimental setup of LWFA-driven muon production (reproduced from Ref.~\cite{Terzani:2024neq}). A drive laser irradiates a plasma gas target to generate electrons, which propagate over a 10 m flight path through a magnetic spectrometer and subsequently produce muons in the downstream electron beam dump. A concrete wall is used to suppress the background. If muons are produced, they will pass through the concrete wall and decay in the scintillators. The presence of muons can be determined by reconstructing the lifetime spectrum from the time difference between the fast and slow signals.}
    \label{fig:lwfa}
\end{figure}

The concept of wakefield acceleration was first introduced by T.~Tajima and J.~M.~Dawson in 1979~\cite{Tajima:1979bn}. They proposed that the ponderomotive force of an intense laser pulse could drive a plasma wake capable of accelerating trapped electrons to high energies over extremely short distances---a process often described as ``surfing''. Limited by the laser technology available at that time, their work was only based on theoretical calculation and simulation.
Substantial experimental progress was achieved in 2004, when three independent groups reported the generation of high-quality, quasi-monoenergetic electron beams with energies around 100 MeV, marking the first clear demonstration of LWFA~\cite{Mangles:2004ta,Geddes:2004tb,Faure:2004tc}. In 2024, A.~Picksley \textit{et al.} reported the acceleration of quasi-monoenergetic electrons up to 9.2 GeV using a 500 TW laser at the BErkeley Laboratory Laser Accelerator (BELLA) facility, representing the highest energy of LWFA electrons reported to date~\cite{Picksley:2024cdd}. As petawatt-class lasers and staged acceleration techniques continue to mature, further increases in beam energy are expected.

High-energy electrons produced via wakefield acceleration can be used for muon generation, and several simulation studies have previously explored this possibility, as reported in Ref.~\cite{Rao:2018njj,Calvin_2023}. In 2025, F.~Zhang \textit{et al.} employed the 1-PW laser of the Shanghai Superintense Ultrafast Laser Facility (SULF) to accelerate electrons to 0.4--1.5 GeV for muon production~\cite{Zhang:2024axy}. The electrons were then transported to a 12 cm-thick lead target, behind which two liquid scintillators shielded by lead were installed. By measuring the time difference between the fast and slow signals of muons and their Michel electrons, then fitting the average muon lifetime, the observation of muon production was confirmed. Geant4 simulations further estimated a muon yield of 0.01 muons per incident electron, representing the first proof-of-principle demonstration of a laser-driven muon source. Subsequently, a research group at LBNL reported a similar experiment at BELLA and observed muons downstream of a 2.1 cm tungsten target as \cref{fig:lwfa}~\cite{Terzani:2024neq}. Furthermore, at the Extreme Light Infrastructure-Nuclear Physics (ELI-NP) facility, experiments were conducted using a 10 cm lead target, in which two dipole magnets transported the produced muons to the detector region~\cite{Calvin:2025huk}. Differing from the previous two studies, this experiment uses silicon pixel sensors read out by Timepix3 instead of scintillators, making it possible to identify muons through reconstructed track length and energy deposition.

Excitingly, laser-driven muon sources have recently progressed from theory to experimental demonstration. This development suggests that large-scale muon beamline facilities (typically extending over several km) could eventually be reduced to room-sized or even tabletop systems. It will offer significant value for applications in particle physics, $\mu$SR, and muon tomography.
Nevertheless, the muon yield achieved in current experiments remains far below that of RF accelerator-based muon sources. Potential strategies to enhance the muon rate may include increasing the charge per laser pulse, increasing the repetition rate of laser systems to the kHz level, and further optimizing the design of the muon capture and transport systems.

\subsection{Muon cooling and acceleration}

Muon beams are typically produced from decayed pions generated by protons interacting with the target. The dynamics in the target region are highly complex, and large-acceptance capture systems are usually required to achieve sufficient muon flux. These factors lead to muon beams with a large phase-space volume, characterized by large emittance, large beam sizes, and significant momentum spread~\cite{Otani:2022hjg}. However, the cooling techniques used for other particles are far too slow compared with the muon lifetime~\cite{H_nsch_1975,M_hl_1980,Parkhomchuk_2000}. Novel methods have been proposed for the phase space quality enhancement~\cite{MICE:2019jkl,Antognini:2021fae,Kamioka:2023xob}, which will be significant for muon colliders, as well as for neutrino factories, $\mu$SR applications, and muonium physics. In recent years, muon cooling techniques have been experimentally demonstrated, with the first reacceleration of a cooled muon beam~\cite{Aritome:2024rlu}.

\subsubsection{Ionization cooling}
\begin{figure}[t!]
    \centering
    \includegraphics[width=0.9\textwidth]{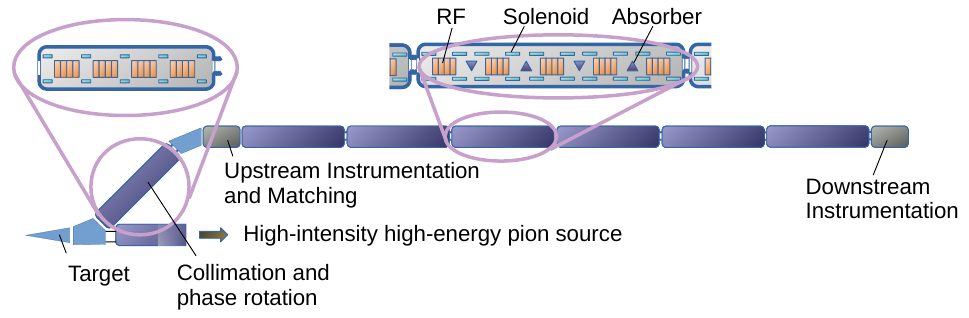}
    \caption{Schematic diagram of ionization cooling (reproduced from Ref.~\cite{Rogers:2023nqc}). Muons lose their energy in the absorber and are then captured and reaccelerated in the RF cavity. The structure is repeated over multiple stages.}
    \label{fig:cooling_mice}
\end{figure}
As plotted in \cref{fig:cooling_mice}, the principle of this method is to reduce the muon energy through ionization losses in a low-Z absorber (such as liquid hydrogen or lithium hydride), then the muons are reaccelerated longitudinally in an RF cavity. By repeating this process, the transverse emittance of the muon beam can be compressed. The Muon Ionization Cooling Experiment (MICE), conducted at the ISIS muon source in the UK, reported the first experimental demonstration of ionization cooling~\cite{MICE:2019jkl}.

However, muons will be multiple-scattered in the absorber, requiring a strong focusing field provided by superconducting solenoids. Moreover, the energy loss in the absorber increases the longitudinal emittance. To avoid the heating, six-dimensional (6D) cooling schemes have been proposed~\cite{Kaplan:2009jd}. These schemes use a bending magnetic field to separate muons with different longitudinal momenta, guiding them through absorbers of specially designed geometry so that higher-momentum muons lose more energy. As a result, the overall momentum spread would be reduced. However, the 6D cooling method could be complicated in transport optics and has not yet progressed beyond simulation.

\subsubsection{Friction cooling}

\begin{figure}[t!]
    \centering
    \includegraphics[width=0.5\textwidth]{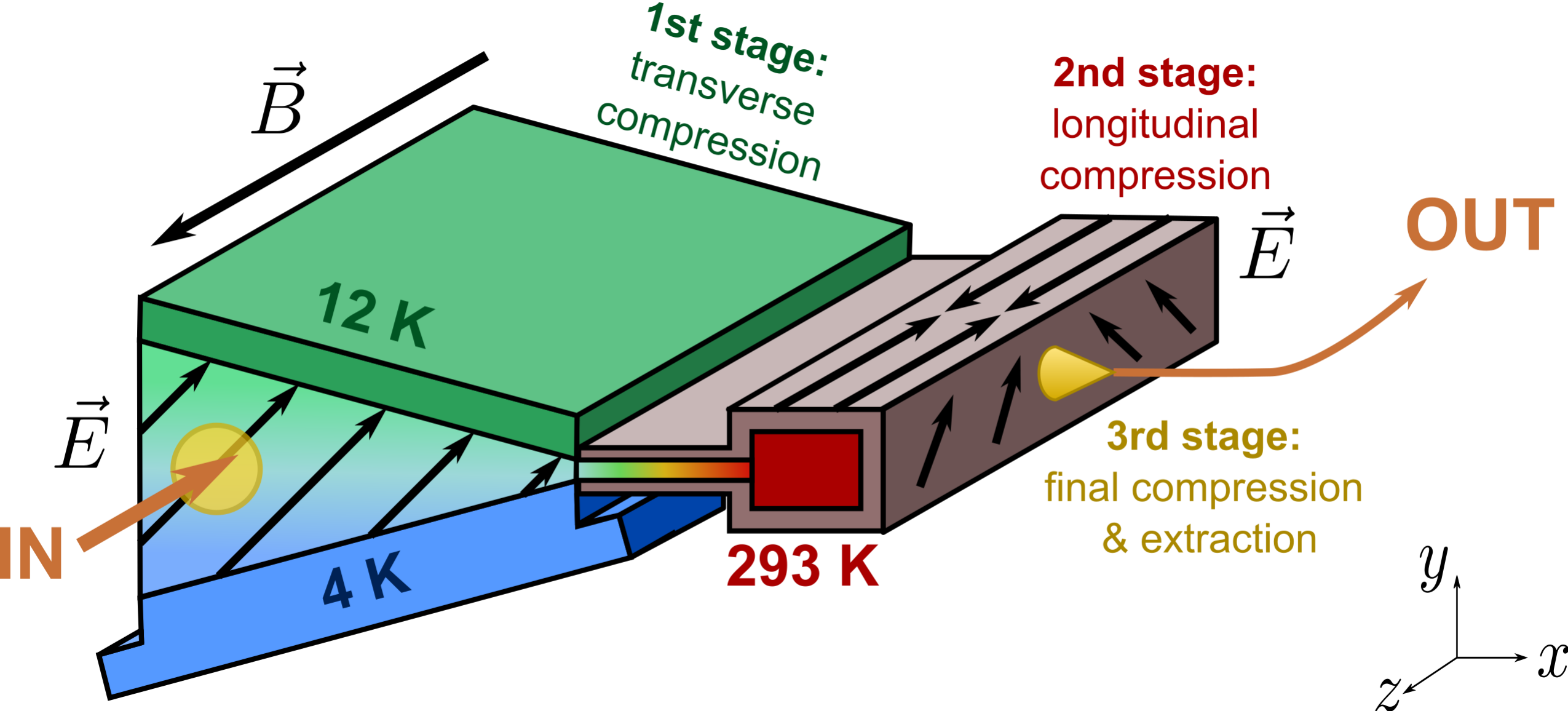}
    \caption{Schematic diagram of friction cooling (reproduced from Ref.~\cite{Antognini:2021fae}). Muons are first degraded in cryogenic helium gas with a density gradient under applied electric and magnetic fields, then further compressed by the electric field, and finally reaccelerated and released into vacuum.}
    \label{fig:cooling_psi}
\end{figure}

To produce low energy muons, PSI developed a slow-muon beamline based on a cryogenic moderator since 1994~\cite{Morenzoni:1994xf,Prokscha:2008zz}. In this method, muons first lose energy rapidly in a cryogenic metal substrate (Al or Ag) with a thickness of $\mathcal{O}(100)$~$\mu$m, and then enter an $\mathcal{O}(100)$~nm cryo-crystals moderator (solid Ne or Ar). Solid rare gases are insulators with large energy gaps ($\mathcal{O}(10)$ eV). In the moderator, muons lose energy through repeated muonium formation and ionization. Once their kinetic energy approaches the band-gap level, the ionization energy loss mechanism will be strongly suppressed, and the muons will remain at this energy until they escape from the moderator~\cite{Bakule:2004xw}. The process lasts only about 10~ps, and depolarization is negligible, so the extracted slow muons will preserve their initial polarization. The moderation efficiency reaches $10^{-5} \sim 10^{-4}$, and the technique is already used in low-energy $\mu$SR~\cite{Morenzoni_2004} and muonium spectroscopy experiments~\cite{Ohayon:2021dec}.

Muons slowed through a cold moderator still exhibit a very large phase space volume. PSI therefore proposed a friction cooling technique, as shown in \cref{fig:cooling_psi}~\cite{Antognini:2021fae}. The muon beam enters a target region filled with cryogenic He gas, where strong electric and magnetic fields are applied. The muon energy rapidly decreases from the MeV scale to the eV scale through collisions with He. Once the energy reduces to the eV range, the electric field becomes dominant. Together with the magnetic field and density gradient of the gas, it induces a drift motion that compresses the broad spatial distribution and focuses the muon beam into a sub-mm size. The beam is then compressed along the $z$-direction at room temperature. Finally, the muons are extracted from the gas into vacuum and re-accelerated along the $-z$ direction, allowing the beam to exit the magnetic field with improved phase space quality. However, multiple scattering with the gas constrains the overall cooling performance. In 2020, the muCool collaboration demonstrated transverse compression, increasing the phase space density by a factor of $10^{10}$ with an efficiency of $10^{-3}$~\cite{Antognini:2020uyp}.

\subsubsection{Muonium laser ionization cooling}\label{sec:M_laser}

\begin{figure}[t!]
    \centering
    \includegraphics[width=0.7\textwidth]{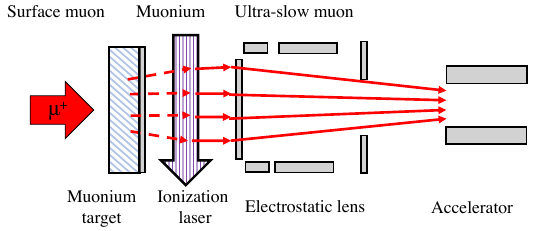}
    \caption{Schematic diagram of muonium laser ionization cooling (reproduced from Ref.~\cite{Kamioka:2023xob}). Muons are stopped in a silica aerogel target to form muonium. After being emitted into vacuum, the muonium is ionized by a laser. The released ultra-slow muons are focused by electrostatic lenses and subsequently reaccelerated.}
    \label{fig:cooling_jparc}
\end{figure}
Injecting surface muons into a tungsten foil heated to high temperature allows muonium to escape into vacuum with a kinetic energy of about 0.2 eV and an efficiency of 1\%~\cite{Nagamine:1995zz}. A pulsed laser with an appropriate wavelength first excites the muonium atoms, and a second laser ionizes the excited state to produce slow muons. The principle of muonium laser ionization cooling method is shown as \cref{fig:cooling_jparc}. The ionization efficiency can reach 100\% in principle, although the polarization of the resulting slow muons is reduced to 50\%. Note that only $\mu^+$ can be slowed by this method, as $\mu^-$ would be captured by nuclei.

In 1995, K.~Nagamine \textit{et al.} achieved the first slow-muon production using laser ionization of muonium, with a yield of 0.07 $\mu^+/$s~\cite{Nagamine:1995zz}. In 2008, P.~Bakule \textit{et al.} implemented the same technique at the ISIS muon source, increasing the yield to 15 $\mu^+/$s with an efficiency of the order of $\mathcal{O}(10^{-5})$~\cite{Bakule:2008zz}. Compared to the LEM beamline at PSI, the USM beamline requires a pulsed muon source and operates with lower yield and polarization, but it provides superior beam quality and achieves lower energies. Fortunately, as introduced by \cref{sec:production}, novel materials have been shown to substantially enhance both the production and emission rates of muonium, such as the currently adopted silica aerogel~\cite{Beare:2020gzr,Beer:2014ooa,Zhao:2023plv}. At present, the Ultra-Slow Muon (USM) beamline at the MUSE muon facility in Japan delivers a yield of 300~$\mu^+/$s~\cite{Kanda:2023gqp}. Future progress can be expected through further developments in both the laser system and the muonium formation materials.

\subsubsection{Muon acceleration}

\begin{figure}[t!]
    \centering
    \includegraphics[width=\textwidth]{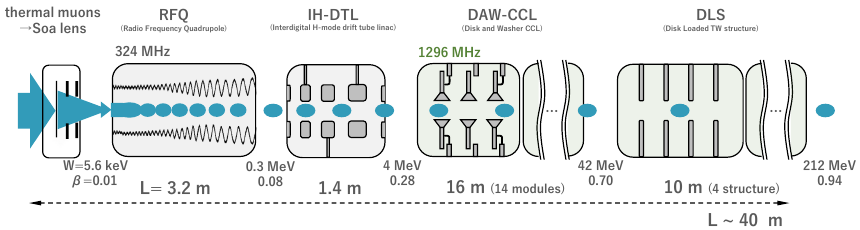}
    \caption{Schematic diagram of the muon linac at J-PARC MUSE (reproduced from Ref.~\cite{Otani:2025jpn}). Muons are first generated by laser ionization described by \cref{fig:cooling_jparc} and then are accelerated in multiple stages by a Radio-frequency Quadrupole linac (RFQ), an Inter-digital H-mode Drift Tube Linac (IH-DTL), and a Disk-And-Washer Couple Cavity Linac (DAW-CCL).}
    \label{fig:muon_linac}
\end{figure}

Muon acceleration is essential for next-generation muon experiments and the future muon collider. Muons are unstable particles with a mean lifetime of only 2.2 $\mu$s, so the cooling or acceleration of muons must be completed within a very short time before decay. The production processes generate muons with large spatial and momentum spreads. However, RF accelerators accept only beams that are narrow and have well-defined velocities. If an unprocessed muon beam is injected into the radio frequency cavity directly, most particles will escape either by hitting the beam pipe or falling out of phase with the RF field. Phase-space compression is therefore required essentially for muon acceleration. The paragraphs above have reviewed available muon cooling techniques, while the area of muon acceleration has recently achieved key progress at the J-PARC MUSE facility~\cite{Bae:2018atj,Otani:2022rbw,Aritome:2024rlu}.

In 2018, S.~Bae \textit{et al.} conducted the first demonstration of muon acceleration at J-PARC MUSE~\cite{Bae:2018atj}. Negative muonium ions ($\text{M}^-$, $\mu^+e^-e^-$) were produced via electron capture as surface muons passed through an aluminum degrader, and were accelerated from 50 keV to 89 keV using a radio-frequency quadrupole linac. Although the experiment served only as a proof of principle and was limited by the large emittance and extremely low yield of $\text{M}^-$, it provided the first experimental evidence that RF structures can accelerate muons. Furthermore, a scheme of a muon linac (up to 212 MeV) based on above techniques as shown in \cref{fig:muon_linac} is also proposed at the H line of MUSE~\cite{Kondo:2022sqb}. In 2025, S.~Aritome \textit{et al.} achieved a technical milestone by demonstrating RF acceleration of slow $\mu^+$ for the first time at J-PARC MUSE~\cite{Aritome:2024rlu}. Muonium laser ionization cooling was employed, followed by collection and focusing with an electrostatic lens system. The muons were then pre-accelerated to 5.7~keV by electrostatic fields and injected into an RFQ, where they were further accelerated to a final energy of 100 keV. The experiment showed that the transverse emittance was reduced by two orders of magnitude relative to the initial surface muon beam.

Notably, the LWFA technique, with its high accelerating gradient, offers a highly promising route for muon acceleration. Several simulation studies have aimed to overcome the dephasing caused by the velocity mismatch between muon beams and the wakefield~\cite{Geng:2024utj,Badiali:2025wsn}. Although these works propose promising new schemes for muon acceleration, significant experimental progress remains to be achieved.

\section{Experimental studies on muonium physics}

\subsection{Muonium production}\label{sec:production}
The quality of a muonium source is a key factor for muonium physics experiments. The formation mechanism of muonium can be summarized as follows: a muon is slowed down to very low energies in the target, ionizes an electron, and subsequently binds to it. Materials suitable for muonium formation usually have low ionization energies and a moderate number of electrons. Achieving a muon beam with a smaller phase space, lower energy, and higher intensity can increase the effective stopping fraction for muonium production targets. The observed muonium yield often depends not only on how much M is formed inside the material, but also on whether the formed M can diffuse out of the material, survive at the surface, and be further detected. Therefore, improving both the muonium formation efficiency and the muonium emission efficiency is equally important. Commonly used production targets include rare gases, solid foils, and porous materials. The superfluid helium is also currently considered as a possible next-generation muonium formation material.

J.~M.~Cassels and R.~A.~Swanson separately found that muons show depolarization in many kinds of materials. They explained this effect by the formation of muonium inside the materials~\cite{Cassels:1957pqg,Swanson:1958zz}. Muonium was first directly observed in a high-pressure argon gas environment by V.~W.~Hughes \textit{et al.} in 1960~\cite{Hughes:1960zz}. Because the muon beams at that time had relatively high momenta, the gas had to be compressed to 40 atmospheres to effectively stop the muons~\cite{Walker_1983}. The muoniums were observed using $\mu$SR methods through measuring its Larmor precession frequency. The muonium formation efficiency in the rare gas can approach nearly 100\%. However, since the electron spin distributes randomly, the muonium polarization is limited to 50\%~\cite{Jungmann:2004sa}. Applying a strong magnetic field can increase the polarization to 100\%. Experimental studies of muonium formation in various gases were carried out~\cite{Hughes:1970zza}. These works built a methodology base for later development of $\mu$SR, from basic atomic physics to a local magnetic probe in condensed matter materials. Later, muonium production in solids~\cite{Gurevich:1971ai} and liquids~\cite{Percival_1976} was also observed. The drawback of this method is that the muonium can hardly emit outside the target. Even in solid rare gas targets, only a very small fraction ($3\times10^{-4}/\mu^+$) of M can escape into vacuum~\cite{Prokscha:2008zz}. For precision measurements such as hyperfine structure splitting or the 1S-2S transition, it is necessary to eliminate the perturbations to the energy levels induced by collisions with the gas, thereby motivating the search for materials capable of emitting muonium into vacuum.

With the invention of surface muon beams~\cite{Pifer:1976ia}, studies of muonium production advanced significantly. Inspired by studies of positronium, several solid targets began to be employed for muonium formation~\cite{Canter_1974}. In 1985, A.~P.~Mills \textit{et al.} observed muonium emission from 99.99\% pure tungsten foil heated to high temperatures. The investigated temperatures ranged from 300 K to 2800 K, achieving a muonium yield in vacuum per muon of 4\% at 2300 K~\cite{Mills:1986zzb}. However, this production method was later found to depend strongly on the purity of the tungsten~\cite{Matsushita_1996}.
Notably, PSI has developed a muonium source based on the slow muons from $\mu$E4 beamline recently. To generate a sufficient number of metastable muonium for improving the precision of $n=2$ Lamb shift and fine structure measurements, the beam-foil method is required~\cite{Andr__1974}. In earlier experiments at TRIUMF and LAMPF, this approach required substantial muon degradation~\cite{C_J_Oram_1981,Badertscher_1992}, which reduced both the M(2S) yield and the overall experimental precision. In 2020, G.~Janka \textit{et al.} employed a 15 nm carbon foil to stop slow muons provided by the LEM beamline, achieving a muonium (1S and 2S) yield in vacuum of 31.8\% for 10~keV incident muons, where M(2S) accounted for 10\% of the total muonium yield~\cite{Janka:2020xky}. This technique is expected to improve the precision of Lamb shift measurements by two orders of magnitude.

D.~R.~Harshman reviewed earlier studies indicating that muonium can escape from the interior of fine oxide grains and is found mainly in the extragranular region and on grain surfaces, and that this behavior is independent of the enviroment temperature~\cite{Harshman_1986}. Among the oxide powders studied, silica ($\mathrm{SiO_2}$) showed the highest muonium fraction~\cite{Kiefl_1979}. It also has a very large specific surface area, which helps the escape of muonium. Therefore, fine-silica-based materials have finally become a main focus in muonium research.
In 1992, W.~Schwarz \textit{et al.} evaluated the performance of silica powder and aerogel targets at PSI and RAL~\cite{Schwarz_1992}. The experimental data were globally fit to a diffusion model to obtain the total yield in vacuum. Silica powder showed a clear advantage: the powder target achieved 8.2\%, whereas the optimized aerogel target got 2.3\%.
Previous measurements indicated a muonium formation efficiency of 60\% in silica powder, reported by R.~F.~Kiefl in 1979~\cite{Kiefl_1979}, while W.~Schwarz \textit{et al.} reported formation efficiency of 40--50\% for aerogels of varying densities.
However, when considering overall performance, silica aerogel proved to be the superior option. Its self-supporting structure eliminates the scattering introduced by the supporting foil required for powder targets, enabling the production of higher-quality beams for precision applications such as laser-ionized ultra-slow muon sources introduced in \cref{sec:M_laser}.
This work not only established the advantages of aerogel as a stable and self-supporting target material but also demonstrated that specialized surface treatments can further enhance its yield.

In 2013, P.~Bakule \textit{et al.} performed further tests in preparation for the J-PARC muon $g-2$ experiment, obtaining a yield of $\sim0.3\%$~\cite{Bakule:2013poa}. This value is an order of magnitude lower than previous results for the main reasons, including:
\begin{enumerate*}[(1)]
    \item the reported efficiency corresponds to direct counts within a specific flight range (10--40 mm), rather than the total yield extrapolated from a diffusion model in other works;
    \item the experiment employed a surface muon beam with a large momentum spread of 5\% (RMS);
    \item the samples did not employ any specialized surface treatment.
\end{enumerate*}
These differences indicate that the reported M vacuum yield depends not only on the counting definition, but is also highly sensitive to the beam phase space (and thus the effective stopping fraction) and to surface treatment.
To facilitate the muonium emission from aerogel targets, G.~A.~Beer \textit{et~al.} reported the first attempt of porous structures in silica aerogel using laser ablation in 2014~\cite{Beer:2014ooa}. The results showed that the muonium yield could reach 3\% in the processed aerogel target with $\sim50\%$ muons stopping rate, thereby demonstrating the feasibility of overcoming diffusion limitations through laser ablation to achieve high-yield muonium emission. Since then, numerous strategies of structural optimization for silica aerogel targets have been proposed~\cite{Beare:2020gzr,Zhang:2022ilj,Zhao:2023plv}.

Interestingly, a series of experiments carried out by Harshman \textit{et~al.} in the late 1980s showed that the surface state of silica can affect the diffusion and trapping behavior of the muonium~\cite{Harshman_1984_1,Harshman_1984_2,Harshman_1986}. Unlike bulk silica, porous silica or nano silica have much more complex surface structures, with a large number of unsaturated siloxane-related surface sites (Si-O-Si). After contact with water during preparation or in air, these sites can easily form silanol groups (Si-OH). Such hydroxyl groups make silica particles hydrophilic and also cause them to agglomerate. In practice, thermal treatment can be used to induce a dehydration-condensation reaction, in which two neighboring hydroxyl groups form a Si-O-Si bond and the excess water is released as vapor. However, once the temperature exceeds 600 °C, the specific surface area of silica decreases sharply and sintering occurs. At the same time, the surface hydroxyl concentration decreases from 8.6 nm$^{-2}$ to 3.6 nm$^{-2}$, and can recover partly after rehydration~\cite{Cheng_2020}. Harshman suggested that the presence of OH suppresses the diffusion of muonium on the surface at low temperature. He found that, as part of the surface OH was removed by vacuum baking, the low-temperature muonium relaxation was greatly reduced. This indicates that muonium was less affected by surface traps. If the muonium formation fraction remained nearly unchanged at the same time, this would usually mean that more muonium could survive, migrate, and desorb. He also studied the effect of other metal impurities, and the results suggested that they may also cause loss of emitted muonium~\cite{Harshman_1986}. Therefore, in addition to improving the M yield in vacuum, the long-term stability of the M production target is also a key issue that must be considered in future high-precision experiments. Systematic tests of the M target are needed, and repeatable pre-treatment procedures and control of the experimental environment should be established. More importantly, the influence of these surface chemical states on muonium emission still needs further study.

\begin{figure}[t!]
    \centering
    \sidesubfloat[]{\includegraphics[width=0.45\textwidth]{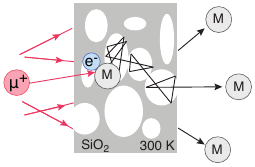}}
    \hfill
    \sidesubfloat[]{\includegraphics[width=0.45\textwidth]{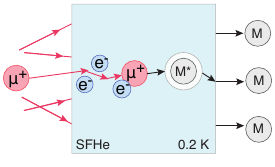}}
    \caption{Schematic diagrams of different approach of muonium formation (reproduced from Ref.~\cite{Soter:2021xuf}). (a) Muons are stopped in porous silica at room temperature to form muonium, which will be emitted into vacuum by thermal diffusion; (b) Muons are stopped in superfluid helium. Due to Pauli repulsion, the muonium will be ejected by the chemical potential of the cavity. Therefore, the emitted muonium is monoenergetic.}
    \label{fig:formation}
\end{figure}

For porous materials above, muonium reaches the vacuum via thermal diffusion, causing broad energy and angular distributions. Taking advantage of its unique properties, superfluid helium (SFHe) has demonstrated potential as a next-generation muonium formation target. When impurity atoms or ions enter SFHe, they are repelled by the Pauli repulsion and form a surrounding cavity inside the material as plotted in \cref{fig:formation}; once reaching the surface, they will be expelled by the chemical potential to achieve an almost monoenergetic emission of $\sim23$ meV~\cite{Saarela_1993,Hayden_1995,Mar_n_1998}. Because the ionization energy of He atoms is far higher than the binding energy of muonium, muonium formation is not observed in either gaseous or liquid He. In 1992, E.~Krasnoperov \textit{et al.} conducted experiments with a $\mu$SR spectrometer on the $\pi$M3 beamline at PSI. The precession signal of triplet-state muonium in SFHe below 1 K was observed, thereby demonstrating that SFHe can serve as a medium for muonium formation for the first time~\cite{Krasnoperov_1992}.

To conduct muonium gravity experiments based on atom interferometry and to further improve the precision of other muonium spectroscopy measurements, PSI plans to develop a cold muonium source based on SFHe~\cite{Taqqu:2010kcz,Soter:2021xuf}. Due to the effects of gravity and surface tension, SFHe can form only nm-scale films, whereas thick targets would suffer beam-induced heating that disrupts the superfluid state and significantly reduces the muonium emission probability. In summary, thick SFHe targets have difficulty in muonium emission, and thin superfluid films are insufficient to effectively stop the incident muons. To address this challenge, the LEMING collaboration has proposed nanostructured targets (such as laser-ablated aerogel and carbon nanotubes) coated with a layer of SFHe. Recently, a demonstration of a cold muonium beam based on SFHe has been reported by the group~\cite{Zhang:2025agk}. They used the $\pi$E1 beamline at PSI, with a muon flux of about $1.5\times10^{4}$, a central momentum of 12.5, and a momentum spread of 1.5\%. With a special beamline setup, the muon beam was bent vertically downward by $30^\circ$ and directed onto a horizontally placed SFHe target, from which muonium was emitted in the vertical direction. A muonium formation efficiency up to 70\% and a vacuum yield of 8.2\% is achieved in this work, which lays the foundation of muonium interferometry and more precise muonium spectroscopy studies.

\subsection{Muonium: charged lepton flavor violation}

\begin{figure}[t!]
    \centering
    \includegraphics[width=0.9\textwidth]{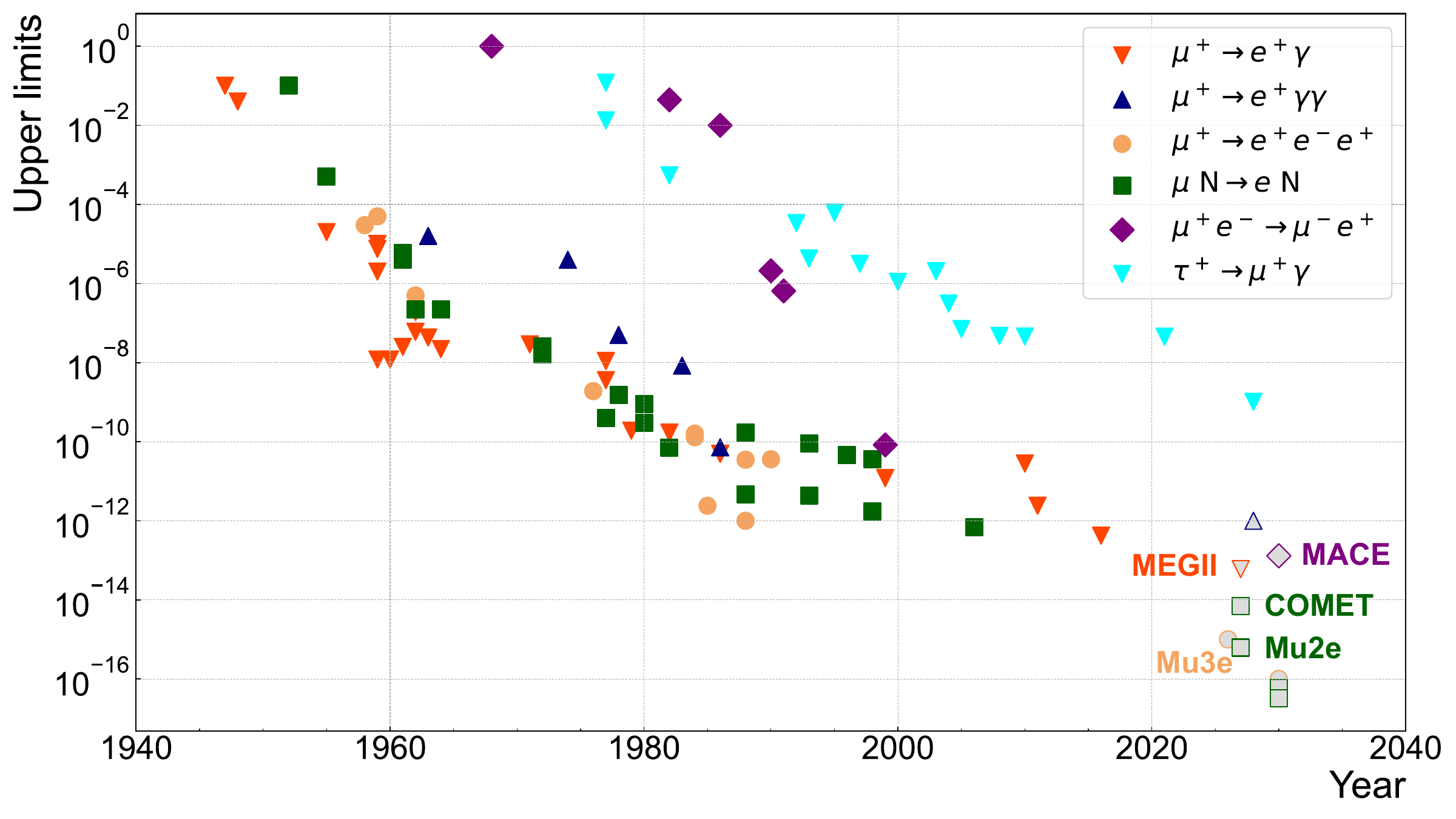}
    \caption{Evolution of experimental upper limits on various charged lepton flavor violation decay processes from the 1940s to the 2040s. Important existing and planned experiments include MEGII, Mu3e, COMET, Mu2e, and MACE, which represent the highest sensitivity in their respective fields. The values used in this figure are summarized from published or expected limits, and tabulated together with a plotting script~\cite{muon_clfv_history_plot}.}
    \label{fig:clfv}
\end{figure}

In the Standard Model, lepton flavor quantum numbers are strictly conserved. Although the discovery of neutrino oscillations has confirmed flavor mixing among neutral leptons~\cite{DayaBay2012,JUNO2015}, the extension of neutrino masses within the Standard Model framework suppresses charged lepton flavor violating (cLFV) processes to the level of $\mathcal{O}(10^{-54})$, making them unobservable. Therefore, the discovery of cLFV in the charged lepton sector would provide direct evidence for new physics beyond the Standard Model. The search for muon CLFV processes has continued for decades, with experiments constantly being proposed, especially the new-generation experiments in recent years (as plotted in \cref{fig:clfv}). The muonium-to-antimuonium conversion process violates the lepton flavor conservation law by two units~\cite {Heeck:2024uiz}, which is complementary to other muon cLFV processes~\cite{Calibbi:2025fzi}, making it a unique probe for the cLFV searching experiment. There might be a correlation between neutrinoless double beta decay and the muonium-to-antimuonium transition in models with a doubly charged scalar~\cite{Fukuyama:2022dhe}.

Similar to the $\text{K}^0$ meson, which consists of first and second generation quarks, muonium consists of first and second generation leptons. Inspired by the $\text{K}^0-\overline{\text{K}^0}$ oscillation mechanism proposed by Gell--Mann and Pais, B.~Pontecorvo first suggested the possibility of muonium-to-antimuonium oscillation, and further neutrino oscillations in 1957, laying the theoretical foundation for modern neutrino physics~\cite{Pontecorvo:1957cp,Bilenky:2013wna}. In 1961, G.~Feinberg and S.~Weinberg reported a study on neutrinoless muon decays, proposing that spontaneous $\text{M}-\overline{\text{M}}$ conversion could exist~\cite{Feinberg:1961zz}. They also calculated the conversion probabilities in various materials~\cite{Feinberg:1961zza}.
Within the framework of four-fermion interaction theory, the conversion probability $P_{\text{M}\overline{\text{M}}}$ can be written as
\begin{equation}
    P_{\text{M}\overline{\text{M}}}\propto\left(\frac{G_{\text{M}\overline{\text{M}}}}{G_{F}}\right)^2~,
\end{equation}
where $G_{\text{M}\overline{\text{M}}}$ stands for the effective coupling constant of $\text{M}-\overline{\text{M}}$ conversion, $G_{F}$ is the Fermi coupling constant.

In 1968, J.~J.~Amato \textit{et al.} conducted the first experimental search for the conversion~\cite{Amato:1968xyq}. In their setup, muonium atoms were formed by stopping muons in argon gas.
If there is a spontaneous conversion, the antimuonium could collide with an argon atom to form an excited muonic atom, which would subsequently emit a characteristic 2P-1S X-ray. The produced X-ray is considered as the signal of $\text{M}-\overline{\text{M}}$ conversion. This experiment set the first upper limit on the $\text{M}-\overline{\text{M}}$ coupling constant, $G_{\text{M}\overline{\text{M}}}<5800G_{F}$ at 95\% confidence level. Later experiments using similar methods have improved the sensitivity, suppressing the upper limit to $G_{\text{M}\overline{\text{M}}}<0.29G_{F}$ (90\%~C.L.)~\cite{Amato:1968xyq}.

In 1991, B.~E.~Matthias \textit{et al.} first implemented an experimental method by detecting the final state of antimuonium decay by coincidence detection~\cite{Matthias:1991fw}. The experiment was conducted at the Clinton~P.~Anderson Meson Physics Facility (LAMPF) in Los Alamos, using a 20 MeV/$c$ collimated muon beam with a momentum spread of $\Delta p/p\approx10\%$ and a duty factor of 6.4\%. A 150 $\mu$m-thick plastic scintillator was placed at the entrance of the detector to count and moderate incoming muons. The muonium production target was made of silica powder, where muons formed muonium atoms that subsequently thermally diffused into the vacuum. Surrounding the target were four layers of multi-wire proportional chambers (MWPC), which measured the Michel electron tracks under a magnetic field of 522 G. Outside the MWPCs, two layers of plastic scintillators and a cylindrical NaI(Tl) scintillator were installed to provide timing and energy information. The other final-state particle, a low-energy positron from the antimuonium, was collected, focused, and accelerated, then transported through a bending magnet and solenoid to a microchannel plate (MCP). The coincidence signal from the Michel electron and the low-energy positron was searched to probe the $\text{M}-\overline{\text{M}}$ conversion event. This experiment set a new upper limit on the coupling constant of $G_{\text{M}\overline{\text{M}}}<0.16G_{F}$, thus $P_{\text{M}\overline{\text{M}}}<6.5\times10^{-7}$ (90\% C.L.)~\cite{Matthias:1991fw}.

\begin{figure*}[!t]
    \centering
    \sidesubfloat[]{\includegraphics[width=0.45\textwidth]{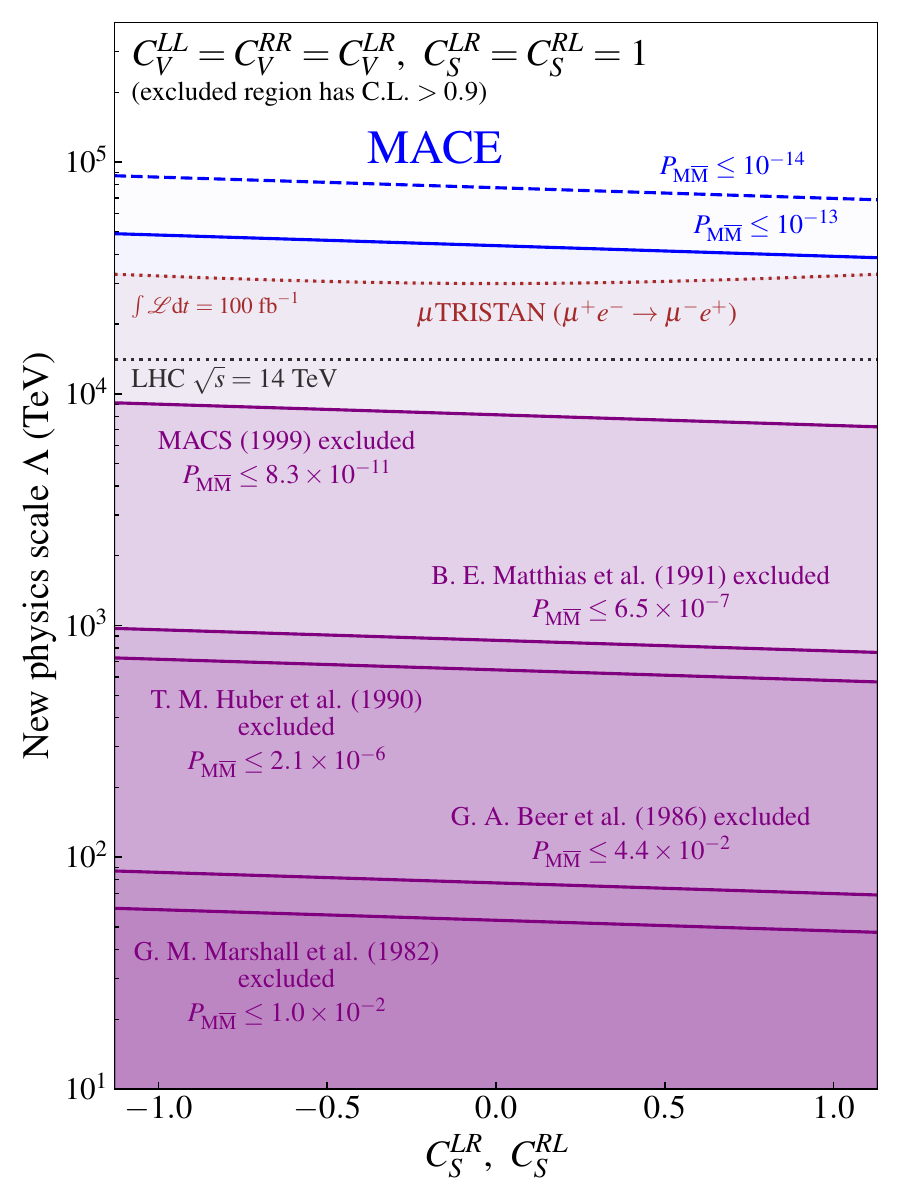}}
    \hfill
    \sidesubfloat[]{\includegraphics[width=0.45\textwidth]{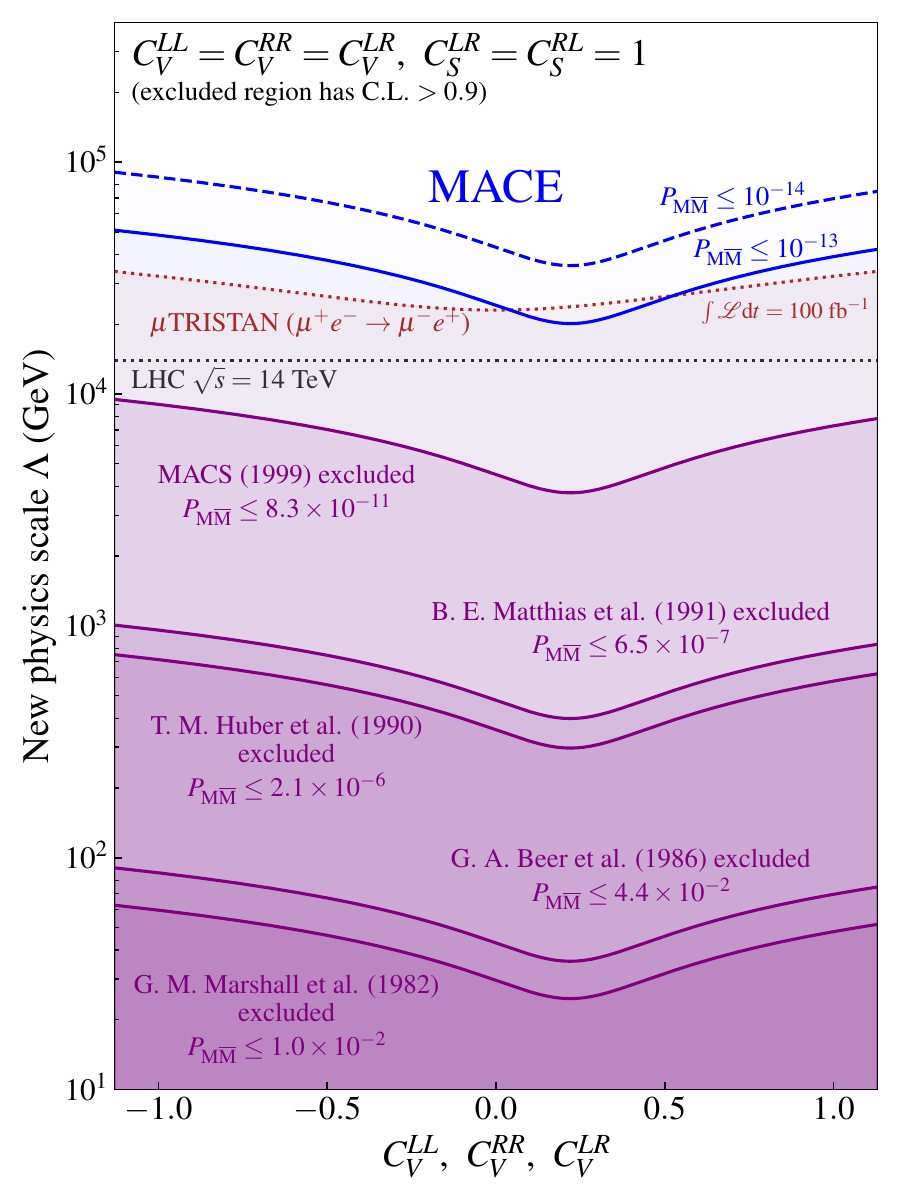}}
    \caption{Operator-dependent energy scale of new physics accessible with experiments searching for $\Delta L_\ell=2$ processes, calculated by Standard Model Effective Field Theory~\cite{Bai:2024skk}. The new physics scales in the same context as all previous $\text{M}-\overline{\text{M}}$ searching experiments and that of the Large Hadron Collider and muon collider $\mu$TRISTAN. (a) The case of fixed vector coupling terms $C^{LL}_V=C^{RR}_V=C^{LR}_V=1$ and equal scalar coupling terms $C^{LR}_S=C^{RL}_S$; (b) The case of fixed scalar coupling terms $C^{LR}_S=C^{RL}_S=1$ and equal vector coupling terms $C^{LL}_V=C^{RR}_V=C^{LR}_V$.}
    \label{fig:MACE_NP_scale}
\end{figure*}

\begin{figure}[!t]
    \centering
    \includegraphics[width=0.9\textwidth]{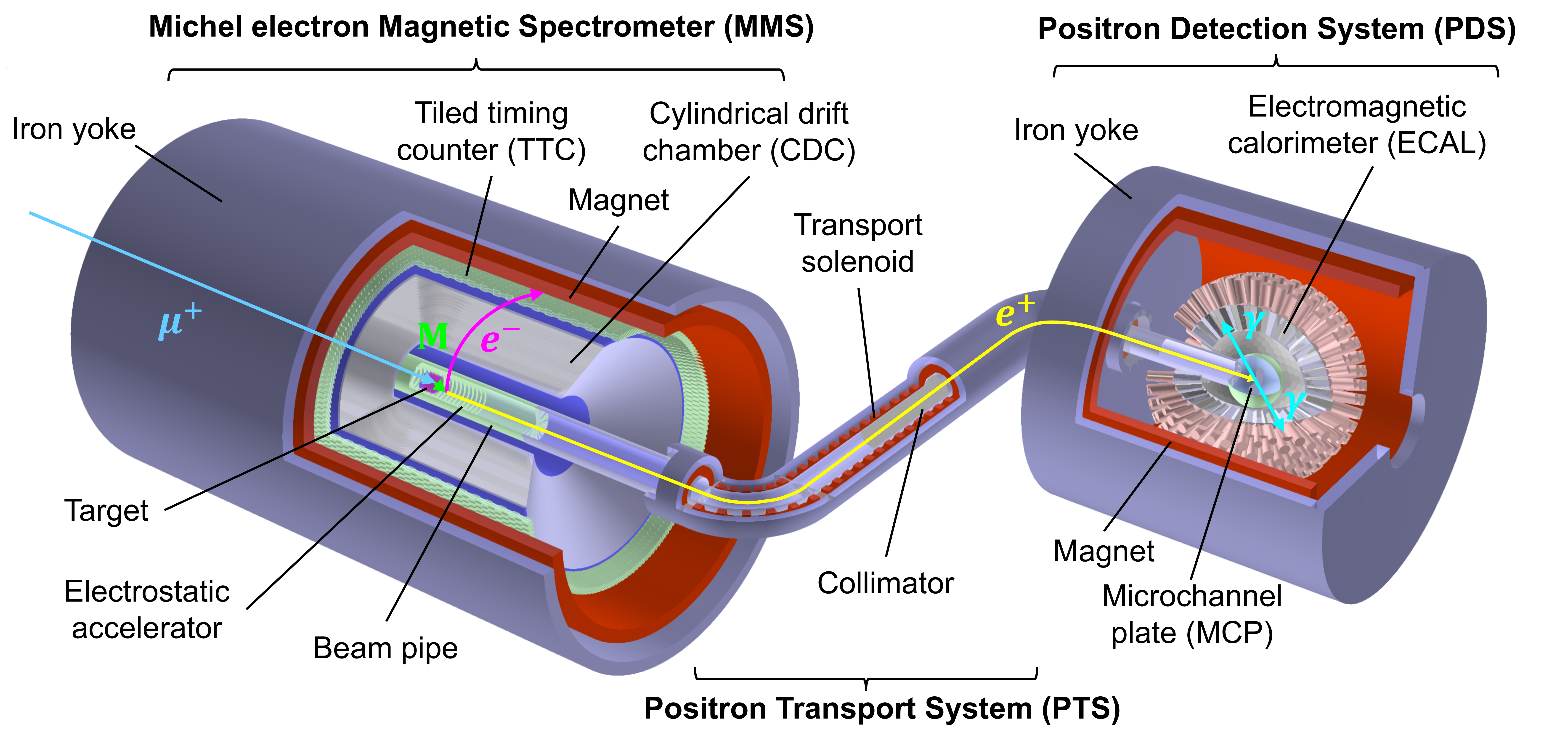}
    \caption{Schematic diagram of Muonium-to-Antimuonium Conversion Experiment. The surface muons are stopped in the target to form muoniums ($\text{M}$), which further diffuse into vacuum and may convert to antimuoniums ($\overline{\text{M}}$). The signal consists of a Michel $e^-$ and an atomic $e^+$ produced by the $\overline{\text{M}}$ decay. The Michel $e^-$ will be detected by the magnetic spectrometer, while the atomic $e^+$ will be accelerated and then transported through a solenoid, which conserves their transverse position. The MCP and ECAL can detect the $e^+$ and its annihilation $\gamma$-rays.}
    \label{fig:mace}
\end{figure}

The Muonium-Antimuonium Conversion Spectrometer (MACS) at PSI adopted a similar scheme~\cite{Willmann:1998gd}. A silica powder target was surrounded by a magnetic spectrometer consisting of five layers of multi-wire proportional chambers (MWPCs) and plastic scintillators to measure the Michel electron tracks. The positron transport system included an electrostatic separator and collimators to suppress the background level. At the end of the transport line, an MCP detector with an MgO coated carbon foil in front was used to enhance the detection efficiency for low-energy positrons. In addition, MACS designed a CsI calorimeter to detect back-to-back 0.511 MeV gamma rays produced by positron annihilation on the MCP, providing further coincidence of positron events. In 1999, MACS reported the updated upper limit of the coupling, which is $G_{\text{M}\overline{\text{M}}}<3.0\times10^{-3}G_{F}$, corresponding to a conversion probability of $P_{\text{M}\overline{\text{M}}}<8.3\times10^{-11}$ (90\% C.L.)~\cite{Willmann:1998gd}. This result remains the most stringent constraint on the $\text{M}-\overline{\text{M}}$ conversion currently.

Taking advantage of the rapid development of particle detection techniques and the construction of muon beams in China, Jian Tang \textit{et al.} have proposed a new experiment aiming to search for the $\text{M}-\overline{\text{M}}$ conversion with a higher sensitivity, known as the Muonium-to-Antimuonium Conversion Experiment (MACE)~\cite{Bai:2022sxq,Zhao:2024qjb,Bai:2024skk,An:2025lws}. \cref{fig:MACE_NP_scale} illustrates the new-physics sensitivity achieved by past experiments as well as that anticipated for the MACE project, while a general and model-independent description is provided by the Standard Model Effective Field Theory (SMEFT). Within this framework, $\Delta L_\ell = 2$ processes including $\text{M}-\overline{\text{M}}$ conversion are governed by the effective Lagrangian
\begin{equation}
    \mathcal{L}_\text{eff} = \frac{1}{\Lambda^2}\left(C^{LL}_V \mathcal{O}^{LL}_V + C^{RR}_V \mathcal{O}^{RR}_V + C^{LR}_V \mathcal{O}^{LR}_V + C^{LR}_S \mathcal{O}^{LR}_S + C^{RL}_S \mathcal{O}^{RL}_S\right),
\end{equation}
where $\Lambda$ is the new physics energy scale, $C^i_j$ are dimensionless Wilson coefficients parameterizing the strength of the new physics interactions, and $\mathcal{O}^i_j$ are the corresponding dimension-6 four-fermion operators involving two muons and two electrons. Given a measured or constrained upper limit on the conversion probability under a 0.1~T magnetic field, the accessible new physics scale is (see Ref.~\cite{Bai:2024skk} for a more comprehensive analysis)
\begin{equation}
    \Lambda \gtrsim \frac{0.02~\text{TeV}}{\left[P^\text{up}_{0.1\,\text{T}}(\text{M}\to\overline{\text{M}})\right]^{1/4}}~.
\end{equation}

The conceptual design of the MACE detector system is as described in \cref{fig:mace}. MACE will utilize a surface muon beam with a momentum of 24 MeV/$c$ and an on-target intensity of $10^8~\mu^+$/s. The detector system consists of a Michel electron magnetic spectrometer (MMS), a momentum-selection positron transport system (PTS), and a positron detection system (PDS). In the conceptual design, subdetectors with higher precision were designed and validated by full simulation~\cite{Zhao:2024qjb,Bai:2024skk,Chen:2024jmg,Lu:2025col}. Particularly, a silica aerogel target is proposed to increase the muonium yield rate by a factor of 2, highly improving the statistical precision~\cite{Zhao:2023plv}. MACE aims to constrain the conversion probability to the $\mathcal{O}(10^{-13})$ level, which is two orders of magnitude better than the previous MACS at PSI. Further improvement on the sensitivity is expected to be achieved by the enhancement of muonium yield of the target and the resolution of each subdetector system.

\subsection{Precise muonium spectroscopy\label{sec:precise_muonium_spectroscopy}}
As a pure leptonic two-body bound system, the energy-level structure of muonium can be understood as the result of adding various small corrections step by step on top of the dominant Coulomb interaction. First, the Coulomb Hamiltonian
\begin{equation}
    H_0=\frac{\mathbf{p}^2}{2\mu}-\frac{\alpha}{4\pi\varepsilon_{0}r}~,
\end{equation}
determines the main energy-level structure of muonium, where $\mu$ is the reduced mass. The corresponding energy eigenvalues are approximately given by
\begin{equation}
    E_n=-\frac{\mu c^2\alpha^2}{2n^2}~,
\end{equation}
This gives the leading frequency scale of the 1S--2S transition. Furthermore, the magnetic dipole interaction between the muon spin and the electron spin leads to the ground-state hyperfine splitting, and the effective Hamiltonian can be written as
\begin{equation}
    H_{\mathrm{hfs}}=A\mathbf{I}_{\mu}\cdot\mathbf{S}_{e},\end{equation}
As a result, the ground state is split into a triplet state and a singlet state, which correspond to the observable ground-state hyperfine transitions. In addition, radiative corrections, such as the electron self-energy and vacuum polarization, cause extra shifts of energy levels that are degenerate in Dirac theory, giving rise to the Lamb shift. Therefore, the main observables in muonium spectroscopy---the 1S--2S transition, the Lamb shift, and the ground-state hyperfine structure---mainly reflect the contributions from the main energy-level structure, radiative corrections, and spin interaction, respectively. Since this system does not contain the nuclear-size effect or the uncertainty from strong-interaction structure that are common in ordinary atoms, its energy levels and transition frequencies can be calculated with high precision within the framework of bound-state quantum electrodynamics. This makes muonium an important platform for testing QED, determining fundamental constants, and exploring fundamental symmetries.

In the following, the present experimental status and future prospects of the muonium Lamb shift, the 1S--2S transition, and the ground-state hyperfine structure will be reviewed. It should also be noted that, as the experimental precision continues to improve in the future, the theoretical model may need to go beyond the present simplified treatment, so that effects like finite muon lifetime broadening can be described more accurately.

\subsubsection{Muonium Lamb shift}
In hydrogen atomic spectroscopy, a discrepancy was observed between the measured energies of the $2S_{1/2}$ and $2P_{1/2}$ levels and the predictions of Dirac theory. In 1947, Lamb and Retherford confirmed the existence of an energy difference between these two levels, now known as the Lamb shift~\cite{Lamb:1947zz}. The discovery of the Lamb shift played an important role in the development of quantum electrodynamics, particularly in advancing the understanding of vacuum polarization and electron self-energy corrections.
Muonium is a hydrogen-like bound state governed purely by quantum electrodynamics (QED), making it a promising system for high-precision tests of the Standard Model and searches for new physics~\cite{Janka:2021xxr}, such as the Lorentz and CPT violation~\cite{Gomes:2014kaa}, dark matter~\cite{Stadnik:2022gaa}, and related topics.

\begin{figure}[t!]
    \centering
    \includegraphics[width=0.9\textwidth]{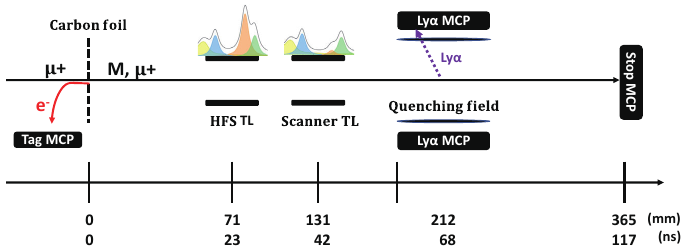}
    \caption{Schematic diagram of the Lamb shift measurement in Mu-MASS experiment (reproduced from Ref.~\cite{Janka:2022pis}). The metastable 2S muoniums are formed in a thin carbon foil. Secondary electrons emitted from the foil are detected by the Tag-MCP. The muoniums pass through two transmission lines (TL). The 2S state is quenched by a grid electric field and decays to the 2P state, then to the ground state. The emitted Lyman-$\alpha$ photons are detected by an MCP. Finally, the remain muoniums are stopped by another MCP. A signal will be determined by triple-coincidence among the Tag, Ly-$\alpha$, and Stop MCPS.}
    \label{fig:mumass}
\end{figure}

Both TRIUMF and LAMPF conducted measurements of the Lamb shift in muonium, reporting results of $1070^{+12}_{-15}$ MHz~\cite{Oram:1983sd} and $1042^{+21}_{-23}$~MHz~\cite{Woodle:1990ky}, respectively. The precision of these measurements was primarily limited by low statistics and high background levels. In 2021, the MuoniuM lAser SpectroScopy (Mu-MASS) experiment as shown in \cref{fig:mumass} at PSI reported the most recent result of $1047.2(2.3)_{\text{stat}}(1.1)_{\text{syst}}$ MHz~\cite{Mu-MASS:2021uou}, representing an order of magnitude improvement comparing to the previous measurement. The experiment utilized surface muons from the $\mu$E4 beamline. Muons were first slowed down to 20 eV by a Ne moderator, then re-accelerated to 7.5 kV and formed muonium atoms in a 10 nm thick carbon foil target. Part of the muonium atoms were excited to the long-lived 2S state. A static electric field of 250 V/cm mixed the 2S and 2P states, relaxing them to the ground state within a few ns, accompanied by the emission of 122 nm Lyman-alpha photons. When microwave radiation resonant with the $2S_{1/2}\to 2P_{1/2}$ transition was applied, it reduced the population of 2S muonium atoms, thereby reducing the Lyman-alpha photon signal. By scanning the microwave frequency around the transition, the Lamb shift can be determined by the centroid energy difference between $2S_{1/2}$ and $2P_{1/2}$ averaged over hyperfine structure.

\subsubsection{Muonium 1S-2S transition}
In 2000, V.~Meyer \textit{et al.} conducted a precision measurement of the 1S-2S transition frequency of muonium at the ISIS pulsed muon source~\cite{Meyer:1999cx}. Muonium was produced by a silica powder target, and was tagged by detecting the Michel positrons with two MWPCs. A key innovation in this experiment was the design of a high-precision laser system, which could minimize the chirping in the laser amplifier, and therefore reported a more precise result than previous measurements. The system utilized a CW Ti:sapphire laser operating at 732~nm, with a tunable pulse frequency (with in 60 MHz interval) controlled by an acousto-optic modulator (AOM). The beam was reflected multiple times near the target, interacting with muonium, and finally getting back to itself. A part of muonium transitioned from 1S state to 2S state by the counter-propagating beam. The same laser field photoionized the 2S state muonium. The released muons were accelerated to 2 keV and transported to an MCP by an electrostatic lens and a bending magnet. Scintillator detectors covering 94\% of the solid angle around the MCP were used to tag Michel positrons. By scanning the laser frequency and recording the MCP counts, the 1S-2S transition frequency of muonium is determined to be 2 455 528 941.0(9.8)~MHz~\cite{Meyer:1999cx}. Additionally, this experiment also measures the 1S-2S transition of deuterium, which turns out to be in good agreement with previous results, further validating the high precision of this system.

\begin{figure}[t!]
    \centering
    \includegraphics[width=0.8\textwidth]{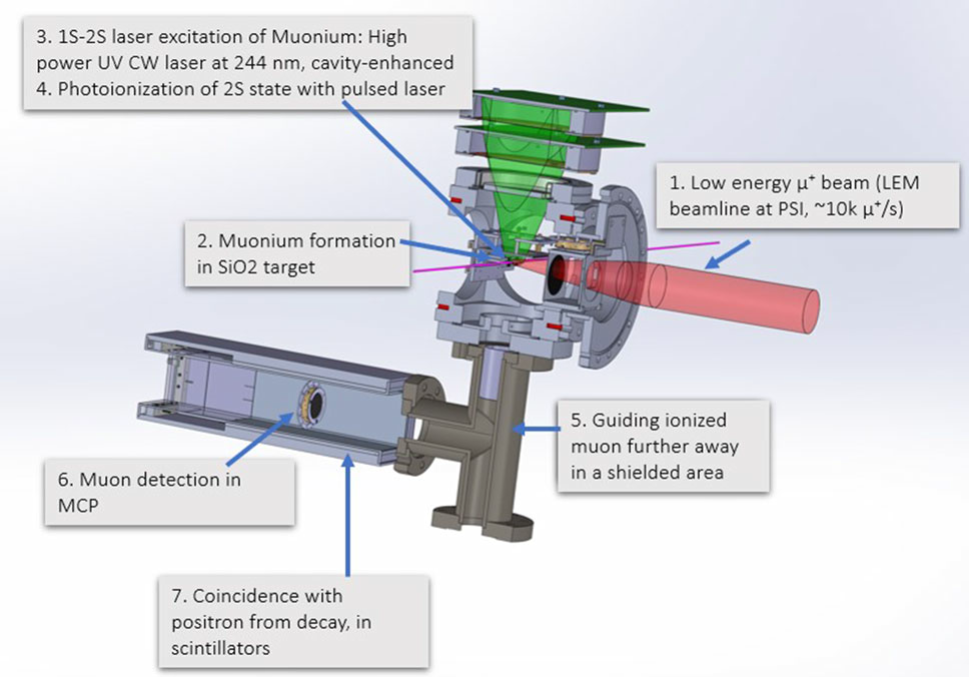}
    \caption{Schematic diagram of the 1S-2S transition frequency measurement in the Mu-MASS experiment (reproduced from Ref.~\cite{Cortinovis:2023zqi}).}
    \label{fig:mumass-1s2s}
\end{figure}

However, the chirping effect of pulsed laser beams still significantly limits the precision of measurement. With development in laser technology, CW laser beams have become available to reduce the chirping effect, leading to a possible improvement of two orders of magnitude. Furthermore, if Doppler shifts and AC Stark shifts can be corrected, the precision could be improved by an additional order of magnitude~\cite{Ohayon:2021dec}. Ref.~\cite{Zhadnov:2023vcr} provides a detailed introduction of the laser system setup. The Mu-MASS experiment at PSI will produce muonium in a porous silica film target at 100 K using low-energy muons from the $\mu$E4 beamline. The experimental setup can be seen in \cref{fig:mumass-1s2s}. Under these conditions, the muonium formation efficiency can reach approximately 20\%. Taking advantage of the advanced laser techniques and a new detection scheme from the positronium 1S-2S spectroscopy, the uncertainty in the transition probability measurement is expected to be reduced to the level of 10 kHz~\cite{Crivelli:2018vfe,Cortinovis:2023zqi}. Additionally, it's worth mentioning that the Advanced Muonium Laser Experiment at Tokai (AmuLET) collaboration has preliminarily reported the first measurement of the $F=0\to F^\prime=0$ transition recently~\cite{UetakeSSP2025}.

\subsubsection{Muonium hyperfine structure}
In 1999, W.~Liu \textit{et al.} reported the most precise measurement of the muonium hyperfine structure interval under the high-field condition~\cite{Liu:1999iz}. The experiment was carried out at the LAMPF. Muons were directed into a strong magnetic field of about 1.7 T. At the center, a copper microwave resonant cavity was filled with high-purity krypton gas (pressure of 0.8 or 1.5 atm) as the muonium production target. In the strong magnetic field, the ground-state energy levels of muonium splits caused by Zeeman effect. A transverse microwave field was applied, and its frequency was finely tuned to match Zeeman transitions $\nu_{12}$ and $\nu_{34}$. The muon spin flips, causing the angular distribution asymmetry of the Michel positrons. The positrons were detected by a scintillator counter downstream of the cavity, and the resonance signal was constructed from the difference in counts with the microwave on and off. In the traditional method, the resonance linewidth is limited by the muon lifetime ($\sim145$~kHz). This experiment applied a resonance line narrowing technique~\cite{Liu:1995hq}. The muon beam was time-modulated by a chopper (beam period with 4 $\mu$s on, 10 $\mu$s off), and Michel positrons were detected in eleven time windows of 0.95 $\mu$s. The ``old" muonium survived several times longer than the muon lifetime. The produced linewidth is three times narrower and stronger in amplitude than in the previous experiments. This innovation greatly improved the precision of measurement, resulting in the muonium hyperfine structure $\Delta \nu=4463.302765(53)$ MHz (12~ppb), the magnetic moment ratio $\mu_\mu/\mu_p=3.18334513(39)$ (120 ppb)~\cite{Liu:1999iz}. These measured values could deduce the $m_\mu/m_e=206.7682670(55)$ (27~ppb), and the fine structure constant $\alpha^{-1}=137.0359963(80)$ (58 ppb)~\cite{Liu:1999iz}.

\begin{figure}[t!]
    \centering
    \includegraphics[width=0.9\textwidth]{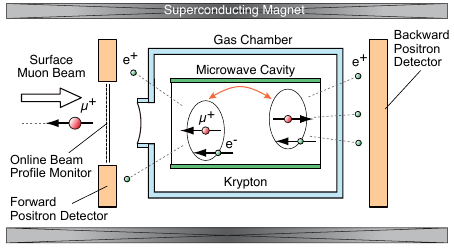}
    \caption{Schematic diagram of the MuSEUM experiment (reproduced from Ref.~\cite{MuSEUM:2025cmo}). The apparatus is surrounded by a superconducting solenoid which provides a high field up to 2.9~T. The incident muon beam profile is recorded by a monitor made of scintillating fibers. The gas chamber is filled with Kr gas to form muonium and contains a microwave cavity to induce muon spin flip. Once the microwave field is applied, the asymmetry of positrons from muon decay is measured by the forward and backward positron detectors (segmented scintillators or silicon strips).}
    \label{fig:museum}
\end{figure}

In 2021, the MuSEUM (Muonium Spectroscopy Experiment Using Microwave) collaboration reported a measurement of the muonium hyperfine structure~\cite{Kanda:2020mmc}. The experiment was conducted at the high-intensity pulsed muon source of J-PARC, as shown in \cref{fig:museum}. Muons were stopped and formed muonium in a krypton gas chamber at 1 bar. The hyperfine structure corresponds to the energy splitting between the spin-singlet state (25\%) and spin-triplet state (75\%) of muonium. An oxygen-free copper resonant cavity generated a microwave field, which could be tuned in the range of 4461--4464.8~MHz by inserting a movable aluminum rod in the chamber. Magnetic shielding was applied to keep a near-zero field. Downstream of the chamber, a two-layer plastic scintillator array coupled to SiPMs was used for coincidence measurements of Michel positrons. The final measured value of the muonium HFS is 4463.302(4) MHz, with a relative precision of 0.9 ppm~\cite{Kanda:2020mmc}. A subsequent study introduced a new method, Rabi-oscillation spectroscopy, which improves the precision by a factor of two compared with conventional frequency-swept spectroscopy. By taking advantage of this more efficient measurement, the result reaches 160 ppb, which is the highest precision under zero magnetic field to date~\cite{MuSEUM:2020mzm}. In the future, the MuSEUM experiment aims to achieve 12 ppb under zero-field conditions and 1.2 ppb under high-field conditions with a higher intensity muon beam~\cite{MuSEUM:2025cmo}.

\subsection{Precise measurement of antimatter gravity}

General relativity is currently the most successful theory of gravitation~\cite{Einstein:1916vd}, with its most recent experimental confirmation coming from the observation of gravitational waves in 2016~\cite{LIGOScientific:2016aoc}. However, the universe is now believed to be permeated by large amounts of invisible dark matter and dark energy. These phenomena remain unexplained. Moreover, the incompatibility between general relativity and quantum field theory suggests that a more complete theoretical framework is still pursued. The weak equivalence principle (WEP) states the universality of free fall, that all forms of matter respond identically in a gravitational field. At present, the equivalence principle has been verified to extremely high precision for both macroscopic and microscopic matter. For example, the MICROSCOPE experiment compared the accelerations of titanium and platinum alloys in a satellite environment, constraining the Eötvös parameter to $\mathcal{O}(10^{-15})$~\cite{MICROSCOPE:2022doy}. In addition, P.~Asenbaum \textit{et al.} reported a measurement using atom interferometry to study the difference between $^{85}$Rb and $^{87}$Rb, achieving $\mathcal{O}(10^{-12})$~\cite{Asenbaum:2020era}. However, due to technical limitations, there have been few experimental measurements of the free fall of antimatter so far.

In general, charged antiparticles such as $p^-$ and $e^+$ experience electromagnetic forces that are orders of magnitude stronger than gravity, making direct gravitational measurements extremely challenging. Neutral bound-state atoms therefore provide an alternative~\cite{Tino:2020dsl}. In 2023, E.~K.~Anderson \textit{et al.} reported measurements of antihydrogen, marking the first observation of the gravitational effect of antimatter~\cite{ALPHA:2023dec}. In the experiment, antiprotons and positrons were combined to form antihydrogen atoms, which were confined in a vertical magnetic trap. Additional magnetic fields were manually applied at the bottom or top of the trap to counteract or enhance the effect of gravity. The results showed that an upward magnetic force equivalent to $1~g$ was required to equalize the escape probabilities from the top and bottom of the trap, corresponding to cancellation of gravity. The result indicates that the hypothesis of no gravitational interaction for antihydrogen is excluded with a probability of $2.9\times10^{-4}$. Further improvements in experimental precision could be achieved by using laser-cooled antihydrogen atoms.

\begin{figure}[t!]
    \centering
    \includegraphics[width=0.9\textwidth]{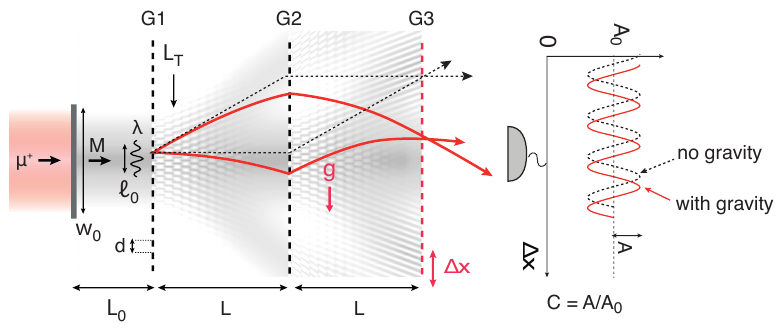}
    \caption{Schematic diagram of the LEMING experiment (reproduced from Ref.~\cite{Soter:2021xuf}). G1, G2, G3 stand for gratings. The pattern of muonium will be recorded by the coincidence detection of Michel $e^+$ and atomic $e^-$ downstream. By adjusting G3 in the vertical direction, the shift of muonium caused by gravity can be derived.}
    \label{fig:leming}
\end{figure}

The majority of the hydrogen atom's rest mass is caused by the strong interaction within its nucleus. In contrast, hydrogen-like purely leptonic bound states, such as positronium ($e^+e^-$) and muonium ($\mu^+e^-$), derive almost all of their mass from the leptons, making them more suitable candidates for gravity measurement. Moreover, muonium is currently the most accessible bound-state atom containing a second-generation lepton. Therefore, muonium has been considered as a promising subject for investigations of antimatter gravity. All current proposals for muonium gravity measurements are based on atom interferometry~\cite{Kirch:2014mna,Phillips:2018twx,MAGE:2018wxk,Soter:2021xuf}. The experimental principle can be summarized as follows (as also can be seen in \cref{fig:leming}):
\begin{enumerate*}[(1)]
    \item muons are stopped in a target and emitted into vacuum as muonium;
    \item muoniums pass through an interferometer consisting of three gratings with a period of $\sim100$~nm, while the de Broglie wavelength of muonium is about $1.6~\mathrm{nm}$;
    \item under gravity, muoniums fall freely through a vertical displacement in flight, causing a vertical phase shift of the interference fringes;
    \item at the end, coincident detection of the Michel positron and the low-energy orbital electron is performed to determine the muonium transmission rate.
\end{enumerate*}
This experimental method demands a very high-quality muonium beam. As a result, current proposals tend to use SFHe, as mentioned in \cref{sec:production}, instead of silica aerogel as the muonium formation target. There are two experiments aimed at the behavior of muonium under gravity, known as the Muonium Antimatter Gravity Experiment (MAGE)~\cite{MAGE:2018wxk} and LEptons in Muonium Interacting with Gravity (LEMING)~\cite{Soter:2021xuf}, which are still in the technical design and demonstration stage.

The LEMING experiment employs a grating period of $d\sim100~\mathrm{nm}$, with an interaction time of about $4~\mu\mathrm{s}$. The interference contrast is expected to be $C>0.3$, and the high-contrast region is about $1$--$2~\mu\mathrm{m}$. Although position stability and vibration control remain highly challenging, LEMING design studies include cryogenic Fabry--Perot probes with a relative-position resolution of about $5~\mathrm{pm}$ for in situ calibration, vibration monitoring, and damping. The piezo scan position has been shown to be reproducible at the level of about $2$--$5~\mathrm{nm}$ over repeated cycles, while measurements with the pulse tube operating indicate a relative displacement below about $10~\mathrm{nm}$ for plates separated by $5~\mathrm{mm}$. The positron will be detected by a cryogenic silicon strip tracker, while the electron will be detected by a superconducting nanowire single-photon detector. The sign of $g$ is expected to be determined less than a day~\cite{soter2025leming_slides}.

\section{Applications of muon beams}

As we discussed in the previous section, muons are widely used in scientific research due to their unique physical properties.
Meanwhile, muon-related applications have been developed with prosperity in recent years.
Before the advent of accelerator muon beamlines, muon-related applications were mainly restricted to muography using cosmic-ray muons.
Nowadays, advanced muon beamlines provide an environment to advance muon-related application technologies and broaden the scope of muon science.
Since the application technologies' principles and setups are highly dependent on the charge of the muon beam, the following section provides an overview of the technologies for positive and negative beamlines, respectively.

\begin{figure}[t!]
    \centering
    \includegraphics[width = \linewidth]{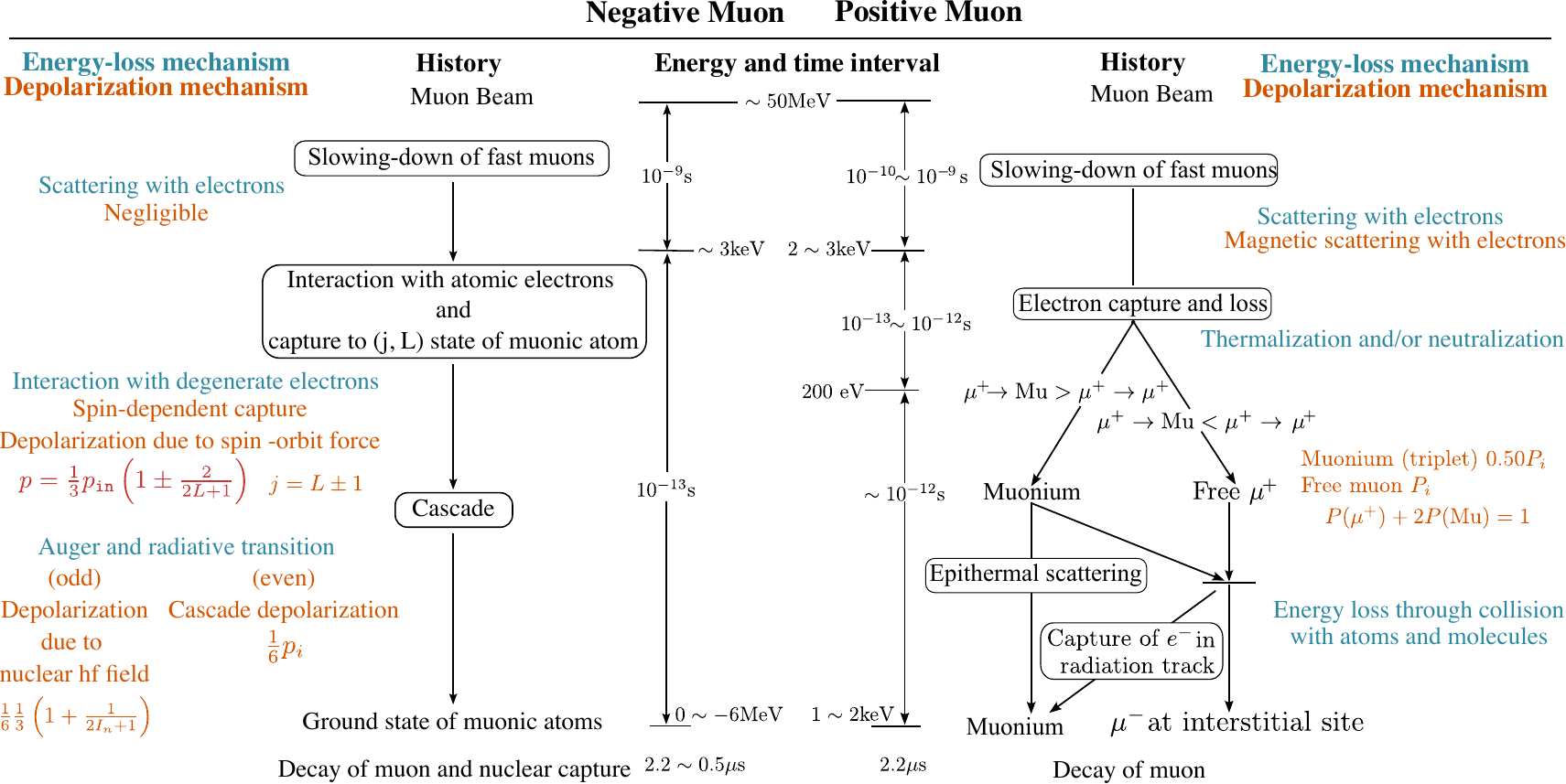}
    \caption{Interaction during the $\mu^\pm$ injection into the material.
        The right part shows the interaction of $\mu^+$, and the left part shows the case for $\mu^-$.
        During the injection, $\mu^+$ can form muonium or remain free while $\mu^-$ will either form a muonic atom or remain free.}
    \label{fig:muon_injection_interaction}
\end{figure}

\subsection{Application of positive muon}

Positive muon beams enable a diverse set of applications, with $\mu$SR being the most prominent.
Developed since the 1980s~\cite{HughesWu1975MuonPhysicsV3}, $\mu$SR has become an indispensable tool for probing the local magnetic properties of materials.
It is now widely utilized in fields such as superconductivity, magnetism, and the study of spin glasses~\cite{de2009muon, bendele2010pressure, coneri2010magnetic, PhysRevB.79.184430}.
Furthermore, recent advancements in accelerator technology have expanded its reach to dynamic processes, including carrier dynamics in organic semiconductors and electron dynamics in chemical reactions~\cite{PhysRevB.110.134504, PhysRevLett.103.147601, cox2009muonium}.
The power of $\mu$SR stems from several key experimental features:
\begin{enumerate*}[(1)]
    \item It directly measures the local magnetic field within a material, even in the absence of an external field.
    \item As an implanted probe, it does not rely on the presence of specific local paramagnetic centers (unlike ESR~\cite{zavoisky1945spin}) or nuclear spins (unlike NMR~\cite{rabi1938new}).
    \item The implanted muon can form a muonium, which serves as a sensitive probe for studying advanced material properties and chemical reactions~\cite{stadnik2023searching, king2010observation, cox2009muonium}.
\end{enumerate*}
To provide a comprehensive overview, our discussion of these applications will be guided by the structure in the monograph by Amato \textit{et al.}~\cite{amato2024introduction}. We will first explore the fundamental physical principles of $\mu^+$ that inspired the $\mu$SR technique, then summarize its representative applications, and finally discuss the crucial role of muonium.

\subsubsection{Basic concepts of \texorpdfstring{$\mu$}{mu}SR}
In $\mu$SR experiments, the measured signals arise from interactions between implanted muons and the local magnetic environment. A brief overview of these interactions shows how muon spin precession and relaxation reflect the system's microscopic magnetic and electronic properties.

After establishing a muon beamline with promising polarization characteristics, the next step is to inject the muon beam into the sample under investigation.
The interaction processes that occur during muon injection into the material are illustrated in~\cref{fig:muon_injection_interaction}.
Assuming an initial muon momentum of $50~\mathrm{MeV}/c$, the muon is rapidly slowed down after entering the material due to multiple scattering with electrons.
This scattering occurs on a timescale of approximately 1~ns and reduces the kinetic energy of muon to about 2--3 keV.
At this point, the muon may form a bound state with electrons or atoms in the material.
For positive muons ($\mu^+$), they will capture an electron to form muonium ($\mu^+ e^-$), while the remaining $\mu^+$ particles will remain unbound and free.
The free $\mu^+$ and muonium typically retain kinetic energies of $1$-$2~\mathrm{keV}$.
In the case of negative muons ($\mu^-$), some will interact with atomic electrons to form muonic atoms.
In contrast, others will remain free and stable, with kinetic energies of a few keV.
For free $\mu^\pm$, depolarization effects are negligible.
However, depolarization becomes significantly more pronounced when muons form bound states such as muonium or muonic atoms.
For the muonic atom consisting of a muon and a non-zero nucleon, the muon will undergo a strong depolarization effect, leading to the loss of most of the initial polarization.
In contrast, the muon in muonium is only slightly depolarized due to the hyperfine interaction between the electron and the muon.
Therefore, $\mu^+$ keeps a stronger polarization than $\mu^-$ when passing through the material, which plays a more significant role in the $\mu$SR technique construction.

\begin{figure}[t!]
    \centering
    \includegraphics[width = .4 \linewidth]{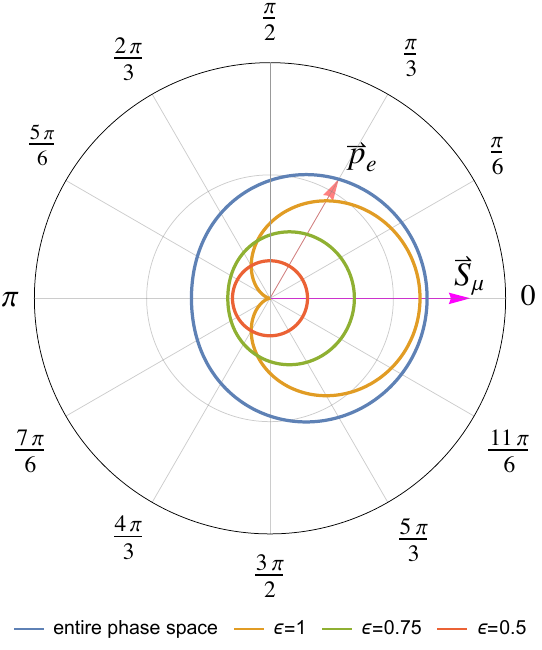}
    \caption{Michel electron spatial distribution as a function of the reduced energy $\epsilon = 2E_e/m_\mu$, where $E_e$ is the total electron energy. The blue curve shows the standard Michel spectrum obtained by integrating over the phase space.
        The yellow, green and orange curves represent Michel spectrum for $\epsilon = 1$, $\epsilon = 0.75$, and $\epsilon = 0.5$, respectively.}
    \label{fig:Michel_spatial_distribution}
\end{figure}

After losing its kinetic energy and stopping in the material, the muon undergoes spin evolution in the local magnetic field until it decays.
During this intermediate period, the primary interaction involves the material's magnetic fields, either intrinsic or externally applied.
To illustrate the basic mechanism, we first consider the case of positive muons ($\mu^+$) and neglect depolarization effects.
The spatial distribution of Michel electrons emitted from a fully polarized muon is given by
\begin{equation}
    N_e(E_e) \propto 1 + \frac{1}{3} \cos \theta_e~,
\end{equation}
where $\theta_e$ is the angle between the muon spin vector $\vec{\sigma}_\mu$ and the Michel electron momentum $\vec{p}_e$.
The coefficient $1/3$ originates from integrating over the Michel electron energy spectrum.
The angular distribution relative to $\vec{\sigma}_\mu$ and $\vec{p}_e$ is plotted in~\cref{fig:Michel_spatial_distribution}.

\begin{figure}[t!]
    \centering
    \includegraphics[width = 0.7\textwidth]{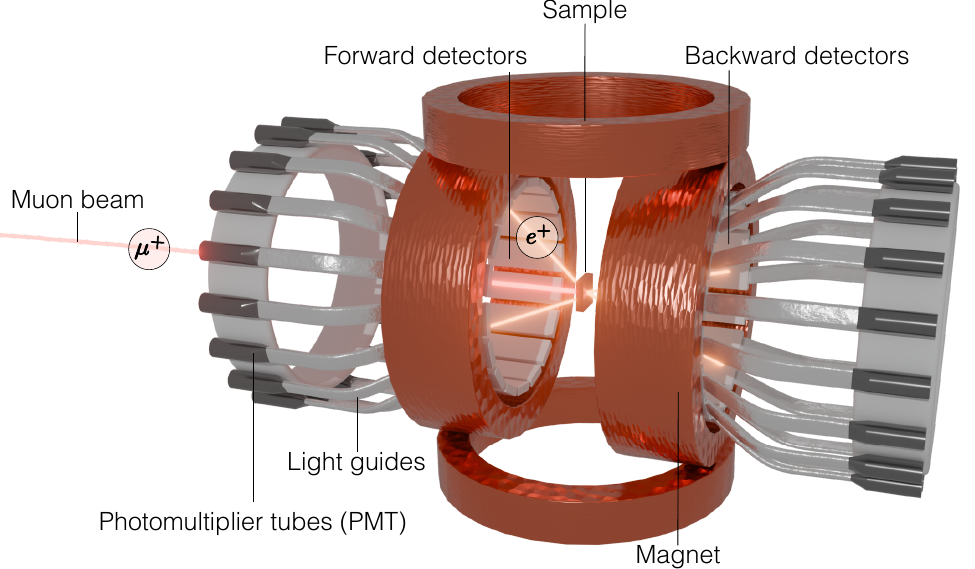}
    \caption{Schematic diagram of a conventional $\mu$SR spectrometer configuration.
        The red trajectory indicates the incident spin-polarized muon beam ($\mu^+$), which stops within the sample. The orange trajectories represent the decay positrons (Michel positrons) emitted anisotropically, which are detected by the scintillator modules with photomultiplier tubes (PMTs). Two pairs of copper Helmholtz coils are arranged to apply external magnetic fields: one pair provides a magnetic field perpendicular to the initial muon spin direction (Transverse Field, TF), and the other for the field parallel to the beam direction (Longitudinal Field, LF).}
    \label{fig:muSR_apparatus}
\end{figure}

Commonly, an array of detectors surrounding the sample measures the number of decay positrons arriving at different locations over time to reconstruct the polarization evolution.
The typical configuration of a $\mu$SR spectrometer is illustrated in~\cref{fig:muSR_apparatus}.
The detector modules are typically arranged symmetrically around the sample to capture positrons emitted in various directions.
The detectors are usually scintillator counters coupled to photomultiplier tubes (PMTs) for high-efficiency positron detection.
Recently, silicon photomultipliers (SiPMs) have also been employed as alternatives to PMTs due to their compact size and insensitivity to magnetic fields~\cite{Amato:2017gps, Kojima:2014qwv, Yang:2025lae}.
By applying different external fields, the muon polarization reveals the magnetic properties of the samples.
In the simple case of a zero magnetic field, the implanted muons' spins do not precess. Consequently, the spatial distribution of the positrons remains static, reflecting only the initial spin polarization.

\begin{figure}[t!]
    \centering
    \includegraphics[width = \linewidth]{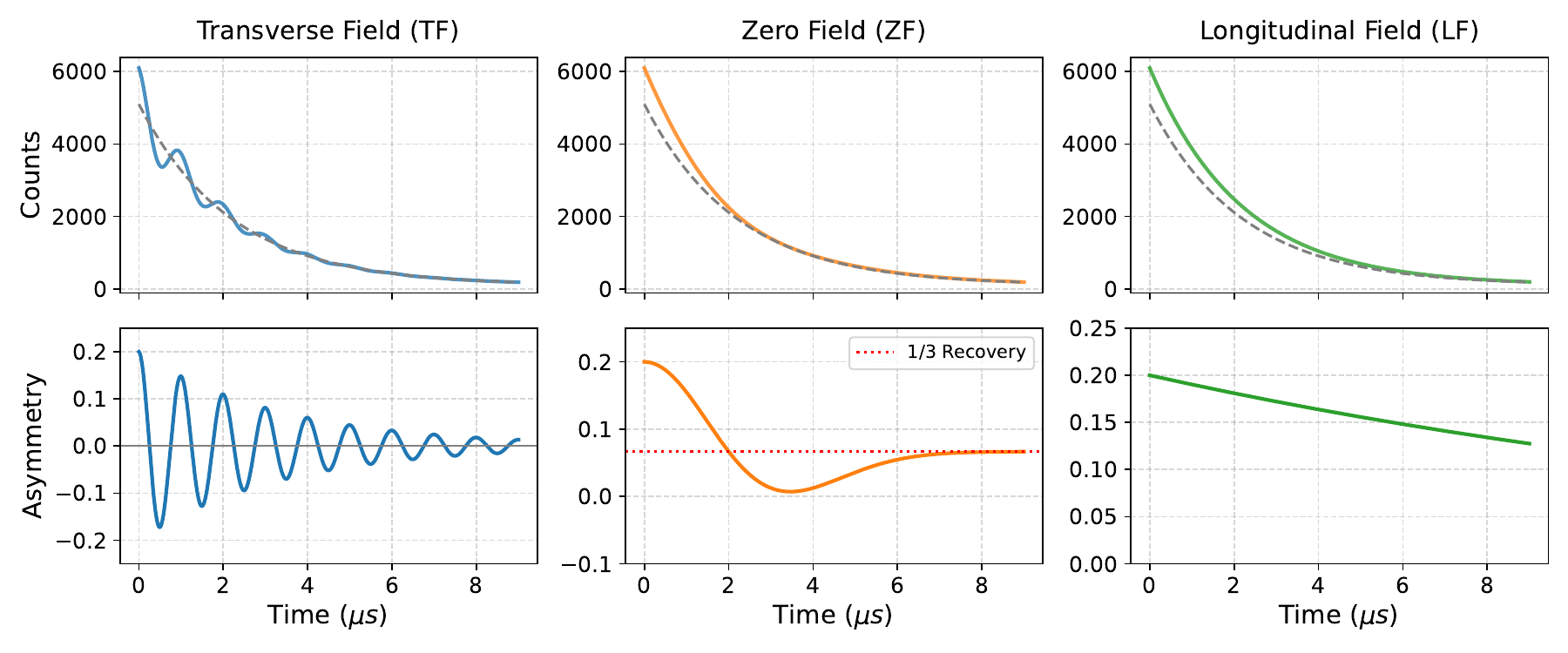}
    \caption{Typical muon spin rotation/relaxation/resonance ($\mu$SR) signals in three types of applied magnetic fields.
        The top row shows raw detector counts $N(t)$ for transverse field (TF), zero field (ZF), and longitudinal field (LF) measurements. The bottom-left plot shows TF-$\mu$SR asymmetry as a function of time, where the signal swings at a rate set by the local field. The bottom-middle plot shows the time evolution of the ZF-$\mu$SR asymmetry, where internal fields point to random directions. The bottom-right plot shows LF-$\mu$SR asymmetry over time, where the external field keeps the muon spin steady.}
    \label{fig:muSR_3_tipical_signal}
\end{figure}

When the local magnetic field is perpendicular to the initial muon spin (or equivalently, to the beamline direction), the muon spin precesses around the field direction, leading to a time-dependent modulation in the Michel electron distribution:
\begin{equation}
    N(\theta, t) = N_0 \exp(-t / \tau_\mu)
    \left(
    1 + A \cos (\theta - \omega_\mu t)
    \right)~,
\end{equation}
where $\tau_\mu$ is the muon lifetime, and the muon spin precession frequency $\omega_\mu$ is related to the local magnetic field $H_0$ via the muon gyromagnetic ratio $\gamma_\mu$:
\begin{equation}
    \omega_\mu = \gamma_\mu H_0 = H_0 (g_\mu e h / 2 m_\mu c)~.
\end{equation}
In practical units, this gives
\begin{equation}
    F_\mu (\text{kHz}) = 13.553 \times H_\mu (\text{G})~.
\end{equation}
$\theta$ stands for the angle between the beamline direction and the direction from the sample pointing to the Michel electron detector.
This transverse-field configuration is one of the most commonly used setups in $\mu$SR experiments (TF-$\mu$SR), as it directly connects the measurable precession frequency with the local magnetic field at the muon site.
The typical TF-$\mu$SR signal is shown in~\cref{fig:muSR_3_tipical_signal} (top left and bottom left).
In other experimental options, such as longitudinal-field and zero-field $\mu$SR, the relaxation of the polarization rather than its precession provides information about the static and dynamic characteristics of the local field distribution.
Together, these complementary configurations allow $\mu$SR to probe a wide range of magnetic phenomena in materials.

In muon spin relaxation measurements, no external magnetic field is applied (zero-field $\mu$SR), or the external field is aligned parallel to the initial direction of the muon spin (longitudinal-field $\mu$SR).
The typical signals for these configurations are shown in~\cref{fig:muSR_3_tipical_signal} (top middle, top right, bottom middle, and bottom right).
In these configurations, the muon spin does not precess coherently as in the transverse-field case. Instead, local magnetic fields at the muon sites cause the spin to relax gradually. These local fields arise from nearby atomic or nuclear moments in the material.
When the local magnetic fields are randomly oriented, the longitudinal component of the muon spin exhibits time-dependent depolarization, described by
\begin{equation}
    \begin{aligned}
         & G_z(t) = \langle \sigma_z (t) \sigma_z (0) \rangle~, \
         & N (\theta, t) = N_0 \exp(-t / \tau_\mu) [1 + A G_z(t) \cos \theta]~,
    \end{aligned}
\end{equation}
where $\sigma_z$ represents the component of the muon spin aligned with its original direction of polarization.
The relaxation function $G_z(t)$ provides insight into both the static characteristics of the local magnetic fields and their temporal fluctuations within the material.
Notably, longitudinal relaxation is observable even in the absence of an external field, making it a sensitive probe of spin dynamics and internal magnetic disorder.

Based on the relaxation measurements, the muon gradually depolarizes under the local magnetic field. The muon spin resonance technique allowed active control of the spin state through an external radio-frequency (RF) magnetic field applied along the initial muon spin direction. The RF field can create Zeeman splitting, but it does not flip the muon spin directly. Instead, when the radio frequency matches the Zeeman energy gap, the muon spin will flip due to the hyperfine interaction. By combining precession, relaxation, and resonance measurements, researchers can fully probe the local magnetic properties of materials. The following sections describe applications of $\mu$SR in condensed matter physics and materials science.

The performance of $\mu$SR spectroscopy highly depends on the intensity and polarization of the muon beam.
Modern accelerator muon sources provide high-quality beams that satisfy precise measurements of spin precession and relaxation.
A spectrometer requires several essential components, including a beam transport system, a controlled sample environment for temperature and magnetic fields, and fast-timing detector arrays.
Table~\ref{tab:key_performance_muSR} presents an overview of key performance parameters for several leading $\mu$SR spectrometers currently operating worldwide. It is not intended as a direct performance ranking or comparison. In particular, it should be noted that the ISIS facility at RAL uses a pulsed muon beam, whereas PSI and other facilities operate with continuous or quasi-continuous beams. These fundamentally different time structures lead to intrinsically different time resolutions, and the parameters listed should be understood within the context of each beam type.
With such instrumentation, $\mu$SR has become a powerful tool to explore a broad spectrum of material properties, ranging from superconductivity to complex magnetic phases.

\begin{table}[t]
    \captionsetup{justification=raggedright,singlelinecheck=false}
    \caption{Key performance parameters of currently operating muon spin spectroscopy instruments.
        The PSI GPS refers to the general-purpose surface-muon spectrometer at PSI, with performance information taken from Refs.\cite{Amato:2017gps, PSI:GPSweb}.
        The ISIS MuSR corresponds to the muon spin spectroscopy instruments at ISIS, with performance details obtained from Refs.\cite{Lord:2011musr,Baker:2023SuperMuSR}.
        The CMMS entry represents the muon spin spectrometers at TRIUMF, with specifications taken from the official TRIUMF documentation~\cite{TRIUMF:CMMS}.}
    \label{tab:key_performance_muSR}
    \centering
    \begin{tabular}{cccc}
        \textbf{}                                                                                  &
        \cellcolor[HTML]{EFEFEF}\textbf{PSI GPS}                                                   &
        \cellcolor[HTML]{EFEFEF}\textbf{ISIS MuSR}                                                 &
        \cellcolor[HTML]{EFEFEF}\textbf{\ TRIUMF CMMS\ }                                             \\
        \hline
        \textbf{Time resolution}                                                                   &
        $\sim$158 ps                                                                               &
        $\sim$700 ps                                                                               &
        $\sim$300--500 ps                                                                            \\
        \hline
        \textbf{\begin{tabular}[c]{@{}c@{}} Magnetic field \\ range\end{tabular}}                  &
        Up to $2.5~\text{T}$                                                                       &
        \begin{tabular}[c]{@{}c@{}} 0--0.3 T (longitudinal) \\ 0--0.06 T (transverse)\end{tabular} &
        0--2 T                                                                                       \\
        \hline
        \textbf{\begin{tabular}[c]{@{}c@{}}Temperature \\ range\end{tabular}}                      &
        $20~\text{mK} \text{--} 300~\text{K}   $                                                   &
        $40~\text{mK} \text{--} 1000~\text{K}$                                                     &
        $2~\text{K}\text{--}300~\text{K}$                                                            \\
        \hline
        \textbf{Event rate}                                                                        &
        $\sim10^7\text{--}10^8$ per hour                                                           &
        $\sim 10^8$ per hour                                                                       &
        $\sim 10^7$ per hour                                                                         \\
        \hline
        \textbf{Asymmetry}                                                                         &
        $\sim 0.25\text{--}0.28$                                                                   &
        $\sim 0.20\text{--}0.25$                                                                   &
        $\sim 0.22\text{--}0.26$                                                                     \\
        \hline
    \end{tabular}
\end{table}

\subsubsection{\texorpdfstring{$\mu$}{mu}SR technique in condensed matter physics}
$\mu$SR is a useful method to characterize solid-state magnetism. The muon acts as a sensitive local probe of the immediate magnetic environment. After implantation, the muon spin precesses and relaxes. These spin dynamics reflect the internal magnetic fields. The magnetic coupling between the muon and nearby electrons governs the observed signal.
This interaction can be described by the following Hamiltonian:
\begin{equation}
    \mathcal{H} = \frac{1}{2 m_e} (\vec{p} + {e \vec{A}})^2 - g_e \mu_B \vec{S} \cdot (\nabla \times \vec{A}) + V(r)~,
\end{equation}
where the first term represents the kinetic energy of the electron in the presence of the vector potential $\vec{A}$.
The second term accounts for the Zeeman interaction between the electron's spin and the magnetic field, while $V(r)$ denotes the potential energy.
The vector potential $\vec{A}$ generated by the magnetic moment of the muon $\vec{\mu}_\mu$ can be written as~\cite{jackson2012classical}:
\begin{equation}
    \vec{A} = \frac{\mu_0}{4 \pi} \frac{\mu_\mu \times \vec{r}}{r^3} = \frac{\mu_0}{4 \pi} \left( \nabla \times \frac{\vec{\mu}_\mu}{r}\right)~.
\end{equation}
From this, the full Hamiltonian describing the muon-electron interaction can be expressed as:
\begin{equation}
    \mathcal{H} = \mathcal{H}_0 + \mathcal{H}' = \left( \frac{\vec{p}^2}{2 m_e} + V(\vec{r}) \right) +
    \left(
    \frac{e}{2 m_e} (\vec{p} \cdot \vec{A} + \vec{A} \cdot \vec{p})
    \right) +
    \left(
    g_e \mu_B \vec{S} \cdot (\nabla \times \vec{A})
    \right)~.
    \label{eq:muon_electron_hamiltonian}
\end{equation}
The first term accounts for the electron's kinetic energy and Coulomb potential.
The second term arises from the electron's orbital motion relative to the muon.
It can be reduced to a magnetic dipole interaction form $\vec{\mu}_\mu \cdot \vec{B}_\text{orb}$, where the orbital magnetic field is given by $\vec{B}_\text{orb} = - \frac{\mu_0}{4\pi} \frac{2 \mu_B \vec{L}}{r^3}$, with $\vec{L}$ representing the electron's angular momentum and $\mu_B$ denoting the Bohr magneton.
The third term describes the Zeeman interaction, which depends on the respective spins of the muon and the electron.
As the dipolar field expression becomes singular at the muon's position ($\vec{r} = 0$), the magnetic field is typically decomposed into the following components to resolve this singularity:
\begin{equation}
    B_i = \frac{\mu_0}{4 \pi} \sum_{j = 1}^3(\nabla_i \nabla_j - \frac{1}{3} \nabla^2 \delta_{ij}) \frac{\mu_{\mu, j}}{r} + \frac{2}{3} \mu_0 \mu_{\mu,i} \sigma(\vec{r})~,
\end{equation}
where the first term is the dipolar field that remains finite at $\vec{r} = 0$, and the second term is the contact field arising from a finite electron spin density at the muon site.
In solid-state materials, electronic interactions are typically complex.
While the orbital contribution to the local field has not been observed experimentally, the dipolar contribution is well established.
This dipolar field can arise from both localized and conduction electrons.
Localized electrons generate dipolar fields through their movement around lattice sites, and the total dipolar field is the sum of all these contributions.
The contact interaction in solids stems from the spin density of conduction or localized electrons at the muon site.
This interaction can be affected by intrinsic magnetic ordering or external magnetic fields.
Additionally, the Ruderman--Kittel--Kasuya--Yosida (RKKY) interaction contributes to the contact term~\cite{ruderman1954indirect, kasuya1956theory, yosida1957magnetic}.
The RKKY interaction represents an indirect exchange interaction mediated by conduction electrons.
In summary, the muon primarily probes local magnetic fields arising from the dipolar and contact fields of localized and conduction electrons.
The spin precession frequency of muons can serve as a sensitive probe of the static and dynamic magnetic properties of solid-state systems.

$\mu$SR spectroscopy provides a unique means to isolate the contact contribution to the local magnetic field, particularly in elemental ferromagnets such as nickel (Ni) and iron (Fe)~\cite{ruegg1981muon}.
In these systems, the contact interaction often dominates the magnetic environment at the muon site, as the dipolar contribution may vanish due to lattice symmetry.
Such measurements yield critical insights into local spin density and interstitial magnetization.
In Ni, the face-centered cubic (FCC) structure causes the dipolar field to cancel out at high-symmetry interstitial sites, allowing for a direct probe of the contact term.
In contrast, while the body-centered cubic (BCC) structure of Fe introduces non-zero dipolar fields at individual sites, the spatial average over equivalent muon stopping positions often results in a vanishing net dipolar contribution.
Furthermore, the sensitivity of $\mu$SR to contact interactions extends its utility to the study of non-centrosymmetric materials, where experimental studies have consistently validated theoretical predictions~\cite{onuorah2018muon, huddart2021intrinsic}.

\begin{figure}[t!]
    \centering
    \sidesubfloat[]{\includegraphics[width=0.45\textwidth]{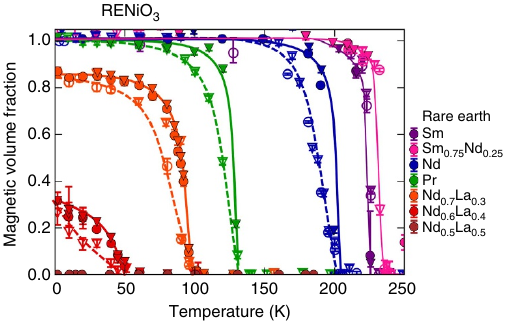}}
    \hfill
    \sidesubfloat[]{\includegraphics[width=0.45\textwidth]{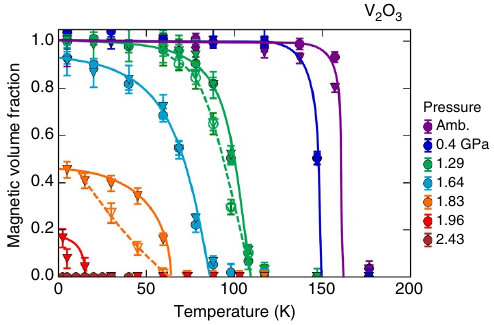}}
    \caption{Temperature-dependent $\mu$SR measurement of magnetic volume fraction in $\text{RENiO}_3$ and $\text{V}_2 \text{O}_3$ (reproduced from Ref.~\cite{frandsen2016volume}).
        (a) Magnetic volume fraction of $\text{RENiO}_3$ under different rare-earth elements as a function of temperature;
        (b) Magnetic volume fraction of $\text{V}_2 \text{O}_3$ under different pressures as a function of temperature.}
    \label{fig:tem-muSR}
\end{figure}

Beyond characterizing local contact fields, the application of $\mu$SR under external magnetic fields is a powerful tool to investigate magnetic volume fractions and phase transitions. The magnetic field is typically a weak transverse field ranging from 5 to 10~$\mathrm{mT}$. A distinct advantage of $\mu$SR is its model-independent nature, which allows for the precise determination of the magnetic fraction without requiring prior assumptions about the magnetic structure of materials.
In a characteristic temperature-dependent measurement, the experiment begins in the paramagnetic state, where the internal field at the muon site is negligible, and the spin precession is governed solely by the external field. As the material is cooled through a magnetic transition, the development of a spontaneous internal field leads to a reduction in the oscillation amplitude (or initial asymmetry) of the $\mu$SR signal. By monitoring this amplitude across a range of temperatures, the evolution of the magnetic volume fraction can be quantitatively determined. Furthermore, the muon spin precession is robust against thermal fluctuations.
Therefore, $\mu$SR is suitable for probing magnetism at mK temperatures. Applications include the study of quantum phase transitions in Mott insulators~\cite{frandsen2016volume}. As shown in~\cref{fig:tem-muSR}, $\mu$SR spectroscopy measures the magnetic volume fraction at various temperatures.
For instance, $\mu$SR spectrometers equipped with pressure cells enable the study of magnetic transitions under high-pressure conditions~\cite{amato2004weak, bourdarot2005hidden}. Due to muons' high sensitivity to weak magnetic fields, even organic magnets and systems with extremely small magnetic moments can be effectively investigated~\cite{le1993searching, blundell1995mu+, andreica2001magnetic}.

\begin{figure}[t!]
    \centering
    \includegraphics[width = .9 \linewidth]{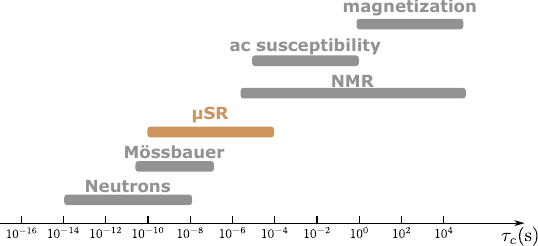}
    \caption{Typical correlation times for various technologies. The orange bar indicates the correlation time range for the $\mu$SR technique.}
    \label{fig:compare_correlation_time}
\end{figure}

With a characteristic time window of $\sim 10~\mu\mathrm{s}$, $\mu$SR is sensitive to magnetic fluctuations with correlation times $\tau_c$ spanning $10^{-4}$ to $10^{-10}~\mathrm{s}$. As shown in \cref{fig:compare_correlation_time}, this extensive range positions $\mu$SR as an important complement to NMR and neutron scattering.
Magnetic fluctuations are typically investigated through two approaches: varying the external field at constant temperature to map the fluctuation spectrum~\cite{aoki2003time, spehling2012magnetic}, or varying the temperature at a fixed field to track dynamical evolution across phase transitions~\cite{kratzer2002musr}. In both cases, the measured muon depolarization rate provides a direct probe of the underlying fluctuation frequency. Beyond scaling behavior, $\mu$SR can resolve fluctuation anisotropy by adjusting the sample orientation relative to the beamline~\cite{hartmann1991asymmetric}. More importantly, the technique can distinguish static from dynamic contributions to the local field, facilitating the study of systems with coexisting orders, such as heavy-fermion compounds~\cite{ran2019nearly, sundar2019coexistence} and uranium-based materials~\cite{de1995absence, keizer1999magnetism, amato2004weak}.

While $\mu$SR provides unique insight into local static and dynamic magnetic behavior, its full potential is often realized when combined with other experimental techniques.
These complementary approaches enable the exploration of fluctuation time scales and ordering phenomena that extend beyond the intrinsic time window of $\mu$SR, providing a more comprehensive understanding of magnetic states in complex materials.
For instance, in materials like $\text{UPt}_3$, conventional techniques detect magnetic ordering~\cite{aeppli1988magnetic}, but $\mu$SR observes no coherent precession signal~\cite{de1995absence}.
This indicates that the magnetic fluctuations occur on timescales faster than those accessible to $\mu$SR (i.e., above 10 MHz), suppressing muon spin precession while still allowing quasi-static magnetic order to be detected by neutron scattering and other instantaneous probes. Furthermore, $\mu$SR is particularly well suited for the investigation of complex magnetic states.
In spin-glass systems, it can track the spin freezing process.
In incommensurate magnetic materials, $\mu$SR can reconstruct the distribution of local fields and infer the magnetic structure~\cite{campbell1994dynamics}.
Field-dependent $\mu$SR experiments can also probe the Knight shift, providing insight into the local magnetic susceptibility and electronic correlations~\cite{camani1979measurement}.
In summary, $\mu$SR offers a diverse and sensitive platform for investigating both static and dynamic magnetic properties in a wide variety of solid-state systems.

Apart from magnetism, $\mu$SR has also been widely applied to the study of superconductivity.
In most cases, superconductivity and magnetism are regarded as mutually exclusive phenomena.
The superconducting state occurs below the critical temperature $T_c$.
In an ideal superconducting state, the local magnetic field at the muon site is zero.
However, this does not imply that a superconductor behaves like an ideal conductor.
Unlike an ideal conductor, which only freezes magnetic flux lines, a superconductor expels them entirely due to the Meissner effect, resulting in a zero internal magnetic field~\cite{meissner1933neuer}.
Additionally, superconductors exhibit intrinsic diamagnetism and therefore repel magnetic fields irrespective of their strength.
The phenomenon of superconductivity was first described by the London equations~\cite{london1935electromagnetic} and later formalized based on Ginzburg-Landau theory~\cite{ginzburg2009theory}.
It was eventually given a microscopic foundation by BCS theory~\cite{bardeen1957theory}.
Throughout the long history of superconductivity research, the critical temperature, denoted $T_c$, has steadily increased with advancements in the field, including the discovery of iron- and nickel-based superconductors.
As of now, the highest reported critical temperature is $203~\mathrm{K}$, which was observed in hydrogen sulfide ($\mathrm{H_3S}$) under a pressure of $150~\mathrm{GPa}$~\cite{drozdov2015conventional}.
The $\mu$SR technique enables the experimental determination of key superconducting parameters, including the critical temperature ($T_c$), the magnetic penetration depth ($\lambda(T)$), the coherence length ($\xi$), and the superconducting gap ($\Delta(T)$).
As previously mentioned, the $\mu$SR technique enables precise probing of the local magnetic field within materials. Consequently, it can effectively study the coexistence and competition between superconductivity and magnetism.

Key superconducting parameters include the magnetic penetration depth $\lambda(T)$ and the coherence length $\xi$.
The magnetic penetration depth $\lambda(T)$ indicates how far an external magnetic field can penetrate a superconductor, reflecting the density of superconducting electrons.
In contrast, the coherence length $\xi$ measures the length scale over which the superconducting order parameter changes significantly, corresponding to the spatial extent of a Cooper pair in BCS theory.
However, its definition may vary across different theoretical frameworks.
In the superconducting state, the external magnetic field rapidly decreases to zero over a very short distance.
Within this distance, referred to as the Ginzburg--Landau length, denoted as $\xi_\text{GL}$, the superconducting order parameter and the density of superconducting electrons transition from zero to their bulk values.
The ratio of the magnetic penetration depth to $\xi_\text{GL}$ is known as the Ginzburg--Landau parameter, $\kappa$.
Based on the value of $\kappa$, superconductors can be categorized into two types:
\begin{itemize}
    \item Type I superconductors, where $\kappa < 1/\sqrt{2}$, have a positive surface energy, resulting in the complete expulsion of the external magnetic field from the material.
    \item Type II superconductors, with $\kappa > 1/\sqrt{2}$, exhibit a negative surface energy. This means that the formation of interfaces between normal and superconducting regions is energetically favorable.
\end{itemize}
Consequently, when the external magnetic field is between the lower and upper critical fields ($B_{c1} < B_\text{ext} < B_{c2}$), magnetic flux enters the material in the form of quantized vortices.
Each vortex carries a single quantum of magnetic flux and features a standard core.
The $\mu$SR technique effectively probes the internal magnetic field distribution within vortices, allowing for a detailed investigation of vortex lattice structures and dynamics in type II superconductors.
Typically, the experimental measurement of vortices is performed using the transverse-field $\mu$SR (TF-$\mu$SR) technique, with a superconductor in the normal state under an external magnetic field.
In this state, the local magnetic field at each muon site is essentially uniform, so all muon spins precess at the same frequency, producing a strong, coherent oscillation signal.
As the sample cools into the superconducting state, vortices begin to form, and the magnetic field becomes spatially inhomogeneous due to the contribution from vortex cores.
The muon spin precession frequency varies slightly depending on each muon's local environment, leading to gradual depolarization of the signal.
Upon further cooling well below the critical temperature, the magnetic inhomogeneity becomes more obvious, leading to rapid muon spin depolarization.
The precession signal broadens significantly, reflecting the wide distribution of local magnetic fields associated with the vortex lattice.
By analyzing the resulting muon precession spectrum, one can extract microscopic information such as the vortex density~\cite{Amato:1997zz}.

Analyzing the muon precession spectrum relies on the fitting method used.
For polycrystalline materials, the local magnetic field distribution is approximately symmetric and can be modeled using a Gaussian function~\cite{bauer2014absence, bauer2010unconventional}.
In this case, the $\mu$SR spectrum is typically fitted with the function:
\begin{equation}
    A_x(t) = A_\text{sample} \exp(\frac{\sigma^2 t^2}{2}) \cos (\gamma_\mu \langle B_{\mu,z} \rangle + \phi) +
    A_\text{bg} \cos(\gamma_\mu \langle B_\text{bg} \rangle t + \phi)~,
\end{equation}
where the first term represents the signal from the sample: $A_\text{sample}$ is the initial asymmetry, $\sigma$ is the Gaussian relaxation rate corresponding to the width of the magnetic field distribution, $\langle B_{\mu,z} \rangle$ is the average local field at the muon site, and $\phi$ is the initial phase.
The second term represents the background signal, which often originates from muons stopping in the sample holder or beamline components.
For high-quality single crystals, where the field distribution is more structured, the spectrum is better described by a sum of multiple Gaussian components~\cite{sonier2000musr, maisuradze2009comparison}:
\begin{equation}
    A(t) = \sum_{i=1}^N A_i \exp(- \sigma^2 t^2 / 2) \cos (\gamma_\mu B_{i} t + \phi)~,
\end{equation}
where each term corresponds to a distinct local magnetic environment.
If a theoretical model of the internal field distribution is available, a full model analysis can be conducted.
This analysis involves fitting the spectrum directly using the model-predicted time evolution of muon polarization.

Using these analytical techniques, $\mu$SR measurements have shown that the emergence of weak internal magnetic fields often occurs alongside the superconducting transition.
This was famously established in high-$T_c$ and iron-based superconductors~\cite{sanna2004nanoscopic, luetkens2009electronic}, and has been more recently identified in the topological Kagome systems~\cite{Mielke2022Nature}.
Moreover, since zero-field $\mu$SR is sensitive to small changes in internal fields ($\sim 10~\mathrm{\mu T}$), it is uniquely suitable for investigating broken time-reversal symmetry (TRSB) in the superconducting state.
Although TRSB superconductors are rare, this phenomenon has been found in a growing number of materials through $\mu$SR, ranging from the pioneering studies on $\text{Sr}_2\text{RuO}_4$ and $\text{UPt}_3$~\cite{Luke:1998394558, Hillier:2009PhysRevLett.102.117007, Aoki:2003PhysRevLett.91.067003} to the recent discoveries in Rhenium-based and Uranium-based heavy-fermion compounds~\cite{Shang2018PRL, Sundar2023PRB, Ghosh2021JPCM}.
Additionally, temperature- and pressure-dependent $\mu$SR experiments allow for the precise determination of critical magnetic fields and the evolution of the order parameter across complex phase diagrams~\cite{guguchia2023tunable}.

\subsubsection{Muonium in \texorpdfstring{$\mu$}{mu}SR}
The applications of $\mu$SR introduced above primarily rely on the properties of free muons.
However, within matter, muons do not always remain in their free state, and form bound states instead.
The bound muon state can either be muonium or a muonic atom.
By detecting their decay products, one can investigate various properties at the atomic scale.
In this section, the fundamental mechanisms of muonium and its applications in $\mu$SR will be introduced.
According to the muonium basic properties which have been introduced in~\cref{sec:precise_muonium_spectroscopy}, the external magnetic field will lead to time-dependent muonium polarization in the material, which can be described as
\begin{equation}
    \vec{P}(t) = \frac{1}{4} \sum_{k,n} \bra{k} \vec{P}(0) \cdot \vec{\sigma}_\mu \ket{n} \bra{n} \vec{\sigma}_\mu \ket{k} e^{-i \omega_{k,n} t}~,
\end{equation}
where $\ket{k}$ and $\ket{n}$ are the hyperfine eigenstates of muonium system, $\omega_{k,n} = (E_k - E_n) / \hbar$ is the transition frequency.
Similar to the hyperfine splitting under the external magnetic field, the specific polarization time evolution depends on the strength and the direction of the external magnetic field.
In a longitudinal magnetic field, the polarization is either along or opposite to the initial polarization direction.
Then, the polarization time evolution can be expressed as
\begin{equation}
    P_z (t) = \frac{1}{2} \left[
        1 + \frac{1}{1 + x_B^2} \left[
            x_B^2 + \cos (\omega_{24} t)
            \right]
        \right]~,
\end{equation}
where $x_B = g_\mu \mu_B^\mu B_\text{ext} / (2 \hbar \omega_0)$ is the dimensionless magnetic field, $\omega_{24}$ is the transition frequency between the muonium states $\ket{2}$ and $\ket{4}$.
The transition frequency can be expressed as $\omega_{24} = \sqrt{\omega_0^2 + {(\omega_e + \omega_\mu)}^{2}} = \omega_0 \sqrt{1 + x_B^2}$, where $\omega_0 = g_e \mu_B B_\text{ext} / \hbar$ is the Larmor frequency.
Therefore, if the external magnetic field is zero, the muonium polarization time evolution is only contributed by the Larmor precession $P_z(t, B_\text{ext}=0) = [1 + \cos (\omega_0 t)] / 2$.
If the external magnetic field is perpendicular to the polarization direction, the muonium will be regularized.
The regularization will lead to a non-oscillation part of the muonium polarization time evolution $P_z^\text{non-osc} = \{1 + [x_B^2 / (1 + x_B^2)]\} / 2$.
If the external magnetic field is transverse, the muonium polarization time evolution is more complicated.
In general, the polarization time evolution under the transverse field can be expressed as
\begin{equation}
    P_x (t) = \frac{1}{2} \left[
        \cos^2 \delta [\cos (\omega_{12} t) + \cos(\omega_{34} t)] + \sin^2 \delta [\cos(\omega_{14} t) + \cos(\omega_{23} t)]
        \right]~,
\end{equation}
where the $\omega_{ij}$ is the transition frequency between the muonium states $\ket{i}$ and $\ket{j}$.
Due to the limited time resolution of the $\mu$SR spectroscopy, the $\ket{1}\to \ket{3}$ and $\ket{2} \to \ket{4}$ transitions do not contribute to the polarization time evolution.
The specific form of the polarization time evolution depends on the strength of the external magnetic field.
Under a very weak magnetic field, which is $x_B \ll 1$, the transition frequencies $\omega_{12}$ and $\omega_{23}$ are nearly degenerate, leading to simplified dynamics.
Therefore, the polarization time evolution is a simple oscillation $P_x(t, x_B \ll 1) = \cos (\omega_\text{Mu}^T t) / 2$, where $\omega_\text{Mu}^T = (\omega_e - \omega_\mu) / 2$ is the effective transition frequency in the transverse field.
If the external magnetic field is slightly increased, typically around $10\text{--}20~\mathrm{mT}$, the hyperfine interaction becomes significant.
However, $x_B$ is still small, the polarization time evolution is no longer a simple oscillation but a combination of two frequencies.
\begin{equation}
    P_x(t, x_B \ll 1) = \frac{1}{2} \left[
        \cos^2 \delta \cos (\omega_{12} t) + \sin^2 \delta \cos(\omega_{14} t)
        \right]~,
\end{equation}
where $\cos \delta$ and $\sin \delta$ are the coefficients of the linear combination of the muonium states.
If a new frequency $\omega_\pm = (\omega_{23} \pm \omega_{12}) / 2$ is defined, the polarization time evolution can be expressed in a more compact form
\begin{equation}
    P_x(t, x_B \ll 1) = \frac{1}{2} \cos (\omega_+ t) \cos (\omega_- t) + \frac{1}{2} \frac{x_B}{\sqrt{1 + x_B^2}} \sin (\omega_+t) \sin(\omega_- t)~.
    \label{eq:muonium_polarization_time_evolution}
\end{equation}
According to the~\cref{eq:muonium_polarization_time_evolution}, the polarization time evolution shows a beating spectrum with two frequencies $\omega_+$ and $\omega_-$.
If the external magnetic field is further increased, typically $x_B \gg 1$.
In this case, the coefficients $\cos \delta \to 1$ and $\sin \delta \to 0$.
The polarization time evolution can be expressed as
\begin{equation}
    P_x(t, x_B \gg 1) = \frac{1}{2} \cos (\omega_{12} t) + \frac{1}{2} \cos(\omega_{34} t)~.
\end{equation}
In this case, the transition frequencies $\nu_{12} = \nu_0/2 - \nu_\mu$ and $\nu_{34} \to \nu_\mu + \nu_0 / 2$.
When $\nu_\mu < \nu_0 / 2$, the sum of two observable frequencies yields the hyperfine coupling of muonium by $\nu_{12} + \nu_{34} = \nu_0$.
If $\nu_\mu > \nu_0 / 2$, the hyperfine coupling constant can be obtained by $\nu_{34} - |\nu_{12}|$.
Except for the hyperfine interaction, the muonium can also interact with the surrounding nuclei.
The nuclear hyperfine interaction splits muonium energy levels, resulting in a complex frequency spectrum with many weak lines that cannot be resolved.
If both the hyperfine interaction and the resulting muonium spin evolution can be treated as isotropic, the system is referred to as isolated muonium.
This situation typically arises in nonconducting materials, where the $\mu^+$ captures an electron at an interstitial site.
Such isolated muonium states are only found when the free-electron density around the muon is sufficiently low or when the muon occupies a highly symmetric lattice site.
According to Cox's work~\cite{cox2003shallow}, the hyperfine constant inside solid state materials is smaller than that of free muonium.
In elemental semiconductors, the hyperfine constant is typically about one-half of the free-muonium value.
More generally, the hyperfine interaction need not be isotropic.
In materials where the electronic environment is anisotropic, the hyperfine coupling is described not by a single constant $A$ but by a tensor $\tilde{A}$.
Such anisotropic muonium states have been reported in several semiconductors through the $\mu$SR technique~\cite{cox2003shallow,cox2009muonium,patterson1988muonium}.

Muonium can be a probe of hydrogen in physical chemistry.
It behaves chemically like hydrogen, while being easier to generate and more stable under extreme conditions.
Its light mass produces an obvious isotope effect, revealing reaction mechanisms such as tunneling-dominated addition reactions and inverse kinetic isotope effects in abstraction processes.
This capability has been demonstrated in classic kinetic studies~\cite{Ghandi:2003muonium, Percival:2007Hkinetics}, and more recently utilized to test quantum reaction rate theories and vibrational bonding phenomena that are inaccessible to heavier isotopes~\cite{Fleming2021PCCP}.
Furthermore, $\mu$SR enables the in situ formation and detection of radicals, providing access to their structures and dynamics through hyperfine interactions.
This remains possible even at high temperatures, in supercritical fluids, or on very short timescales where conventional EPR techniques struggle.
The method applies to a wide range of chemical environments, from gas-phase kinetics to soft matter.

As we discuss in this section, $\mu$SR is a sensitive method for detecting the magnetic properties of materials.
In addition, several procedures have been developed to precisely estimate the muon stopping point.
These include MUESR~\cite{Herak:2013xxx}, DFT+$\mu$~\cite{Moller:2013vqa, Bernardini:2013xxx, Moller:2013hideandseek, Bonfa:2016xxx}, and a density functional theory-based technique using Ab Initio Random Structure Searching~\cite{Sturniolo:2020xxx, Liborio:2018xxx}.
These methods have been successfully applied across a wide range of systems, from inorganic and molecular magnets to organic molecules.
Meanwhile, silicon pixel detectors are being employed to develop the next generation of $\mu$SR detectors.
Multilayers of silicon pixel detectors allow reconstruction of the muon stopping point and improve the maximum muon counting rate~\cite{Isenring:2025gdk, Mandok:2025sen}.
Overall, the combination of $\mu$SR with modern computational techniques and advanced detector technology provides a unique synergy.
It allows subtle local phenomena to be captured while enabling rational interpretation of spin-relaxation and rotation data at higher event rates.

The applications of positive muon beams clearly demonstrate the advantages of $\mu^+$SR and muonium-based spectroscopy in probing electronic, magnetic, and chemical environments.
However, the full potential of muon-beam techniques extends beyond positive muons.
Negative muons, through atomic capture and subsequent nuclear processes, offer element-specific analysis, local chemical information, and access to material systems where $\mu^+$SR is not feasible.
Therefore, after summarizing positive-muon techniques, it is natural to look ahead and introduce negative-muon methods, such as $\mu^-$SR and muonic X-ray emission (MIXE), to provide a more complete picture of the capabilities of muon-beam applications.

\subsection{Application of negative muon}
Negative muons are produced by the decay of negative pions, the charge conjugates of positive pions.
Generally, negative muon beams have lower intensities than positive ones.
On the one hand, negative pion production in proton-nucleus interactions is inherently less efficient. On the other hand, a portion of the negative pions is captured by target nuclei before decaying, further reducing the yield.
In the material, the $\mu^-$ particles lose energy through Coulomb excitation and ionization of target electrons until their velocity becomes comparable to that of atomic electrons.
Following this slowing-down process, the $\mu^-$ may remain free or be captured by a nucleus to form a muonic atom.
The $\mu^-$SR technique is based on measuring the $\mu^-$ polarization, while the muon-induced X-ray emission (MIXE) technique utilizes the X-rays emitted from the muonic atom to provide information about the nucleus and the surrounding electronic structure.
This section provides an overview of $\mu^-$SR and MIXE, along with their respective applications.

\begin{figure}[t!]
    \centering
    \includegraphics[width = .4 \linewidth]{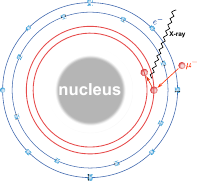}
    \caption{Schematic diagram of the cascade process of a muonic atom. The red circle represents the muon and its orbits, while the blue circle represents the electron and its orbits. The wavy line symbolizes the X-ray photon emitted during the muon cascade process.}
    \label{fig:MIXE_technique}
\end{figure}

During implantation, the $\mu^-$ loses most of its energy through electromagnetic interactions with the target material.
Once sufficiently slowed, it is captured by the nucleus, forming a muonic atom and immediately entering a rapid cascade down to the 1S ground state as shown in \cref{fig:MIXE_technique}.
Using the Bohr model~\cite{PhysRev.177.164, PhysRevC.71.058501}, this cascade can be described in terms of atomic radii, where the radius of the orbit with quantum number $n$ is given by
\begin{equation}
    r_{n, \mu} = \frac{4 \pi \varepsilon_0 \hbar^2 n^2}{\bar{m}_{\mu A} Z e^2} = \frac{\hbar n^2}{\bar{m}_{\mu A} Z c \alpha} \simeq \frac{m_e}{\bar{m}_{\mu A}}r_{n, e}~,
\end{equation}
where the $A$ and $Z$ are the mass number and atomic number of the nucleus, $\bar{m}_{\mu A} = m_\mu m_A / (m_\mu + m_A) \sim m_\mu$ is the reduced mass of the muonic atom, $c$ is the speed of light, $\alpha$ is the fine structure constant, and $r_{n, e} = n^2 a_0$ is the Bohr radius of the electron.
The energy of the muonic atom can be expressed as
\begin{equation}
    E_{n, \mu} = \frac{\bar{m}_{\mu A} Z^2 e^4}{{(4 \pi \varepsilon_0)}^2 2 n^2 \hbar^2} = -\frac{1}{2} \frac{{(Z \alpha)}^2 \bar{m}_{\mu A} c^2}{n^2} \simeq \frac{\bar{m}_{\mu A}}{m_e} E_{n, e}~,
\end{equation}
where the energy $E_{n,\mu}$ is about 207 times larger than the corresponding electron energy $E_{n,e}$.
During the cascade process, the $\mu^-$ rapidly transitions to the lower atomic energy levels; the primary energy emission is in the form of characteristic X-rays or, in some cases, Auger electrons. The energies of these muonic X-rays typically range from several $\mathrm{keV}$ up to a few $\mathrm{MeV}$, and are highly element-specific due to their dependence on the atomic number $Z$. This property makes muonic X-ray emission a powerful tool for elemental analysis and depth profiling in materials science.
Analogous to the empirical Moseley's law~\cite{moseley1913xciii}, which describes the relationship between atomic number and X-ray emission energies for electronic transitions, the energies of muonic X-ray transitions can be similarly expressed as
\begin{equation}
    E_{i\to f,\mu} = \frac{\bar{m}_{A \mu}}{m_e} R_y {(Z - S_{\text{scr},\mu})}^2 \left(
    \frac{1}{n_f^2} - \frac{1}{n_i^2}
    \right)~,
\end{equation}
where $R_y = 13.6~\mathrm{eV}$ is the Rydberg energy, $S_{\text{scr},\mu}$ is the screening constant for a muonic atom, and $n_i$ and $n_f$ are the principal quantum numbers of the initial and final states, respectively.
With the Muon-Induced X-ray Emission (MIXE) technique, one can measure the X-ray spectrum emitted by the target material for elemental analysis.
During the cascade process, $\mu^-$ will be strongly depolarized due to the spin-orbit coupling and the hyperfine interaction with the non-zero nuclear spin.
The remaining polarization of the $\mu^-$ is typically very low, which makes the construction of the $\mu^-$SR technique not as efficient as the $\mu^+$SR.

After the cascade process, the $\mu^-$ will stop in the 1S state of the muonic atom.
In this state, the $\mu^-$ is extremely close to the nucleus, which leads to a significant increase in the probability of muon capture by the nucleus.
In the primary capture process, which is governed by the weak interaction, $\mu^-$ is captured by a bound proton in the nucleus~\cite{measday2001nuclear,mukhopadhyay1977nuclear}.
\begin{equation}
    \mu^- + p \to n + \nu_\mu~.
\end{equation}
The capture probability increases with the nuclear charge $Z$, for $Z \gtrsim 7$ approximately as $Z^4$.
Various theoretical and experimental studies have investigated the capture process of $\mu^-$ by the nucleus and provided corrections to the capture probability law.
After capture, the proton is converted into a neutron, leaving the nucleus in an excited state.
The excited nucleus will then be de-excited by emitting an X-ray, and the characteristic X-ray can also be used for elemental analysis.
Due to the significant increase in capture probability, the lifetime of $\mu^-$ is considerably shortened compared to its lifetime in free decay.
The reduced lifetime restricts the available time window for the muon to interact inside the material, which makes the construction of the $\mu^-$SR technique not as efficient as the $\mu^+$SR technique.
Based on the above discussion, the muonic atom is less suitable for $\mu$SR, but it can be used for elemental analysis.

\subsubsection{$\mu^-$SR technique}
In $\mu^-$SR, through the Michel electron detection, one actually measures the polarization of the muon captured by the nucleus.
The strong depolarization during the cascade process results in a very low polarization of the $\mu^-$.
Consequently, the $\mu^-$SR technique presents more challenges in measurement and analysis compared to the $\mu^+$SR technique.
However, the $\mu^-$SR presents some unique advantages in the study of hydrogen-related dynamics.

In material studies involving hydrogen diffusion and desorption, $\mu^+$ and muonium are not ideal probes.
This is because muonium shares many properties with hydrogen, leading to diffusion and desorption behaviors that are comparable.
Consequently, $\mu^+$SR measurements cannot reliably distinguish whether the observed effects are due to muonium or hydrogen.
In contrast, $\mu^-$SR can be used to study the hydrogen properties in materials.
First, $\mu^-$ captured by the nucleus will not be affected by the muon diffusion.
Second, for the nucleus with no spin, the muon polarization is only affected by the surrounding hydrogen spin and is sensitive to possible hydrogen diffusion.
Using the zero-field, transverse-field, and longitudinal-field $\mu^-$SR techniques, Sugiyama \textit{et al.}~\cite{PhysRevLett.121.087202} have successfully studied the hydrogen diffusion in $\mathrm{MgH_2}$.
By coincident detection with characteristic X-ray emission, the $\mu^-$SR technique can also distinguish the contributions of different elements.

\subsubsection{Muon Induced X-ray Emission (MIXE) technique}
MIXE is a nondestructive technique for elemental analysis, particularly in materials science.
By utilizing high-energy X-rays emitted during the muon-nucleus interactions, MIXE enables depth-dependent characterization of elemental composition.
These features make MIXE potentially suitable for specialized applications in fields such as biology and archaeology, where nondestructive analysis is crucial.

First, the muon has a strong penetrating power due to its relatively high mass, allowing it to reach deeper into a material than electrons or conventional X-rays.
Second, the muonic X-rays emitted during the atomic cascade have significantly higher energies than electron-transition X-rays.
High-energy photons can escape from the material more easily, increasing the efficiency of the MIXE technique.
For MIXE measurements, the emitted muonic X-rays are typically recorded using high-purity germanium (HPGe) detectors, which provide both the energy and the arrival time of the photons.

\begin{figure}[t!]
    \centering
    \sidesubfloat[]{\includegraphics[scale=0.45]{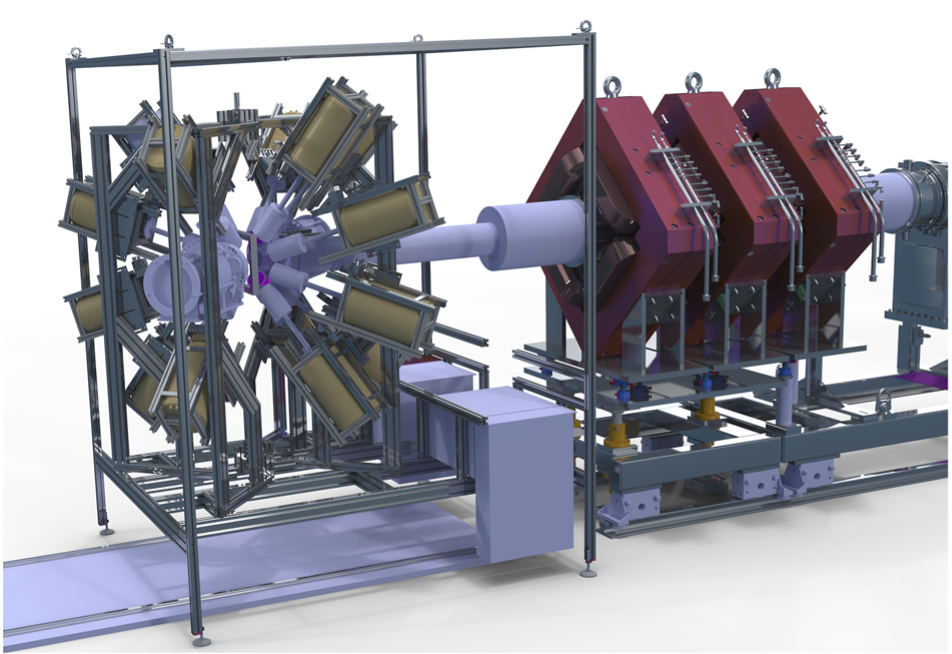}}
    \sidesubfloat[]{\includegraphics[scale=0.45]{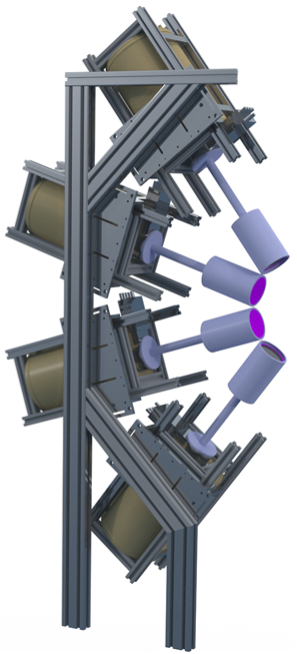}}
    \caption{Schematic diagram of the MIXE spectrometer operating at PSI. Panel (a) shows the support frame, HPGe detector, and sample system;
        Panel (b) shows the single detector arm structure of the spectrometer (reproduced from Ref.~\cite{Gerchow:2022zcw}).}
    \label{fig:PSI_MIXE}
\end{figure}

A schematic diagram of the PSI MIXE spectrometer is shown in~\cref{fig:PSI_MIXE}, illustrating a representative configuration of the spectrometer frame, detection unit, and sample environment.
The surrounding HPGe array ensures a high energy resolution for muon-induced X-rays.
The characteristic X-ray energies enable element identification, while their intensities provide quantitative information on elemental composition.

With the MIXE technique, we can obtain the elemental composition and distribution in the rare and precious materials.
For example, in the study of the formation and evolution of carbonaceous asteroid Ryugu~\cite{doi:10.1126/science.abn8671}, the MIXE technique has been used to analyze the elemental composition of the samples returned by the Hayabusa2 mission.
Meanwhile, the MIXE technique can also be used to study the elemental composition of the battery materials~\cite{brown_depth_2018,ninomiya_non-destructive_2022,aramini_using_2020,rossini_new_2023}.
The MIXE technique is now widely used across various fields and is expected to play an even more important role in the future.

\subsection{Perspectives on muon applications}
Recent progress on muon source and detector technology has broadened the application of muon science from a limited field to a research platform covering varying research areas.
Positive muons have become an essential tool for the study of local magnetism, superconductivity, and chemical reaction mechanisms, represented by the $\mu$SR and muonium chemistry.
Since the muon spin precession can reflect the weak internal magnetic field and dynamic fluctuation, it fills in the gap left by other spectroscopic techniques such as NMR and neutron scattering.
On the other hand, negative muon techniques, such as MIXE and $\mu^-$SR, offer complementary capabilities, enabling non-destructive elemental analysis and the investigation of hydrogen diffusion in complex environments.

With the construction of next-generation muon facilities, the beam intensity and quality will be further improved.
These superior beam performances will bring further challenges and opportunities for the muon applications.
This evolution will not only refine our understanding of condensed matter systems but also open up novel avenues in archaeology, battery research, and fundamental physics.
Meanwhile, the high-intensity muon beams will bring new challenges for the detector technology.
For example, the high counting rate may lead to the pile-up effect, which will significantly degrade the detector performance.
Therefore, the development of new detector technology is essential to fully exploit the advantages of high-intensity muon beams.
In conclusion, the combination of advanced beams and next-generation detectors promises a bright future for muon applications.

\section{Summary and outlook}
This article reviews the main experimental efforts in muonium physics and other progress in the context of muon beams, covering both historical measurements and ongoing projects. A brief overview of existing muon facilities worldwide and future construction plans is also provided. Muon is characterized by a mass (105.66 MeV$/c^2$) 200 times larger than that of an electron and a relatively long lifetime (2.2~$\mu$s), and it can be easily produced in the laboratory with high intensity and polarization. During the interaction with certain material, a muon can form muonium, which is a bound state largely free from quantum chromodynamic effects, making it an ideal subject for precise tests of QED. Moreover, experiments involving intense muon beams are sensitive to new physics, such as the muonium-to-antimuonium conversion, a $\Delta L_\ell=2$ cLFV process. The discovery of cLFV would provide unambiguous evidence of new physics beyond the Standard Model and may shed light on fundamental questions such as the mechanism of neutrino mass, non-standard neutrino interactions, and the origin of matter-antimatter asymmetry in the Universe. Additionally, muons and muonium serve as powerful probes of the intrinsic magnetic properties of materials.

Research on various scenarios involving accelerator muon beams has continued for nearly half a century. The demand for high-intensity muon beams is rapidly increasing. Therefore, several countries, especially China, are currently planning and constructing next-generation muon facilities that can drive future experiments to higher precision. In summary, muonium, as a unique system, can support physics motivations across particle physics, nuclear physics, and materials science. Understanding of fundamental symmetries in nature also drives innovation in beam technology, particle detection, and advanced computing. By leveraging world-class infrastructure, muon beams towards muonium physics across the road exemplifies how fundamental research can stimulate technological progress and international collaboration. With the construction of new muon facilities, the studies of muonium are expected to constantly provide exciting contributions to the particle physics community and the related areas.

\section*{Acknowledgments}
The authors are especially grateful to Klaus P. Jungmann, Yoshitaka Kuno, and Jian Zhang for the substantial time and effort they devoted to revising this manuscript. Their insightful comments and careful revisions played an important role in improving the structure, clarity, and overall quality of the paper. The authors would like to acknowledge the MACE Working Group for their invaluable discussion, and in particular, Shihan Zhao and Guihao Lu for their contributions to the preparation of several figures. This work was supported by the National Natural Science Foundation of China under Grant Nos. 12347105 and 12075326, Guangdong Basic and Applied Basic Research Foundation under Grant No. 2025A1515010669, the Natural Science Foundation of Guangzhou under Grant No. 2024A04J6243, and Fundamental Research Funds for the Central Universities (23xkjc017) in Sun Yat-sen University. This research was supported by the Munich Institute for Astro-, Particle and BioPhysics (MIAPbP) which is funded by the Deutsche Forschungsgemeinschaft (DFG, German Research Foundation) under Germany's Excellence Strategy--EXC-2094--390783311. J.~T. is grateful to the organization of the invited seminar by Kim-Siang Khaw at Tsung-Dao Lee Institute in Shanghai JiaoTong University and the Southern Center for Nuclear-Science Theory (SCNT) at the Institute of Modern Physics in the Chinese Academy of Sciences for hospitality. The present article would not have been possible without the collaboration and multiple discussions with the students. It is a great pleasure to thank them all.

\bibliographystyle{elsarticle-num}
\bibliography{bib}

\end{document}